\documentclass{article}

\usepackage{arxiv}

\usepackage[utf8]{inputenc} 
\usepackage[T1]{fontenc}    
\usepackage[hidelinks]{hyperref}       
\usepackage{url}            
\usepackage{booktabs}       
\usepackage{amsfonts}       
\usepackage{nicefrac}       
\usepackage{microtype}      
\usepackage{amsmath,bm}
\usepackage{cleveref}       
\usepackage{lipsum}         
\usepackage{graphicx}
\usepackage{natbib}
\usepackage{doi}
\usepackage{placeins}

\title{Top-of-atmosphere radiation over the last millennium reconstructed from proxies}

\date{March 22, 2025}

\newif\ifuniqueAffiliation

\ifuniqueAffiliation 
\author{ \href{https://orcid.org/0000-0000-0000-0000}{\includegraphics[scale=0.06]{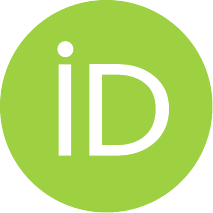}\hspace{1mm}David S.~Hippocampus}\thanks{Use footnote for providing further
		information about author (webpage, alternative
		address)---\emph{not} for acknowledging funding agencies.} \\
	Department of Computer Science\\
	Cranberry-Lemon University\\
	Pittsburgh, PA 15213 \\
	\texttt{hippo@cs.cranberry-lemon.edu} \\
	\And
	\href{https://orcid.org/0000-0000-0000-0000}{\includegraphics[scale=0.06]{orcid.pdf}\hspace{1mm}Elias D.~Striatum} \\
	Department of Electrical Engineering\\
	Mount-Sheikh University\\
	Santa Narimana, Levand \\
	\texttt{stariate@ee.mount-sheikh.edu} \\
}
\else
\usepackage{authblk}

\newbox{\orcid}\sbox{\orcid}{\includegraphics[scale=0.06]{orcid.pdf}} 
\author[1]{%
	\href{https://orcid.org/0009-0003-3206-3735}{\usebox{\orcid}\hspace{1mm}Dominik~Stiller\thanks{Corresponding author: Dominik Stiller, \texttt{dstiller@uw.edu}}}%
}
\author[1]{%
	\href{https://orcid.org/0000-0001-8486-9739}{\usebox{\orcid}\hspace{1mm}Gregory J.~Hakim}%
}
\affil[1]{Department of Atmospheric and Climate Science, University of Washington, Seattle, WA, USA}
\fi

\renewcommand{\headeright}{}
\renewcommand{\undertitle}{}
\renewcommand{\shorttitle}{Top-of-atmosphere radiation over the last millennium reconstructed from proxies}

\begin{document}
\maketitle

\begin{abstract}
	Earth's energy imbalance at the top of the atmosphere is a key climate system metric, but its natural variability is poorly constrained by the short observational record and large uncertainty in coupled climate models. While existing ocean heat content reconstructions offer a longer perspective, they cannot separate the contributions of shortwave and longwave radiation, obscuring the underlying processes. We extend the energy budget record into the pre-industrial period by reconstructing the top-of-atmosphere radiation and related surface variables over the last millennium (850–2000~CE) by using data assimilation to combine proxy data and dynamics from a coupled climate emulator. Validation reveals skill in the reconstructed radiation fields, especially in the tropics. Results show a familiar last-millennium cooling trend, which coincides with persistent heat loss and a reduction in upper-ocean heat content. The cooling trend differs by season and latitude, and is associated with radiative anomalies suggestive of an eastward shift in Indo–Pacific convection. Following large volcanic eruptions, ocean heat content anomalies persist for 10–20 years on average, supporting previous evidence that the cooling trend was forced by decadally-paced eruptions. The reconstruction also reveals that the current rate of energy gain is unprecedented relative to the period before 1850.
\end{abstract}


\section{Introduction}

Earth's energy imbalance (EEI) at the top of the atmosphere (TOA) is a fundamental climate system metric, governing Earth's total heat content and constraining global temperatures, the hydrological cycle, sea levels, and ice cover~\citep{Schuckmann2016}. Much of the EEI is forced, over the last millennium mostly by volcanic eruptions, solar irradiance, and changes in Earth's orbit, and more recently by anthropogenic greenhouse gases and aerosols. In addition, there is internal, unforced variability in the energy budget at all timescales. Seasonal variability is primarily due to extratropical storms and the Madden--Julian oscillation, while the El~Niño--Southern Oscillation (ENSO) explains most interannual variability~\citep{Trenberth2015}. However, decadal to centennial variability in the energy budget remains poorly understood~\citep{Wills2021,Trenberth2014}. This complicates, for example, the interpretation of recent albedo trends~\citep{Goessling2025,Mauritsen2025,Hodnebrog2024} and the warming hiatus~\citep{Trenberth2013a,Xie2016}, both of which are not reproduced well by coupled climate models~\citep{Olonscheck2024,Raghuraman2021,Medhaug2017,Fyfe2013}. A better understanding of natural variability in Earth's energy budget is also essential to isolating the contribution of anthropogenic forcing in observations, which informs the attribution of historical climate change~\citep{Lean2018} and helps narrow uncertainty in climate sensitivity~\citep{Sherwood2020}.

Our understanding of EEI variability is limited by the short satellite observational record of around 25 years~\citep{Loeb2024}, which prohibits the investigation of low-frequency variability and the response to rare, episodic forcings. Even during this well-observed satellite period, discrepancies and uncertainties between EEI estimates are large~\citep{Hakuba2024}. Additionally, it is difficult to disentangle forced and internal variability of Earth's energy budget due to the strong anthropogenic greenhouse gas forcing and the uncertain aerosol forcing during the observed period. Therefore, climate models have been used to investigate aspects of the energy budget like radiative feedbacks~\citep{Sherwood2020} and decadal variability~\citep{Palmer2014,Brown2014,Zhou2016}, more recently through coordinated experiments like CFMIP and CERESMIP~\citep{Webb2017,Schmidt2023}. However, even coupled models of the CMIP6 generation~\citep{Eyring2016} produce a large spread in TOA radiation, on the order of $5$ to $20\,\mathrm{W\,m^{-2}}$ for most components~\citep{Wild2020}, and systematically underestimate the radiative response to surface warming~\citep{Olonscheck2024}. In addition, the specified forcings and simulated internal variability may not be accurate~\citep{Lean2018,Lucke2023,Fyfe2021}. Other studies have extended the energy budget record by reconstructing the ocean heat content (OHC), which is storing much of the energy gained in recent decades~\citep{Schuckmann2023}. For example, \citet{Zanna2019} and \citet{Wu2025a} derive the global-mean OHC over the historical period, while \citet{Gebbie2019} reconstruct the gridded OHC over the Common Era. Further OHC reconstructions from the Last Glacial Maximum to present are reviewed in \citet{Gebbie2021}. The OHC provides a valuable perspective since it is a more reliable indicator of global change than surface temperatures on interannual to decadal timescales~\citep{Palmer2014,Allison2020}. However, the OHC perspective on the energy budget, particularly in the global mean, obscures the atmospheric processes that cause heat content changes, and the role of clouds and sea ice in mediating them. Rather, a top-of-atmosphere perspective that separates shortwave (SW) and longwave (LW) components is needed to improve the process understanding of energy budget variability.

\begin{figure}[t]
      \centering
      \includegraphics{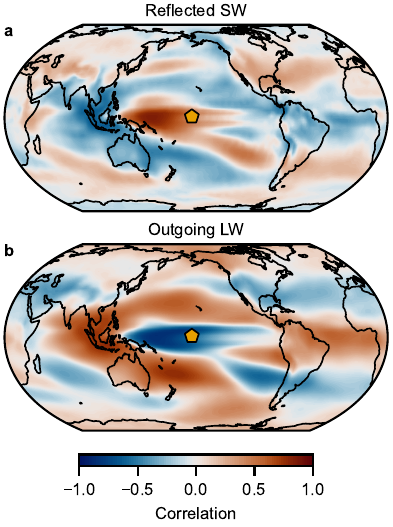}
      \caption{Annual-mean correlation between the SST at a point in the Niño 3.4 region (yellow pentagon; e.g., location of a hypothetical coral proxy) and the TOA (a) reflected SW and (b) outgoing LW radiation globally. Correlations are from last-millennium simulations over 850--1850, averaged over five models using the Fisher z-transformation.}
      \label{fig:sst_toa_corr_example}
\end{figure}

Here, we reconstruct Earth's energy budget over the last millennium (850--2000 CE) from paleoclimate proxies. The reconstruction is seasonally and spatially resolved, and partitions the TOA radiation into SW and LW components. We use data assimilation, which combines proxy observations with model dynamics, and has been used successfully to reconstruct past climates~(\citealp[e.g.,][]{Goosse2010,Bhend2012,Steiger2014,Steiger2018,Perkins2021,Hakim2016,Franke2017,Osman2021,Erb2022,Valler2022,Valler2024,Judd2024,Meng2025,Cooper2025}; see \citealp{Smerdon2023} for an overview). Our method exploits the covariance of TOA radiation and OHC with surface temperatures, to which the proxies are sensitive. For example, if a proxy in the Pacific Warm Pool indicates warmer sea surface temperatures (SSTs), the atmospheric response there is to enhance deep convection, which mediates an increase in reflected SW and a decrease in outgoing LW radiation (Fig.~\ref{fig:sst_toa_corr_example}; e.g., \citealp{Dong2019}). Similarly, changes in temperature gradients can affect the overturning atmospheric circulation and tropospheric stability, which affects SW radiation through changes in subtropical low clouds~\citep[e.g.,][]{Mackie2025,Ceppi2017,VanLoon2025b}. Besides clouds, snow/ice cover and solar surface heating play a role, and establish a temperature--radiation relationship over much of the globe~\citep{Trenberth2015}. In addition to this physical underpinning, \citet{Loeb2020} demonstrate that the current generation of atmosphere-only climate models can skillfully simulate TOA radiation when SST and sea ice are prescribed; both of these variables can be reconstructed from proxies.

We take three perspectives on the TOA energy balance. The first perspective involves a partition of the EEI into insolation $S$, reflected SW radiation (RSR), and outgoing LW radiation (OLR):
\begin{equation*}
      \text{EEI}^\downarrow = S^\downarrow - (\text{RSR}^\uparrow + \text{OLR}^\uparrow).
\end{equation*}
By our chosen sign convention, EEI is positive downwelling (energy gain), while RSR and OLR are positive upwelling (energy loss). Separating the SW and LW radiation fields allows us to infer the physical processes governing the EEI, such as clouds.

The second perspective on the TOA energy balance emphasizes the relation to the global-mean surface temperature $T$ in a one-layer energy balance model~\citep{Geoffroy2013,Gregory2016}:
\begin{equation}
      \label{eq:eei-decomposition-ohc}
      \text{EEI} = C \frac{dT}{dt} + \gamma T,
\end{equation}
where EEI is the global-mean TOA energy imbalance, $C$ is the heat capacity of the upper ocean, and $\gamma$ is the heat uptake coefficient of the deep ocean. The product $CT$ is the upper-ocean heat content, which we also reconstruct. On seasonal timescales, such as for the initial response to volcanic cooling, before the upper ocean has adjusted significantly, the deep-ocean heat uptake $\gamma T$ is relatively small. We can then approximately compare OHC and EEI by differentiation or integration. On interannual and longer timescales, the assumption of $\gamma T \ll C (dT/dt)$ breaks down~\citep[e.g.,][]{Jeevanjee2025}.

A third perspective on EEI involves the forcing--feedback framework~\citep[e.g.,][]{Sherwood2015}:
\begin{equation}
      \label{eq:eei-decomposition-rff}
      \text{EEI} = R(T) + F_\text{nat} + F_\text{anthro},
\end{equation}
where $R(T)$ is the radiative response (also called feedback), and the forcing is separated into natural ($F_\text{nat}$; i.e., volcanic and solar) and anthropogenic ($F_\text{anthro}$) components. Both forcing and response correlate with temperature-sensitive proxies. Forcing relates to near-surface temperature through its cumulative effect on ocean heat content, and the response through temperature-mediated processes such as Planck and cloud feedbacks~\citep{Forster2021}. Thus, the energy imbalance both causes and results from temperature changes. Our method, targeting the pre-industrial last millennium, restricts the reconstruction to the sum of $R(T) + F_\text{nat}$, without anthropogenic forcing.

The remainder of the paper is organized as follows. We first describe the data assimilation method (Section~\ref{sec:methods}). After validating this method in pseudoproxy experiments and against instrumental datasets (Section~\ref{sec:validation}), we present our reconstructed energy budget over the last millennium (Section~\ref{sec:reconstruction}), then illustrate applications to the natural energy budget variability and the response to volcanic eruptions (Section~\ref{sec:discussion}). Section~\ref{sec:conclusion} presents the conclusions.

\section{Methods}
\label{sec:methods}

We use online paleoclimate data assimilation (DA) to reconstruct gridded climate fields at seasonal resolution. DA combines proxy information with model forecasts in a statistically optimal way \citep[e.g.,][]{Kalnay2024}. Our DA system comprises three interacting components: (1) linear inverse models (LIMs; \citealp{Penland1995}), trained to emulate the last-millennium climate dynamics of CMIP6 models, are used for forecasting; (2) proxy system models (PSMs; \citealp[e.g.,][]{Dee2016,Evans2013}) map from the climate variables predicted by the model to proxy values (e.g., from local SST to coral $\delta^{18}$O); and (3) an ensemble Kalman filter (EnKF; \citealp{Evensen1994}) that blends proxy data with model forecasts weighted by their uncertainties, spreading information across space and climate variables. Together, these components provide an estimate of the evolving state of the climate system by continuously updating a model forecast, which maintains memory of the climate from past proxies, with proxy observations during the forecast period.

Specifically, the interaction of the three components proceeds through a forecast--update cycle. A LIM produces a forecast ``prior'' given an initial state. The first initial state is a random draw from the model climatology, whereas subsequent initial states come from the cycle's update step described next. The PSMs map this prior state to the expected proxy values, and the EnKF compares these expected values against the actual proxy observations, updating the prior state to produce a posterior estimate. This completes the cycle, which continues in time by using the posterior state as the new initial state to forecast the next season. The methodology for our seasonal-resolution DA is similar to \citet{Meng2025} and based on the prior work by \citet{Perkins2021} and \citet{Hakim2016}.

\subsection{Forecasts using linear inverse models}
We use LIMs as efficient emulators of the dynamics and statistics of the last-millennium climate as simulated in five CMIP6 models. Using such emulators allows us to run large ensembles and to incorporate the dynamics of multiple climate models. The LIM dynamics have the form~\citep{Penland1995,Penland1996}
\begin{equation*}
      \frac{d\mathbf{x}}{dt} = \bm{\mathsf{L}} \mathbf{x} + \bm{\mathsf{S}} \bm{\eta},
\end{equation*}
where $\mathbf{x}$ is the state vector, $\bm{\mathsf{L}}$ is a matrix representing the linear dynamics, $\bm{\mathsf{S}}$ is the noise amplitude matrix, and $\bm{\eta}$ is a vector of independent Gaussian white noise with unit variance. Together, $\bm{\mathsf{L}}$ and $\bm{\mathsf{S}}$ encode, for example, how surface temperatures and TOA radiation are related. During LIM training, the deterministic dynamics $\bm{\mathsf{L}}$ are derived from a lag-time linear regression, and the noise amplitude $\bm{\mathsf{S}}$ follows from a fluctuation--dissipation relation such that the long-term covariance statistics of the training data are reproduced. During forecasting from one season to the next, each ensemble member starts from a different initial condition and uses a different realization of $\bm{\eta}$. The full equations for determining the LIM matrices and for ensemble forecasts are presented in the Supplemental Information Text S1.

The state vector $\mathbf{x}$ represents seasonal anomalies of atmospheric and oceanic climate variables. During LIM training, the state vector is constructed from last-millennium simulations. During DA, the goal is to estimate the posterior $\mathbf{x}$ by blending the LIM forecast (i.e., the prior $\mathbf{x}$) and proxy information. We include 2-m surface air temperature (SAT), sea surface temperature (SST), TOA energy imbalance (EEI), TOA reflected SW radiation (RSR), TOA outgoing LW radiation (OLR), ocean heat content of the upper 300~m (OHC300), Arctic sea-ice concentration (SIC), and Antarctic SIC. We integrate the OHC over 300~m, which is the depth required to close the seasonal energy budget~\citep{Johnson2023} and is sufficiently deep to remove surface noise~\citep{Allison2020}. The OHC also imparts memory to the LIM forecasts since anomalies persist longer than in atmospheric fields.

The LIM training data, consisting of last-millennium simulations over 850--1850, are regridded to the same 2°$\times$ 2° grid, converted to anomalies relative to their own 850--1850 climatology, then linearly detrended by season to remove the orbital precession signal. Any low-frequency trend in the reconstruction is therefore driven by the assimilated proxies.

Because the physical gridded state exhibits significant spatial correlation at seasonal resolution, we perform a large dimensionality reduction prior to LIM training. We accomplish this by first subtracting the global means from the gridded state, then calculating the empirical orthogonal functions (EOF) of the area-weighted gridded residuals. Explicitly separating the global means preserves their variance, which is particularly beneficial for spatially noisy fields such as RSR. We then construct the state vector from the global means and truncated principal components (PCs):
\begin{displaymath}
      \mathbf{x} = [
            \tilde{\mathbf{x}}_{1+20}^\text{SAT};
            \tilde{\mathbf{x}}_{1+20}^\text{SST};
            \tilde{\mathbf{x}}_{1+15}^\text{EEI};
            \tilde{\mathbf{x}}_{1+15}^\text{RSR};
            \tilde{\mathbf{x}}_{1+10}^\text{OLR};
            \tilde{\mathbf{x}}_{1+15}^\text{OHC300};
            \tilde{\mathbf{x}}_{10}^\text{SICn};
            \tilde{\mathbf{x}}_{10}^\text{SICs}
            ],
\end{displaymath}
where semicolons denote vertical stacking, and $\tilde{\mathbf{x}}_{1+n}^\text{var}$ represents the global mean, followed by the leading $n$ PCs for variable $\text{var}$. For Arctic and Antarctic sea ice (SICn and SICs), we do not separate the global mean and use only the leading 10 PCs. The global mean is standardized by its temporal standard deviation, and the PCs are standardized by the square root of their retained variance after truncation. The truncation ranks $n$ were selected subjectively based on the cumulative variance explained, trading off reconstruction fidelity and degrees of freedom required for LIM training. To ensure a fair comparison across model priors, we use the identical number of PCs for each, despite variations in the actual variance explained (Fig.~S1). Mapping this EOF-space state vector $\mathbf{x}$ back to the physical gridded state amounts to a simple linear transformation.

We train LIMs on the CMIP6 last-millennium simulations (``past1000'' and ``past2k''; \citealp{Jungclaus2017}) from the following five models: MPI-ESM1-2-LR, CESM2-WACCM-FV2, MRI-ESM2-0, EC-Earth3-Veg-LR, and MIROC-ES2L~\citep{Mauritsen2019,Danabasoglu2020,Yukimoto2019,Doscher2022,Hajima2020}. These are the ``model priors,'' hereafter referred to as MPI, CESM, MRI, EC-Earth, and MIROC. Last-millennium simulations from other models exist but either lack necessary variables, like SIC, or have not been published. Previous reconstructions like \citet{Perkins2021} and \citet{Meng2025} train LIMs on CMIP5 models. However, only since the CMIP6 generation are these models able to faithfully simulate TOA radiation, mainly due to better representations of the low-cloud SW effect~\citep{Loeb2020}. We exclude Antarctic sea ice from the MIROC-based LIM during our analysis due to its Antarctic SIC low bias~\citep{Hajima2020}.

LIM forecast skill from season to season is important to propagate proxy information through time and to provide accurate priors. Forecast tests show skillful forecast correlations for the global means ($r > 0.8$ for all fields at one season lead time; Fig.~S2). The LIM tends to be slightly underdispersive, with ensemble spread underestimating forecast error in the ensemble mean (not shown). Spatial skill varies from $r \approx 0.5$ for SAT and OHC300, to $r \approx 0.3$ for radiation fields, with locally better skill in the tropical Pacific. Spatial correlation is primarily lost due to EOF truncation rather than lack of LIM skill, as evident from the reduced correlations of the initial condition. Conversely, the correlations for the global mean are close to unity, highlighting the value of handling it separately in the state vector.

Training on last-millennium simulations means that our reconstruction is most representative for the 850--1850 period. During the historical period (1850 onwards), the energy imbalance is dominated by anthropogenic forcing, for which the relationship with surface temperatures is different than the one learned during the last millennium. Our reconstructed EEI therefore only includes the response and natural forcing, even during the historical period.

\subsection{Data assimilation using an ensemble Kalman filter}
The Kalman filter is a DA algorithm that optimally combines a prior state, such as a forecast, with available proxy observations, weighted by their uncertainties. In an EnKF, the prior and posterior distributions of the state are approximated as normal distributions using samples~\citep{Evensen1994}. Multiple flavors of the EnKF exist. Here, we use a serial ensemble square root filter~\citep{Whitaker2002}.

The EnKF combines the prior ensemble mean $\overline{\mathbf{x}}_b$ and a proxy observation $y$ into the posterior ensemble mean $\overline{\mathbf{x}}_a$:
\begin{equation}
      \label{eq:kf-update}
      \overline{\mathbf{x}}_a = \overline{\mathbf{x}}_b + \bm{\mathsf{K}} (y - \bm{\mathsf{H}} \overline{\mathbf{x}}_b),
\end{equation}
where $\bm{\mathsf{K}}$ is the Kalman gain matrix, and $\bm{\mathsf{H}}$ is the forward operator for estimating the observation from the prior. The individual ensemble members are updated such that the posterior covariance (i.e., their spread), estimates the uncertainty consistently with respect to the non-ensemble Kalman filter. The Kalman gain matrix encodes covariances between the observation and the state, which allows us to estimate any field that covaries with surface temperature. The full equations for determining $\bm{\mathsf{K}}$ and for updating the ensemble members are presented in the Supplemental Information Text S2.

Seasonally resolved proxies are assimilated in the season they represent. Assimilating annually resolved proxies is complicated by the fact that the proxy seasonality can span multiple seasons. Our DA strategy in this case is to update all seasons within this time window when the proxy can be estimated by the ensemble (i.e., once the end of the window is reached), similarly to \citet{Meng2025}. For example, a proxy with seasonality MAMJJA is assimilated during the JJA step to update the MAM and JJA values. The update to the last season of the window (JJA) then informs future seasons through the LIM forecast, while updates to past seasons are not propagated forward in time. To perform the EnKF update of multiple seasons simultaneously, we use the time-averaging algorithm from \citet{Huntley2010}. This algorithm uses the time mean over multiple seasons as $\overline{\mathbf{x}}$ in Eq.~(\ref{eq:kf-update}), then adds the deviations of each season around that prior time mean to the posterior time mean.

We perform DA directly on the state vector (i.e., in EOF space) rather than in physical space. This improves computational and storage efficiency because both LIM and EnKF operate in the same low-dimensional EOF space. However, operating in EOF space prevents the use of covariance localization \citep[e.g.,][]{Anderson2012}. Instead, we mitigate sample error by using an ensemble that is much larger than the low-dimensional state vector, and we account for model error by running separate reconstructions with five distinct LIMs. To estimate the proxy value in Eq.~(\ref{eq:kf-update}), the observation operator $\bm{\mathsf{H}}$ must map the state vector to the physical temperature at the proxy location, and subsequently to the proxy value using the PSM~\citep[e.g.,][]{Hakim2022}.

SIC is physically bounded between zero and one, but the Gaussian distributions of the EnKF do not enforce this constraint. To obtain valid values, we convert SIC anomalies to absolute values by anchoring them to a 1961--1990 climatology, then clip SIC values falling outside the $[0, 1]$ interval. The climatology, obtained from \citet{Cooper2025}, is based on the multi-model mean from eight CMIP6 historical simulations before 1979, and on satellite observations thereafter.

The LIMs from each of the five model priors are used separately in the DA algorithm, allowing us to sample structural error in the training data, such as differences in covariance relationships. We repeat the reconstruction procedure 20 times for each LIM using a Monte Carlo approach. In each iteration, we assimilate a random sample of 80\% of the available proxy data. This allows us to sample over uncertainty in the proxy error estimates by randomly removing proxies that may have an outsized impact~\citep{Tardif2019}. The remaining 20\% of proxies in each iteration are used for independent validation of reconstruction skill in the pre-instrumental period. Each of the 20 iterations uses 400 ensemble members. We reduce the full 8000-member ensemble per model prior to 400 members by subsampling 20 members from each iteration. The analysis below is performed on the resulting 2000-member multi-model ensemble derived from reconstructions with all five model priors.

\subsection{Proxy system models}

The EnKF requires PSMs that estimate the proxy values from the current state, corresponding to $\bm{\mathsf{H}} \overline{\mathbf{x}}_b$ in Eq.~(\ref{eq:kf-update}). We use linear univariate PSMs that are calibrated over the instrumental period. Such statistical PSMs allow for estimating the proxy error needed for DA from regression residuals, while sacrificing little skill compared to process-based PSMs for most proxy types~\citep{Dee2016,Sanchez2025}. Tree-ring width (TRW) proxies, however, may be poorly modeled by a univariate regression in temperature due to the confounding influence of moisture~\citep{Dee2016}. We address this by removing TRW proxies that are primarily moisture-sensitive.

The linear model takes as input the temperature (SAT for terrestrial, SST for marine proxies) at the nearest gridpoint. For seasonally resolved proxies, we fit one PSM for each season. For annually resolved proxies, we objectively determine the proxy seasonality based on the Bayesian information criterion (as in \citealp{Tardif2019}), then fit a single PSM that takes the mean over those seasons as input. Seasonal DA allows us to explicitly update seasonal temperatures rather than the annual-mean temperature, whereas annual DA can suffer from seasonality biases~\citep{Lucke2021}.

The PSMs are calibrated on GISTEMP v4~\citep{Lenssen2024,GISTEMPTeam2025} and ERSSTv6~\citep{Huang2025a,Huang2025}. We truncate these temperature datasets to the same EOF basis used for each LIM to include representativeness error in the estimated observation error. We remove GISTEMP data before 1900 since there are spatial discontinuities that may affect the calibration. We also remove calibration data after 2000 to avoid the divergence problem, i.e., the decoupling of proxy values from local temperature, such as is common for tree-ring proxies~\citep[e.g.,][]{DArrigo2008}. The full calibration procedure is described in the Supplemental Information Text S3.

\subsection{Proxy network}

We assimilate proxies included in the PAGES2k~\citep{PAGESConsortium2017} and CoralHydro2k~\citep{Walter2023} databases, and add the Palmyra coral record from \citet{Dee2020}. Proxy types that are likely not temperature-sensitive are removed from PAGES2k. This includes moisture-sensitive TRW proxies, defined as correlating more strongly with local drought conditions~\citep{Dai2004} than temperatures. All coral proxies from PAGES2k are removed to avoid duplicates with CoralHydro2k, and duplicates of the \citet{Dee2020} record are removed from CoralHydro2k. For paired Sr/Ca--$\delta^{18}$O proxies in CoralHydro2k, we only use the Sr/Ca proxy, which is not confounded by salinity. We also remove data from Central Pacific coral $\delta^{18}$O proxies after 1970 since they may be affected by a salinity anomaly~\citep{Tierney2015}.

We average all subseasonal proxies to seasonal resolution, and remove proxies that have a time resolution longer than one year since we cannot assimilate them easily. Proxies are excluded if their overlap with the calibration period is too short ($<$ 25 years), their calibration correlation is too low ($< 0.10$), or their error autocorrelation is too high ($> 0.90$; errors are assumed uncorrelated in time in the Kalman filter). The resulting proxy network comprises 382 records, most of which derive from tree rings (58\%) and corals (30\%). The proxies have broad spatial coverage, but few are located in the ocean interior, particularly outside the tropics (Fig.~S5). The proxy network becomes increasingly sparse toward the beginning of the last millennium, and there are only four seasonally resolved proxies before 1500. Importantly, since annual proxies are sensitive to specific seasons, and the LIM propagates information through proxy-sparse seasons, the algorithm yields seasonal climate information even when there are no seasonal proxies.

\subsection{Skill metrics}

We validate the reconstruction skill in pseudoproxy experiments, against instrumental datasets, and against withheld proxies. Our primary validation metrics are the Pearson correlation coefficient $r$ and the coefficient of efficiency~\citep{Nash1970}
\begin{equation*}
      \text{CE} = 1 - \frac{\sum_i (v_i - x_i)^2}{\sum_i (v_i - \bar v)^2},
\end{equation*}
where $x_i$ is the reconstruction time series, $v_i$ is the verification time series, and $\bar v$ is the time mean of $v_i$. Compared to $r$, the CE thus also accounts for differences in signal amplitude and mean, measured relative to the verification variability.

The statistical significance of correlations is determined using the random-phase test, a nonparametric test that generates surrogate time series with the same spectral density~\citep{Ebisuzaki1997}. This accounts for spectral properties such as autocorrelation, which reduces the effective sample size~\citep[e.g.,][]{Bretherton1999}. For p-values, we use a significance level of $\alpha=0.05$ with the null hypothesis that $r = 0$. The alternative hypothesis is $r > 0$ if reconstruction skill is compared, and $r \neq 0$ for physical correlations between variables. In some figures and tables, we denote highly significant correlations ($p \leq 0.01$) with **, significant correlations ($p \leq 0.05$) with *, and non-significant correlations ($p > 0.05$) with n.s.

\section{Validation of reconstruction skill}
\label{sec:validation}

We validate our method using pseudoproxy experiments (PPEs) and compare the real-proxy reconstruction against instrumental datasets. PPEs help to establish an upper bound on reconstruction fidelity in an idealized scenario~\citep{Smerdon2012}, while instrumental validation assesses the real reconstruction, albeit over a limited time period and against uncertain verification sources~\citep[e.g.,][]{Thorne2026,Chan2025}.

\subsection{Realistic pseudoproxy experiment}
\label{sec:ppe-realistic}

PPEs are particularly useful to assess the feasibility of reconstructing new target variables such as TOA radiation. A climate model simulation serves as ``truth,'' sampled only by sparse and noisy pseudoproxies that mimic the characteristics of the real proxy network. The challenge is analogous to the real problem, with the benefit that skill can be easily quantified~\citep{Smerdon2012,Steiger2014}.

\begin{figure*}[t]
      \includegraphics{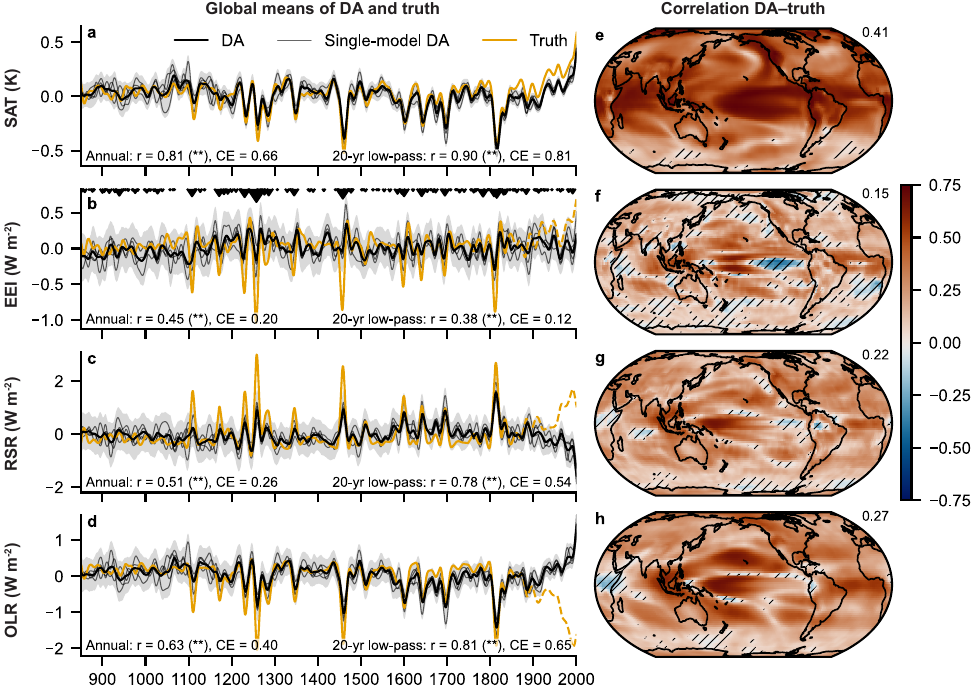}
      \caption{Comparison of the pseudoproxy imperfect-model reconstruction with the truth simulation. (a--d) Global-mean time series and their correlation coefficients~r over 850--1850. The reconstruction (black) and the truth (yellow) are anomalies relative to 850--1850 with a 20-yr low-pass filter. Radiation after 1850 is dashed since it includes anthropogenic forcing, which we cannot reconstruct. DA is the mean of the four single-model reconstructions with the MPI, CESM, MRI, and EC-Earth model priors. Carets indicate volcanic eruptions, scaled by their volcanic stratospheric sulfur injection, from the eVolv2k~\citep{Toohey2017} and the CMIP7 historical volcanic forcing datasets (\citealp{Aubry2025a}; small eruptions with a VSSI below $0.5\,\mathrm{Tg\ S}$ removed). All correlations of global means are significant as denoted by (**). Shading denotes the 5th--95th percentile range. (e--h) Spatial correlation of annual anomalies over 850--1850 between reconstruction and truth. Correlations that are not significantly positive ($\alpha = 0.05$) are hatched. Numbers in the top right corners are global means of the spatial correlation.}
      \label{fig:ppe_panels}
\end{figure*}

For our realistic PPEs, truth consists of the concatenated last-millennium and historical simulations from the MIROC model. We then reconstruct MIROC fields using LIMs trained on the MPI, CESM, MRI, and EC-Earth model priors. This constitutes an imperfect-model experiment, which includes error from a mismatch in LIM dynamics and the EOF basis relative to the truth simulation, analogous to the real reconstruction where the true dynamics are unknown. For each real proxy, a pseudoproxy is drawn from the truth simulation using the real PSM with added noise consistent with the estimated error. The pseudoproxies mimic the real proxies in location, temporal availability, seasonal sensitivity, signal-to-noise ratio, and temperature field sensitivity (SAT or SST).

Results show that the global-mean reconstructions for the fields of interest correlate significantly with the truth simulation (Fig.~\ref{fig:ppe_panels}a--d). For SAT, decadal variability is tracked well (annual $r = 0.8$), and the truth is generally within the ensemble spread. Modern warming is somewhat underestimated, possibly because patterns of warming are different in the MIROC model compared to the LIMs (Fig.~\ref{fig:ppe_panels}a). The reconstruction is also mostly skillful for the TOA radiation fields, although correlations and CE are lower for EEI than for RSR and OLR. Large excursions in TOA radiation coincide with major volcanic eruptions, although the magnitude is underestimated, particularly in the EEI. The variance of the EEI in the PPE is similar to the real reconstruction, suggesting that the PPE may be representative of the real problem. Arctic SIC validates well on sea-ice area and spatial patterns, in contrast to Antarctic SIC, which shows low correlations and a large spread across model priors (Fig.~S3). Reconstruction skill for all fields is significantly reduced in the PPE before 1100, when fewer than 70 proxies are available, none of them in the tropical Pacific.

The PPE also demonstrates spatial reconstruction skill over the last millennium (Fig.~\ref{fig:ppe_panels}e--h). Correlations with the truth simulation are positive in most regions, although they are smaller over the Southern Ocean, which is relatively sparse in proxies. Like for the global mean, annual skill is highest for SAT and lowest for EEI.

\subsection{Pseudoproxy experiment for forcing and response}
\label{sec:ppe-fr}

We run further PPEs to test if our DA system is able to reconstruct both the response and the natural forcing, despite their distinct lead/lag relationships with temperature~\citep[e.g.,][]{Proistosescu2018}. In these PPEs, we separate the forcing and response in the state vector. The truth and training data come from CMIP6 ``hist-nat'' simulations~\citep{Gillett2016}, with forcing derived from ``piClim-histnat'' simulations~\citep{Pincus2016}, of the CanESM5~\citep{Swart2019} and NorESM2-LM~\citep{Seland2020} models.

We find skillful reconstructions of the response, natural forcing, and 300-m OHC (Fig.~S4, Table~S1). Correlations are particularly high for the response and OHC. While forcing skill is relatively high in perfect-model experiments, it is not always significant in imperfect-model experiments. The magnitude of volcanic forcing is greatly underestimated, by up to a factor of five, similar to the realistic PPE experiment (Fig.~\ref{fig:ppe_panels}b). The lead/lag relationships of the response and forcing in the models are reproduced by the reconstruction (Fig.~S4d), which is critical for proxy information to inform the energy imbalance through DA. Natural forcing also includes solar variability, although its magnitude is below the noise level of the reconstruction. We conclude that reconstructing both response and natural forcing is possible, but with a much reduced magnitude for volcanic forcing. We hypothesize that OHC plays a critical role in linking the energy budget and radiation to temperatures through proxies.

\subsection{Instrumental validation of real-proxy reconstruction}

Validation against independent datasets allows us to assess the real reconstruction. The SAT, SST, and OHC reconstructions are highly correlated with instrumental products and other reconstructions (Figs.~S7 and S9a,b). The Arctic SIC reconstruction is skillful at annual and seasonal scales when compared to satellite observations (Figs.~S8 and S9c--f), but Antarctic SIC is not. Further, unassimilated proxies (20\% withheld for validation), which are predicted from the reconstruction and compared directly to proxy values, correlate significantly in the network mean, indicating skill over the whole last millennium (Fig.~S6).

The Earth Radiation Budget Experiment (ERBE) satellite observations of TOA radiation overlap with our reconstruction over 1985--1999. We validate against the DEEP-C v5 dataset~\citep{Liu2020,Liu2022}, which combines these satellite observations with atmospheric reanalyses and atmosphere-only model simulations. We also validate against the ERBE WFOV Edition 4.1 datasets~\citep{ERBEScienceTeam2020a}, which are direct, recently recalibrated observations but lack complete spatial and temporal coverage.

\begin{figure*}[t]
      \includegraphics{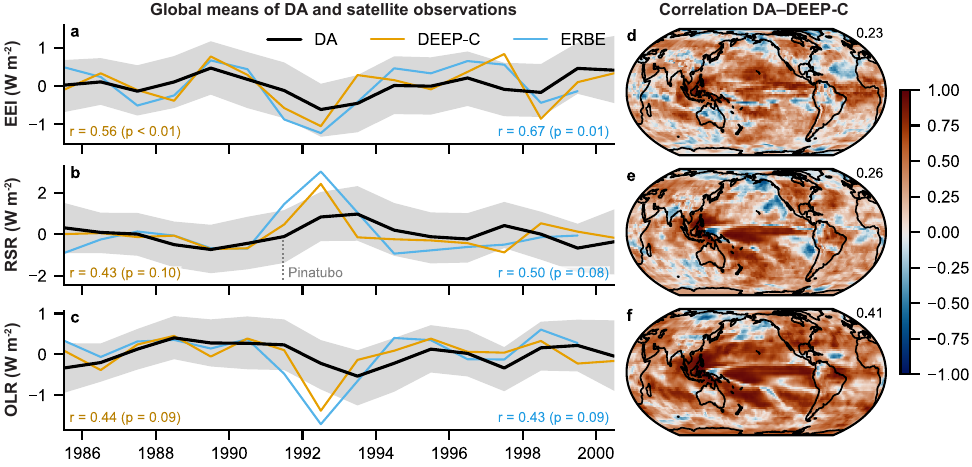}
      \caption{Annual TOA radiation anomalies from our reconstruction, the DEEP-C combined product, and ERBE satellite measurements. The linear trend over 1985--1999 has been removed. Shading denotes the 5th--95th percentile range. (a--c) Global-mean time series and their correlation coefficients~r. (d--f) Spatial correlation of reconstruction with DEEP-C. Numbers in the top right corners are global-mean values. Significance of spatial correlations is not indicated since most are non-significant due to the small effective sample size ($n_\text{eff} \approx 10$ to $15$ for global means).}
      \label{fig:deepc_panels}
\end{figure*}

Reconstructed global-mean TOA radiation qualitatively tracks interannual variability in the DEEP-C and ERBE data, with positive but not always significant correlations of $r > 0.4$ (Fig.~\ref{fig:deepc_panels}a--c). Most of this variability represents the radiative response to internal variability, but there is a forced signal from the 1991 Mt. Pinatubo eruption. As in the PPEs (Section~3\ref{sec:ppe-realistic},\ref{sec:ppe-fr}), this volcanic forcing is underestimated. Much of the skill comes from the tropics and subtropics, likely mediated by ENSO, but large regions in the mid-latitudes and polar regions also have positive correlations (Fig.~\ref{fig:deepc_panels}d--f). The reduced skill at higher latitudes is likely the result of a weaker temperature--radiation relationship, mainly due to weather noise from extratropical storms in TOA radiation~\citep[e.g.,][]{Trenberth2015}. The reconstructed seasonal TOA radiation is also skillful, but with correlations slightly less than the annual values (Fig.~S10). Compared to the PPE, the correlations of the global mean are similar, but the skill mostly comes from the tropics, while it is more uniform in the PPE. The correlations are not robust due to the short overlap period and high temporal autocorrelation.

\begin{figure}[t]
      \centering
      \includegraphics{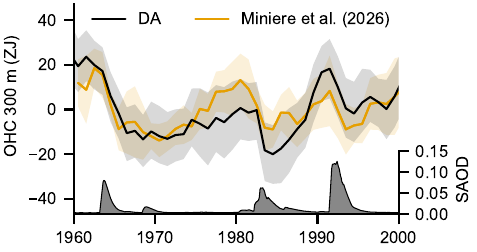}
      \caption{Annual upper-ocean heat content from our reconstruction and the \citet{Miniere2026} compilation of instrumental datasets. The linear trend over 1960--2000 has been removed to emphasize interannual to decadal variations as in \citet{Church2005}. Only instrumental datasets that span the full period are used, leaving a compilation of eight. Stratospheric aerosol optical depth (SAOD) at 550 nm is from the CMIP7 historical volcanic forcing dataset~\citep{Aubry2025a}. Shading denotes the 5th--95th percentile range.}
      \label{fig:timeseries_historical_ohc}
\end{figure}

Our reconstruction also correlates well ($r = 0.72$, $p \leq 0.01$) with decadal variations in instrumental estimates of the upper-ocean heat content (Fig.~\ref{fig:timeseries_historical_ohc}). For comparison, we use the \citet{Miniere2026} compilation of instrumental datasets, derived primarily using in-situ temperature profiles. Most of the decadal variability in OHC appears forced by volcanic eruptions. The recovery of OHC following the 1963 Agung eruption took multiple decades, possibly prolonged by the 1968 Fernandina eruption, unforced decadal variability, and tropospheric aerosol forcing. We conclude that our reconstruction of the energy budget has many skillful aspects, despite the spatiotemporal sparseness of the proxy network and the complex relationship between surface temperature and TOA radiation.

\section{Reconstruction over the last millennium}
\label{sec:reconstruction}

Our reconstruction consists of gridded, seasonal climate fields over 850--2000 CE. Results labeled as DA are the combined ensemble of the five single-model reconstructions. Annual means are taken from December to the following November. Uncertainties refer to the very likely range (5th to 95th ensemble percentile, i.e., the 90\% median credible interval).

\subsection{Global means of temperature and radiation}

\begin{figure*}[p]
      \centering
      \includegraphics{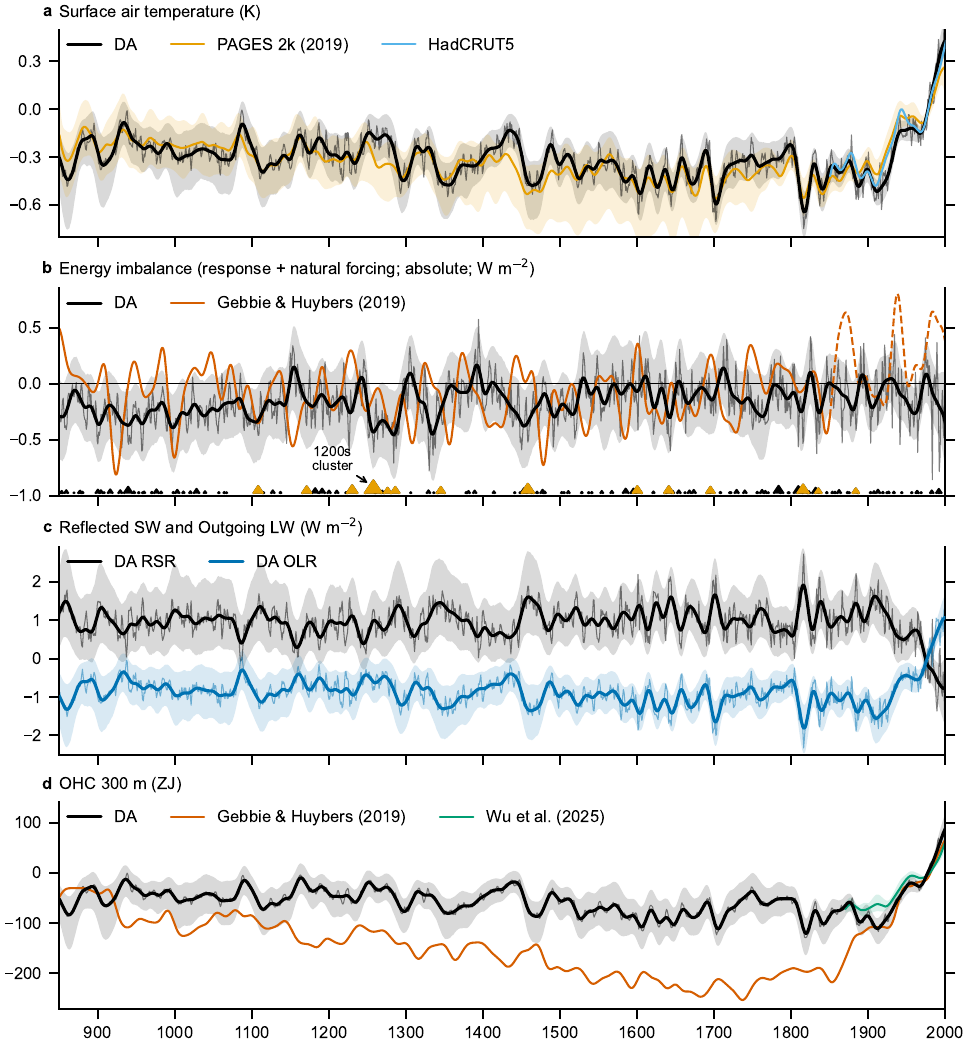}
      \caption{Global-mean time series over the last millennium. DA refers to the combined ensemble of the multi-model reconstructions. Shading denotes the 5th--95th percentile range. Bold lines have a 20-yr low-pass filter; thin, faint lines are annual values. (a) SAT anomalies relative to 1961--1990, comparing to the multi-method ensemble from \citet{PAGESConsortium2019} and the HadCRUT5 instrumental dataset~\citep{Morice2021}. (b) Absolute energy imbalance (rather than anomalies), representing the radiative response and natural forcing. The orange line shows the rate of OHC change from \citet{Gebbie2019}, expressed as equivalent EEI and divided by 90\%, which is the fraction of energy imbalance absorbed by the ocean~\citep{Schuckmann2023}. \citet{Gebbie2019} data after 1850 is dashed since it includes anthropogenic forcing, which we cannot reconstruct. Carets indicate volcanic eruptions, scaled by their volcanic stratospheric sulfur injection, from the eVolv2k~\citep{Toohey2017} and the CMIP7 historical volcanic forcing datasets (\citealp{Aubry2025a}; small eruptions with a VSSI below $0.5\,\mathrm{Tg\ S}$ removed). Those marked in orange are composited in Fig.~\ref{fig:composite_volcano}. (c) RSR and OLR, constituting the EEI in (b). (d) Upper-ocean heat content anomalies, compared to the reconstructions by \citet{Gebbie2019} and \citet{Wu2025a}.}
      \label{fig:timeseries_gm}
\end{figure*}

The reconstructed global-mean SAT has a cooling trend over much of the last millennium, followed by warming over the 1900s (Fig.~\ref{fig:timeseries_gm}a). The SAT transition from a warm Medieval Climate Anomaly (MCA; c.~800--1200) into a colder Little Ice Age (LIA; c.~1300--1850) is a feature often found in last-millennium proxy reconstructions~\citep[e.g.,][]{Mann1999,Esper2002,Mann2009, PAGESConsortium2019} and simulations~\citep[e.g.,][]{Fernandez-Donado2013,Otto-Bliesner2016,Ljungqvist2019}. Based on linear regression over 850--1850, we find a cooling trend of $-0.18\,\mathrm{K\,kyr^{-1}}$ (see also Fig.~\ref{fig:trend_zm_panels}a). There is also considerable multidecadal variability, some of which coincides with large volcanic eruptions; we will consider these events in greater detail in Section 4\ref{sec:volcanic}. The global-mean SST and OHC300 are highly correlated with SAT ($r > 0.8$; OHC shown in Fig.~\ref{fig:timeseries_gm}d, SST in Fig.~S11d). Our reconstruction has a similar mean value, multidecadal variability ($r = 0.85$, $p \leq 0.01$ at 20-yr timescales), and trend as the multi-method ensemble from \citet{PAGESConsortium2019}. This is also the case when comparing to LMR v2.1~\citep{Tardif2019} and PHYDA~\citep{Steiger2018} in Fig.~S11a. The NH extratropical land summer temperatures are at the cold end of comparable reconstructions~\citep[e.g.,][]{Wilson2016,Buntgen2021}, and colder than instrumental temperatures before 1950, while the SH temperature agrees well with \citet{Neukom2014}, as shown in Fig.~S11b,c.

\begin{figure}[p]
      \centering
      \includegraphics{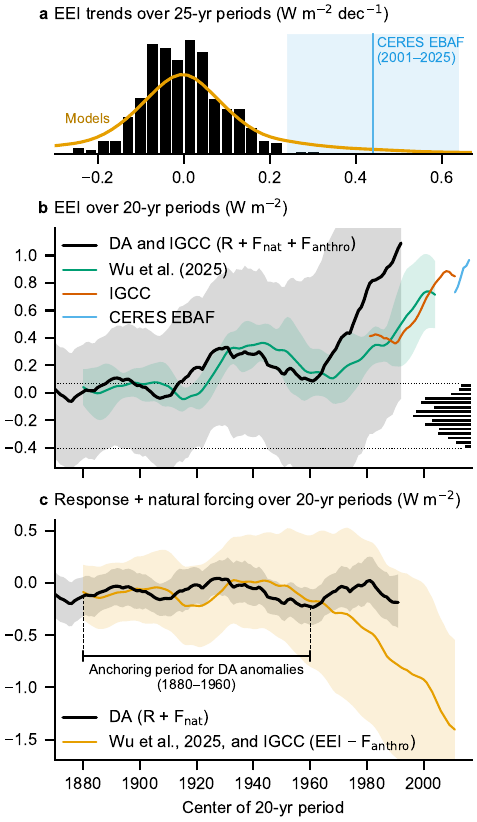}
      \caption{Recent energy imbalance in the context of pre-industrial variability. The EEI reconstructed in this study is compared to CERES EBAF Ed4.2.1 data~\citep{Doelling2025,Loeb2018}, and to the 20-yr sliding window trend in the heat content reconstructions of IGCC~\citep{Forster2025,Smith2025} and \citet{Wu2025a}. The OHC from \citet{Wu2025a} has been divided by 90\%, which is the fraction of energy imbalance absorbed by the ocean~\citep{Schuckmann2023}. Shading denotes the 5th--95th percentile range, assuming independent errors when adding or subtracting time series. (a) Energy imbalance trends over 25-yr periods, which is the number of full years in the CERES record. The histogram shows the distribution of 25-yr trends in the ensemble-mean EEI from our reconstruction over 850--1850. The yellow line shows the kernel density estimate of 25-yr trends in the five CMIP6 last-millennium simulations. The trend from CERES EBAF is shown in blue, with an uncertainty of $\pm 0.20\,\mathrm{W\,m^{-2}\,dec^{-1}}$ as in \citet{Raghuraman2021}. (b) Absolute energy imbalance over 20-yr periods. The black line is the sum of the reconstructed EEI and the anthropogenic forcing from IGCC. The vertical histogram shows the distribution of 20-yr-mean, ensemble-mean EEI from our reconstruction over 850--1850, and dotted lines show its minimum and maximum. (c) Response and natural forcing over 20-yr periods. The yellow line is the EEI from \citet{Wu2025a} minus the anthropogenic forcing from IGCC, to which we anchor our EEI anomalies in order to obtain absolute values.}
      \label{fig:timeseries_historical_eei}
\end{figure}

The absolute EEI governs changes in the energy budget (e.g., a positive EEI implies energy gain), whereas the reconstruction provides anomalies. To obtain absolute values, we anchor the reconstructed EEI anomalies, which contain the response and natural forcing, to an independent climatology of the total EEI minus the anthropogenic forcing, based on 20-yr running means with centers during 1880--1960. The equivalence follows from rearranging Eq.~(\ref{eq:eei-decomposition-rff}):
\begin{align*}
      \underbrace{R + F_\text{nat}}_\text{DA} = \underbrace{\text{EEI}}_\text{Wu et al. (2025)} - \underbrace{F_\text{anthro}}_\text{IGCC},
\end{align*}
where the EEI from \citet{Wu2025a} is the rate of change of reconstructed ocean heat content, and the anthropogenic forcing from IGCC (Indicators of Global Climate Change, an update to metrics from the IPCC AR6 report; \citealp{Forster2025,Smith2025}) is estimated using empirical relationships. Our anomalies and the anchor timeseries have similar variability over the anchoring period of 1880--1960 (Fig.~\ref{fig:timeseries_historical_eei}c).

The cooling trend over the last millennium coincides with energy loss (Fig.~\ref{fig:timeseries_gm}b), manifesting as a negative energy imbalance over 850--1500. After that, the energy imbalance is close to zero. The only other proxy-based estimate of the last-millennium energy budget is the OHC reconstruction from \citet[][version OPT-0015]{Gebbie2019}. At multidecadal timescales, our reconstruction has a weakly negative correlation ($r = -0.10$, $p = 0.73$ at 50-yr timescales over 850--1850) with the \citet{Gebbie2019} estimate (Fig.~S12a). The disagreement increases before 1100, although our reconstruction is likely less reliable during this early period, as discussed in Section~3\ref{sec:ppe-realistic}. Over 1100--1850, our mean imbalance ($-0.13 \pm 0.13\,\mathrm{W\,m^{-2}}$) agrees well with theirs ($-0.12\,\mathrm{W\,m^{-2}}$), and consequently, the integrated energy imbalance also agrees well (Fig.~S12b), indicating a loss of about $1700\,\mathrm{ZJ}$ over 850--1850. After 1850, our reconstructions diverge because \citet{Gebbie2019}'s estimate also includes anthropogenic forcing. Further, their prescribed SSTs show stronger and earlier warming than ours (Figs.~\ref{fig:timeseries_gm}d and S11d).

The global-mean RSR and OLR are dominated by multidecadal variability, often associated with volcanic eruptions (Fig.~\ref{fig:timeseries_gm}c). RSR is anticorrelated with SAT (annual $r = -0.78$, $p \leq 0.01$), while OLR has a strong positive correlation (annual $r = 0.97$, $p \leq 0.01$). These correlations between temperature and TOA radiation are about $20\%$ stronger than in last-millennium simulations, likely since our method depends on temperature to reconstruct radiation. In contrast, the correlation of EEI with SAT is weakly negative (annual $r = -0.21$, $p = 0.02$), while it is weakly positive in models. RSR has more variance than OLR at all timescales.

\subsection{Context for recent energy imbalance}
\label{sec:context-recent-eei}

We consider the recent energy imbalance and its trend in the context of pre-industrial, natural variability. The EEI trend over 25-yr periods during 850--1850 in our reconstruction ranges from $-0.26$ to $+0.32\,\mathrm{W\,m^{-2}\,dec^{-1}}$ (Fig.~\ref{fig:timeseries_historical_eei}a). The distribution of these pre-industrial EEI trends provides context for the modern EEI trend. The trend over the 25-yr CERES satellite period (2001--2025) as provided in CERES EBAF data~\citep{Doelling2025,Loeb2018} is $+0.44\,\mathrm{W\,m^{-2}\,dec^{-1}}$, far exceeding pre-industrial values. Even a trend of $+0.24\,\mathrm{W\,m^{-2}\,dec^{-1}}$, corresponding to the lower uncertainty bound of the CERES EBAF trend, is in the 99th percentile of pre-industrial variability in our reconstruction. Last-millennium simulations have a slightly wider range of EEI variability, likely due to a stronger response to volcanic eruption. The CERES trend of $+0.44\,\mathrm{W\,m^{-2}\,dec^{-1}}$ is in the 97th percentile of pre-industrial variability from last-millennium simulations.

EEI ranges from $-0.41$ to $+0.07\,\mathrm{W\,m^{-2}}$ over 20-yr periods during 850--1850 in our ensemble-mean reconstruction  (Fig.~\ref{fig:timeseries_historical_eei}b). We compare this pre-industrial distribution to the modern EEI estimated from CERES EBAF satellite data, the IGCC heat inventory~\citep{Forster2025,Smith2025}, and the OHC reconstruction by \citet{Wu2025a}. In these three modern datasets, the EEI over all 20-yr periods starting after 1915 exceeds any pre-industrial 20-yr-average EEI, although the EEI around 1960 almost falls within the pre-industrial distribution. Considering not just the distribution of the ensemble mean but also all ensemble members, the maximum pre-industrial EEI is $+0.82\,\mathrm{W\,m^{-2}}$. With this more conservative threshold, the EEI for all 20-yr periods starting after 1990 is unprecedented in the pre-industrial period.

While our reconstruction does not contain the anthropogenic forcing, we can add its estimate from IGCC to our reconstructed EEI. The sum then agrees well with the total EEI from \citet{Wu2025a} over 1880--1960, although there appears to be a 10-yr lag (Fig.~\ref{fig:timeseries_historical_eei}b). After 1960, the response in our reconstruction is likely underestimated (Fig.~\ref{fig:timeseries_historical_eei}c).

\subsection{Sea ice}

Sea ice influences high-latitude TOA radiation through surface albedo, lapse rate, and cloud feedbacks~\citep{Jenkins2021}. We therefore evaluate Arctic and Antarctic sea-ice area (SIA), calculated as the sum of the products of SIC and grid-cell area in each hemisphere~\citep[e.g.,][]{Notz2014}. While Arctic sea-ice reconstructions over the last millennium exist~\citep[e.g.,][]{Brennan2022,Meng2025}, Antarctic sea ice has so far been elusive~\citep{Thomas2019}, and we caution that our Antarctic reconstruction shows much lower skill than the Arctic based on pseudoproxy experiments (Fig.~S3) and instrumental validation (Figs.~S8 and S9).

\begin{figure*}[b]
      \includegraphics{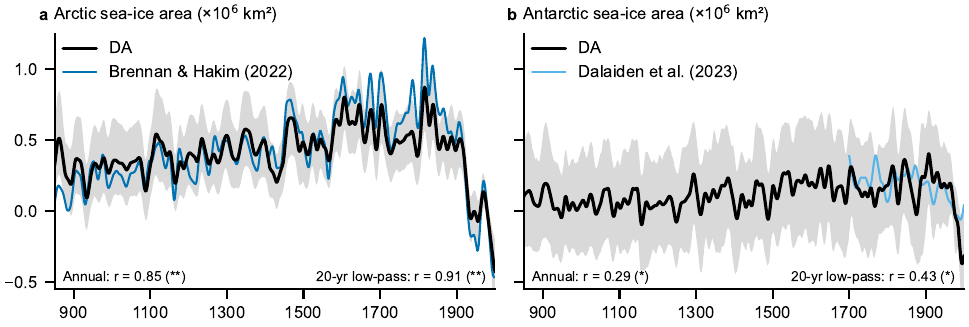}
      \caption{Annual-mean sea-ice area over the last millennium. Shading denotes the 5th--95th percentile range. We show anomalies relative to 1961--1990 with 20-yr low-pass filter and compare to the reconstructions from \citet{Brennan2022} for the Arctic and \citet{Dalaiden2023} for the Antarctic.}
      \label{fig:timeseries-siarea}
\end{figure*}

SIA in both hemispheres increased with the cooling trend over the last millennium (Fig.~\ref{fig:timeseries-siarea}). Arctic SIA agrees closely with \citet{Brennan2022}, which is a proxy-based reconstruction using offline DA (no model forecasts). Over 850--1850, the annual-mean Arctic SIA grows by $0.3 \times 10^{6}\,\mathrm{km^2}$, mostly in the Barents Sea, while $0.9 \times 10^{6}\,\mathrm{km^2}$ are lost over 1850--2000 (Figs.~S8 and S13). The peak SIA during the LIA is slightly lower than in \citet{Brennan2022}. Antarctic SIA has less variability and more uncertainty across model priors when compared to Arctic SIA. Interannual to decadal variability in Antarctic SIA over 1700--2000 correlates significantly with \citet{Dalaiden2023}, another proxy-based reconstruction using offline DA. There is no clear trend in Antarctic sea-ice loss over 1850--2000 (Fig.~S8) in our reconstruction, \citet{Cooper2025}, and \citet{Dalaiden2023}.

\subsection{Last-millennium trends}
\label{sec:results-lmtrends}

The global-mean cooling trend from the MCA into the LIA is a prominent feature in our reconstruction. Associated with it are regional variations, which we summarize here in terms of zonal-mean trends, based on a linear regression over 850--1850; spatial trends are shown in Fig.~S14.

\begin{figure*}[p]
      \includegraphics{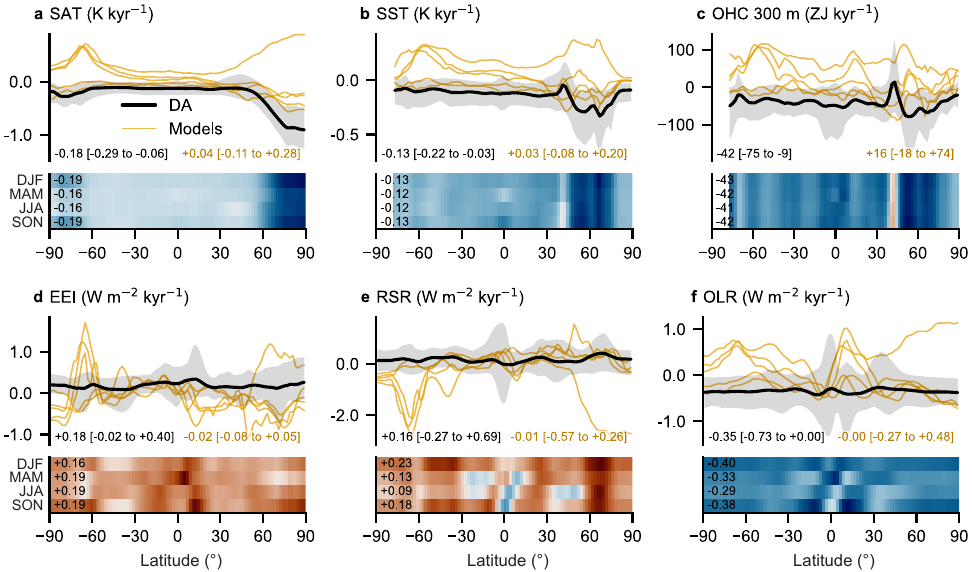}
      \caption{Zonal-mean trends over 850--1850. Each panel shows (top) the annual mean and (bottom) the seasonal means. Yellow lines correspond to the CMIP6 last-millennium simulations from the MPI, CESM, MRI, EC-Earth, and MIROC models. Numbers denote the global-mean trend (reconstruction in black with 5th--95th percentile range, models in yellow with min--max range). Shading denotes the 5th--95th percentile range.}
      \label{fig:trend_zm_panels}
\end{figure*}

\begin{figure*}[p]
      \includegraphics{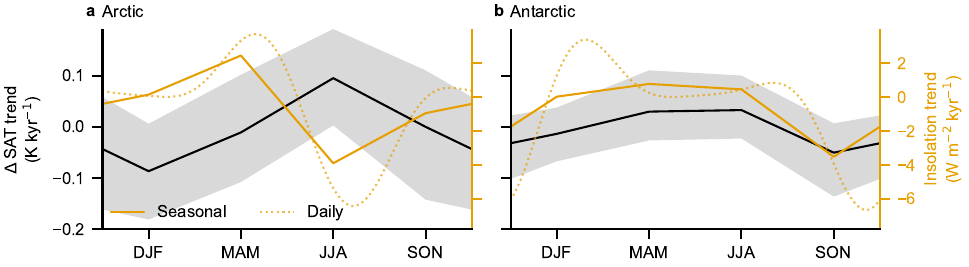}
      \caption{Seasonal high-latitude ($>$ 60\textdegree) temperature trends over the last millennium for the Arctic and Antarctic. Trends are expressed as departures in seasonal temperature trends from the annual-mean trend over 850--1850. The high-latitude ($>$ 60\textdegree) insolation trend (yellow) from \citet{Berger1991} is shown as daily (dotted) and seasonal-mean (solid) values. Shading denotes the 5th--95th percentile range.}
      \label{fig:timeseries-highlattemp}
\end{figure*}

The global-mean cooling trend over 850--1850 is $-0.18\,\mathrm{K\,kyr^{-1}}$, with a 5th--95th percentile range of $-0.29$ to $-0.06\,\mathrm{K\,kyr^{-1}}$ (Fig.~\ref{fig:trend_zm_panels}a). In the multi-method reconstruction from \citet{PAGESConsortium2019}, the mean cooling trend is $-0.30\,\mathrm{K\,kyr^{-1}}$, with a 5th--95th percentile range of $-0.51$ to $-0.03\,\mathrm{K\,kyr^{-1}}$. In comparison, this trend in the five CMIP6 last-millennium simulations ranges from $-0.11$ to $+0.28\,\mathrm{K\,kyr^{-1}}$. Coupled models thus tend to simulate last-millennium weak cooling or warming, contradicting proxy evidence. Cooling is stronger in the Northern Hemisphere (NH; $-0.22\,\mathrm{K\,kyr^{-1}}$) than in the Southern Hemisphere (SH; $-0.14\,\mathrm{K\,kyr^{-1}}$) in our reconstruction. While almost all regions cool from the MCA into the LIA, polar amplification is evident in the SAT trend. Arctic amplification, measured by the ratio of the 67\textdegree N--90\textdegree N trend to the global-mean trend in SAT, is around $4.4$, about $15\%$ higher than in recent decades~\citep{Rantanen2022}.

Trends in SST and OHC are closely related to each other (Fig.~\ref{fig:trend_zm_panels}b,c), with cooling at all latitudes, particularly in the Pacific and North Atlantic. However, there is significant uncertainty in the magnitude of reconstructed cooling in the northern mid-latitudes across model priors. Warming occurs around the Kuroshio--Oyashio Extension and the Gulf Stream (Fig.~S14b).

Trends in TOA radiation have a more complex spatial structure that varies by season (Figs.~\ref{fig:trend_zm_panels}d--f). The OLR trend is strongly negative and spatially defined by the Indo--Pacific convective region, whereas the RSR trend is positive and has large local trends at all latitudes. Together, this makes the global-mean EEI trend positive (toward energy gain), until the absolute EEI is approximately zero on centennial timescales after 1500. The RSR and EEI trends are not significantly different from zero, and vary strongly by location and season (Fig.~S14d--f).

At high latitudes ($>$ 60\textdegree), seasonal differences in temperature trends are most pronounced (Fig.~\ref{fig:timeseries-highlattemp}). The Arctic cooling trend is most enhanced in DJF and most subdued in JJA, which increases the amplitude of the seasonal cycle in Arctic temperatures during the last millennium. In contrast, Antarctic cooling is most enhanced in SON and most subdued in MAM and JJA, reducing the seasonal cycle amplitude. The seasonal temperature trends are compared with high-latitude insolation trends, which differ between the hemispheres. Extremes of insolation trends are smoothed in the seasonal averages, which obscures lead--lag relationships at seasonal timescales. However, a lag of up to one season is evident for the Arctic.

\subsection{Volcanic eruptions}
\label{sec:volcanic}

We composite pre-industrial tropical eruptions with a volcanic stratospheric sulfur injection (VSSI) of at least $9\,\mathrm{Tg\ S}$ (similar to Mt. Pinatubo in 1991) and a latitude equatorward of 30\textdegree. For volcanic events within 10 years of each other, we only use the second eruption. Eruption dates, latitude, and VSSI are taken from the eVolv2k version 4 database~\citep{Toohey2017,Sigl2024}, which is derived from ice core records.

\begin{figure*}[t]
      \includegraphics{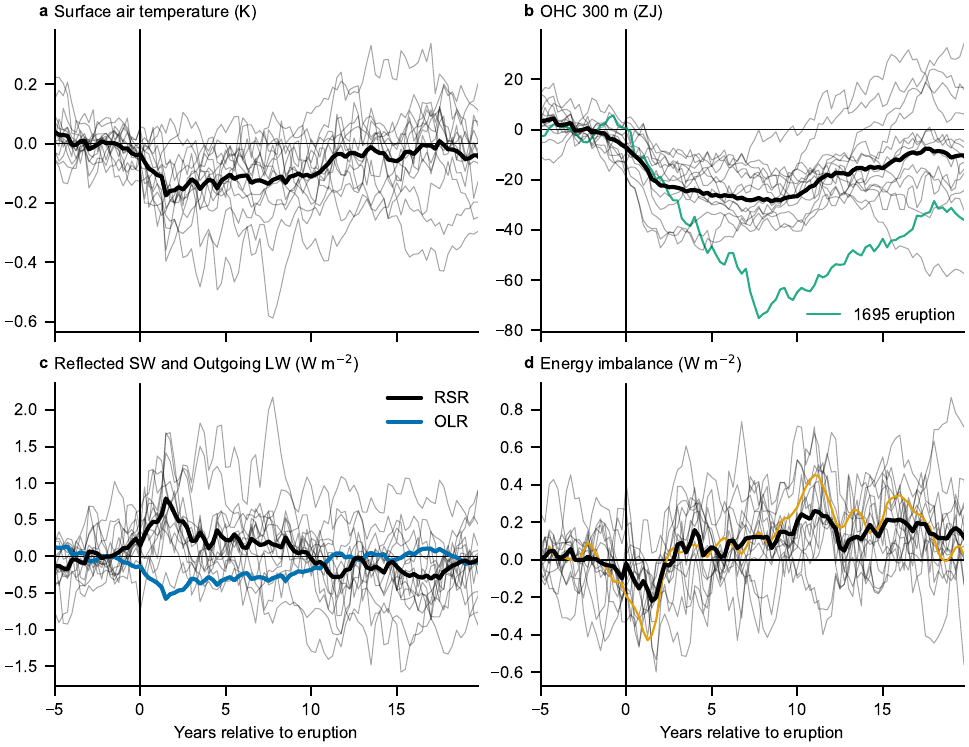}
      \caption{Composite analysis (also known as superposed epoch analysis) of global-mean seasonal anomalies for large tropical volcanic eruptions over 850--1850 (n = 14). Eruptions have a VSSI of at least $9\,\mathrm{Tg\ S}$, a latitude equatorward of 30\textdegree, and are marked in orange in Fig.~\ref{fig:timeseries_gm}a. Anomalies are computed relative to the 5 years prior to each eruption, and only the second eruption is used for volcanic double events. The green line in (b) corresponds to the 1695 eruption. The orange line in (d) is the composite of the time derivative of OHC300, rescaled by Earth's ocean fraction (71\%) to obtain the equivalent EEI and smoothed with an annual running mean. This is expected to differ from the time derivative of the composite of OHC300 in (b) since the pre-eruption climatology is different.}
      \label{fig:composite_volcano}
\end{figure*}

There is a clear global cooling signal, albeit with large scatter across individual eruptions, peaking one to two years after the eruption at $-0.17\,^\circ\mathrm{C}$ (NH: $-0.24\,^\circ\mathrm{C}$, SH: $-0.12\,^\circ\mathrm{C}$) on average and fully recovering after 10--15 years (Fig.~\ref{fig:composite_volcano}). The stronger NH cooling, particularly in summer (Fig.~S15a), is possibly due to proxy distribution and seasonality, but model simulations also suggest a hemispherically asymmetric response~\citep{Pauling2021}. Cooling is within the spread of the proxy reconstruction and model composite by \citet{He2026}. OHC300 is reduced by around $25\,\mathrm{ZJ}$ on average, most rapidly during the first two years after the eruption. This heat loss is coincident with enhanced RSR (up to $+0.8\,\mathrm{W\,m^{-2}}$), a small reduction in OLR (up to $-0.6\,\mathrm{W\,m^{-2}}$) and negative EEI of up to $-0.2\,\mathrm{W\,m^{-2}}$. EEI becomes positive after 3--5 years, and the OHC recovers over the course of a decade. Sea-ice expansion has a temporal response similar to surface temperature, and there is no discernible impact on ENSO (Fig.~S15), consistent with other reconstructions and models~\citep{He2026,Zhu2022,Dee2020,Dee2022}. Compared to coupled-model simulations, the volcanic response in all fields is muted and time-lagged (Fig.~S16). Relative to the reconstruction, model-simulated cooling is three times larger, and tTOA radiation anomalies are 10--20 times larger.

Multiple large eruptions spaced a few years apart can initiate periods of prolonged OHC loss (Fig.~S17; \citealp{Gupta2018}). For example, the 1693 and 1695 eruptions caused a decade of OHC loss, peaking at $70\,\mathrm{ZJ}$ (green line in Fig.~\ref{fig:composite_volcano}b). The recovery from these two eruptions took 30 years due to five additional, smaller eruptions.

\subsection{Consistency between OHC and EEI}

Approximately 90\% of the heat absorbed due to a non-zero EEI is stored in the ocean~\citep{Schuckmann2023}. Recently, 42\% of the accumulated OHC has been stored in the upper 300 m~\citep{Pan2025}. Therefore, a strong correlation between the EEI and the rate of change of the upper-ocean heat content (dOHC300/dt) is expected, particularly at seasonal to annual timescales, as described by Eq. (\ref{eq:eei-decomposition-ohc}). This can provide insight into the connection between Earth's energy budget at the TOA and at the surface, but also serves as a consistency check for the DA procedure. We multiply dOHC/dt by Earth's ocean fraction (71\%) to obtain the equivalent EEI over the total surface area. The derivative is calculated using a central time difference as in \citet{Trenberth2014}.

Reconstructed EEI and dOHC300/dt are correlated annually at $r = 0.64$ ($p \leq 0.01$; Fig.~S18). The volcanic response is similar in shape and timescale, but EEI is weaker than dOHC300/dt (Fig.~\ref{fig:composite_volcano}d). At seasonal timescales, the correlation is $r = 0.50$ ($p \leq 0.01$), but dOHC/dt leads EEI by about one season. With increasing timescale beyond four years, the two quantities decorrelate (Fig.~S18e).

\section{Discussion}
\label{sec:discussion}

Our reconstruction of Earth's energy budget over the last millennium represents a new application of paleoclimate DA. After establishing the reconstruction skill and its limitations, this new dataset allows us to consider several aspects of natural energy budget variability, constrained by proxy information and model dynamics.

\subsection{Reconstruction skill and limitations}
\label{sec:discussion-feasibility}

The reconstruction presented here is the first proxy-constrained dataset of TOA radiation over the last millennium. We use the fact that radiation covaries with surface temperature, to which the proxies are sensitive. However, the proxy information is temporally and spatially sparse, and the link to TOA radiation is mediated by complex and noisy physical processes. Therefore, we rely on pseudoproxy experiments, instrumental validation, and internal consistency to establish TOA radiation reconstruction skill. As in previous reconstructions, surface variables generally agree well with instrumental products and other reconstructions (Figs.~S7--S9 and S11), and withheld proxies for independent validation over the whole last millennium have a significantly positive network-mean correlation (Fig.~S6). However, Antarctic SIC is likely unreliable, and NH extratropical land summer temperatures are too cold.

The energy imbalance at interannual timescales can be reliably reconstructed. This follows from the PPEs (Figs.~\ref{fig:ppe_panels} and S4) and the satellite comparison (Fig.~\ref{fig:deepc_panels}). Interannual variability is dominated by the radiative response to unforced surface temperature variability~\citep{Trenberth2015}, to which our proxies are sensitive. Indeed, our PPEs showed that the radiative response can be reconstructed accurately from proxies (Fig.~S4 and Table~S1), explaining the high interannual EEI skill. The magnitude of volcanic forcing, another important but sporadic source of interannual variability, however, is greatly underestimated. This is the case both in PPEs (Figs.~\ref{fig:ppe_panels} and S4) and in the real reconstruction (Fig.~\ref{fig:composite_volcano}). Nonetheless, a volcanic signal is present and the relationship to other variables is physically consistent. In particular, our DA system is able to capture the complex lead/lag relationship of forcing and response with temperature (Fig.~S4d; \citealp{Proistosescu2018}).

The constituent RSR and OLR of the energy imbalance have good global-mean skill in the PPE (Fig.~\ref{fig:ppe_panels}). In the satellite comparison, correlations are not significant, although this is very likely a result of the short overlap period (Fig.~\ref{fig:deepc_panels}). Spatial OLR skill is higher than spatial RSR skill. This is not surprising since OLR covaries strongly with SAT through the Planck response and is spatially coherent, especially in association with ENSO and other modes of internal variability. RSR has slightly lower correlations because it is generally noisier at seasonal timescales due to extratropical storms and the Madden--Julian oscillation~\citep[e.g.,][]{Trenberth2015}. In the PPE, the EEI skill is lower than RSR and OLR, possibly because the EEI is the small residual of two large quantities.

Consistency between the TOA radiation and OHC, even though not explicitly enforced by the DA system, is further evidence of reconstruction skill (Figs.~S18 and \ref{fig:composite_volcano}d). EEI and dOHC300/dt should correspond since much of the energy imbalance is absorbed into the ocean~\citep{Schuckmann2023,Palmer2014,Allison2020}. Indeed, their seasonal and annual correlation are moderately high ($r = 0.5$ to $0.6$), with a correlation peak at a timescale of 4 years ($r = 0.7$). The correlations between EEI and dOHC/dt are similar for instrumental estimates, and would likely be higher if atmospheric heat storage were considered~\citep{Hakuba2024,Mayer2026}. Up to timescales of 20 years, the correlation is above $0.6$, with a trend toward decorrelation for longer timescales. We speculate that the multidecadal correlation might be higher if OHC were to include the deeper ocean. At seasonal timescales, the EEI lags dOHC/dt by one season (Fig.~S18d). This may be a result of the more direct covariance of OHC with proxies, while EEI mostly follows from LIM dynamics, or an indication that much of the interannual variability in EEI results from internal fluctuations in OHC rather than external forcing (cf. \citealp{Proistosescu2018}).

One caveat of our method is that we can only reconstruct the radiative response and natural forcing, but not the anthropogenic forcing dominating the historical period. This is because the temperature--radiation relationship in the LIM is learned from last-millennium simulations, which do not include a strong greenhouse gas forcing signal since they end in 1850. A nonlinear reconstruction framework may be required to capture relationships that change over time and with forcing type. However, if we add an estimate of the anthropogenic forcing to our reconstructed EEI, the sum agrees well in magnitude and phasing with an OHC-based estimate of the total EEI (Fig.~\ref{fig:timeseries_historical_eei}). Only after 1960 do they diverge as a result of an underestimated radiative response in our reconstruction.

The reconstruction skill diminishes as proxies become more sparse toward the early part of the last millennium. Additionally, there may be a drop in skill before 1100, when no coral proxies are available to constrain the tropical Pacific. This drop is evident in PPEs but not in the withheld-proxy correlations. Further, our reconstruction has a last-millennium warming trend over North America and parts of continental Asia, particularly in summer (Fig.~S14a). This trend may be spurious and would explain the negative correlations for some TRW proxies before 1600 (Fig.~S6d). During the development of our DA system, we also experimented with other PSM calibration datasets and periods. While the presented version agrees best with instrumental temperatures, other versions tended to have more temperature variance, resulting in a colder LIA and a steeper last-millennium cooling trend. We suggest that PSM calibration uncertainties should be considered in future DA reconstructions, for example, by sampling calibration datasets and periods across Monte Carlo iterations, and by making use of recent instrumental temperature ensembles~\citep[e.g.,][]{Chan2024,Lenssen2024}.

We conclude that the reconstruction is skillful for surface fields, the radiative response associated with internal variability, and the timing of volcanic eruptions, despite the spatiotemporal sparseness of the proxy network and the complex relationship between temperature and radiation. However, the magnitude of volcanic forcing is greatly underestimated, and anthropogenic forcing is not represented. Most reliable are global means at interannual to decadal timescales after 1100, although even seasonal fields have considerable skill. The reconstruction fidelity is highest in the tropics and subtropics.

\subsection{Earth's energy budget over the last millennium}

Reconstructions and CMIP5 models agree that there was a global-mean cooling trend over the last millennium, although the patterns of cooling may not have been globally synchronous~\citep{PAGESConsortium2019,Atwood2016,Neukom2019}. This cooling was driven primarily by volcanic forcing, with a potential minor role of greenhouse gas and solar forcing~\citep{Schurer2014,Buntgen2020,Wanner2022,Esper2012}, then amplified by positive feedbacks~\citep{Atwood2016}. Since the climate system is overall stable due to the strongly negative Planck feedback, it responds to cooling with reduced OLR. Indeed, the OLR decreased by $0.35\,\mathrm{W\,m^{-2}\,kyr^{-1}}$, while the RSR only increased by $0.16\,\mathrm{W\,m^{-2}\,kyr^{-1}}$, which explains the increasing EEI trend of $0.18 \pm 0.21\,\mathrm{W\,m^{-2}\,kyr^{-1}}$, although it is not significantly different from zero (Fig.~\ref{fig:trend_zm_panels}). To sustain the cooling trend, the EEI needs to remain negative on average despite the increasing trend, leading to a slowdown of cooling over centuries. Indeed, both our reconstruction and the OHC reconstruction of \citet{Gebbie2019} have a 1100--1850 mean EEI of $-0.1\,\mathrm{W\,m^{-2}}$, and a total integrated energy loss over 850--1850 of $1600$ to $1800\,\mathrm{ZJ}$ (Fig.~S12). For comparison, Earth has gained about $600\,\mathrm{ZJ}$ over the much shorter 1880--2020 period~\citep{Wu2025a}.

While the anthropogenic origin of the modern global warming is well-established~\citep{Masson-Delmotte2021}, it is merely a ``symptom'' of a positive energy imbalance~\citep{Schuckmann2016}. By extending the record of energy budget variability to 850, our reconstruction reveals that the current rate of energy gain, which is the fundamental driver of global climate change~\citep{Mauritsen2025,Schuckmann2016,Trenberth2014}, is unprecedented in the last millennium. Specifically, the energy imbalance over any 20-yr period after 1990, and possibly already after 1915, exceeds any 20-yr average over 850--1850. Moreover, the 2001--2025 energy imbalance trend is unprecedented among all 25-yr EEI trends in our reconstruction, and is in the 97th percentile of last-millennium model variability (Section~4\ref{sec:context-recent-eei}). 

By comparing the spatial structure of the radiation trends (Fig.~S14d--f), we can speculate on the physical processes involved. The OLR trend is negative almost everywhere, presumably due to the Planck feedback. Superimposed is an El Niño-like response, with an eastward shift of the Indo--Pacific convection region toward the Central Pacific~\citep{Hartmann2016}, causing negative OLR trends in regions with enhanced deep convection and positive OLR trend over the West Pacific Warm Pool. The shift is likely driven by west Pacific cooling, which moves the ascending branch of the Walker circulation to the east (Fig.~S14b). In addition to this shift, the EEI has trends in the subtropical east Pacific, likely associated with the subtropical stratus decks. There is also a clear-sky RSR contribution from anomalous Antarctic sea-ice area (Fig.~S13a).

\Citet{Gebbie2019} provide the only other proxy-based reconstruction of the last-millennium energy budget. They propagate SSTs into the ocean interior using a constant circulation model. Annual-mean proxy SSTs from the Ocean2k database~\citep{McGregor2015} are used before 1870, and instrumental winter SSTs after. The SSTs are prescribed as global-mean values rather than spatially resolved fields, but the reconstruction is constrained by basin-wide averages of ship-based vertical temperature profiles in the 1870s. Their SSTs are systematically colder over the last millennium and start to warm about 150 years before most instrumental datasets and our reconstructed SSTs (in 1750 rather than 1900; Fig.~S11d), although recent corrections to instrumental data may support a colder 19th century~\citep{Chan2024,Chan2025}. The cooler SSTs also imply a lower OHC (Fig.~\ref{fig:timeseries_gm}d), but the pre-1850 energy imbalance is not affected by such a mean offset. Our EEI reconstruction is not significantly correlated with theirs, but the mean EEI is similar (Fig.~S12). The lack of correlation and the offset in pre-industrial SSTs may be due to differences in proxy network (e.g., the lack of tropical Pacific proxies in Ocean2k), the prescription of global-mean rather than spatially resolved SSTs, the method of proxy-to-temperature conversion, and the blending of winter and annual SSTs. Ocean2k, on which \citet{Gebbie2019} is based, may not provide a suitable dataset for annual values in temperature units, rather than the 200-yr-binned standard deviations of proxy units presented in \citet{McGregor2015}. However, by resolving the deep ocean, \citet{Gebbie2019} may be able to better constrain the energy imbalance on long timescales.

\subsection{Seasonality in last-millennium trends}

The last-millennium trends of all fields differ consistently by season. The seasonal cycle in the global-mean cooling trend ranges from $-0.16\,\mathrm{K\,kyr^{-1}}$ in MAM and JJA to $-0.19\,\mathrm{K\,kyr^{-1}}$ in SON and DJF (Fig.~\ref{fig:trend_zm_panels}). High-latitude temperature and insolation trends have a lag correlation, with higher insolation corresponding with warmer temperatures (Fig.~\ref{fig:timeseries-highlattemp}). The lag of around one season may be a reconstruction artifact, but it could also be a real property of the high-latitude climate, e.g., due to thermal inertia. In the Arctic seasonal cycle, the annual maximum temperature is nearly synchronous with maximum insolation, but the annual minimum temperature lags insolation by around two months~\citep{Donohoe2020}.

While the annual-mean insolation is approximately constant over the last millennium, there are latitude-dependent seasonal trends in insolation due to axial precession (cf. Fig.~2d in \citealp{Lucke2021}). Proxies will reflect these seasonal trends based on their location and seasonality; seasonal DA explicitly accounting for this fact is therefore is essential to separating seasonal and annual-mean trends.

\subsection{Climate response to volcanic eruptions}

Large volcanic eruptions are the dominant pre-industrial climate forcing~\citep{Buntgen2020}. Clusters of them are thought to be the main cause of the LIA~\citep{Miller2012,Bronnimann2019,Gupta2018} and previous cold periods~\citep{VanDijk2024}. Sea-surface cooling after eruptions is initially damped by the upper ocean and absorbed into the deeper ocean~\citep{Gregory2016}. The cold anomalies at depth re-emerge over subsequent years, extending surface cooling well beyond the volcanic aerosol residence time of 1--2 years. In our volcanic composite, OHC does not recover until 10--20 years after the eruption (Fig.~\ref{fig:composite_volcano}b). A decadal pacing of large eruptions would thus allow heat loss to accumulate and persist~\citep{Gupta2018,Zhong2011}. Our results show OHC anomalies of up to $80\,\mathrm{ZJ}$ for volcanic clusters during the late 1200s, late 1400s, early 1700s, and early 1800s (Fig.~S17). \Citet{Miller2012} hypothesize that the late 1200s cluster (annotated in Fig.~\ref{fig:timeseries_gm}b), with major eruptions in 1276 and 1286, initiated the cooling trend leading to the LIA. Indeed, the 20-yr period with the lowest EEI over the last millennium is 1272--1291 ($-0.4\,\mathrm{W\,m^{-2}}$), with a minimum in 1284 ($-0.7\,\mathrm{W\,m^{-2}}$). This is decades after the 1257 Samalas eruption, which had an exceptional magnitude but a relatively weak climate response (Fig.~S17a; \citealp{Vidal2016,Guillet2017}). The 20-yr period with the largest loss of OHC300 is 1803--1822 ($-0.4\,\mathrm{W\,m^{-2}}$; Fig.~S17d), including the 1815 Tambora eruption, and the recovery from this cluster marks the end of the LIA~\citep{Bronnimann2019}. This early 1800s cluster complicates the definition of a pre-industrial baseline~\citep[e.g.,][]{Thorne2026} and the initialization of historical simulations in 1850~\citep{Ballinger2026}. 

OHC anomalies in our reconstruction differ from model simulations (Figs.~\ref{fig:composite_volcano}, S16, and S17). Simulated ocean cooling peaks 3 years after the eruption and recovers immediately, while the reconstructed cooling peaks at a lower amplitude. In contrast to the simulations, the reconstructed onset of recovery varies by eruption, occurring 3--10 years post-eruption and resulting in the flatter composite (Fig.~\ref{fig:composite_volcano}b). The reconstructed response is likely smoothed by biological memory, or persistence, in TRW proxies, which form the bulk of our network~\citep{Fritts1976,Esper2015,Lucke2019,Zhu2020}. This results in a muted and time-lagged volcanic response, which has long been noted in the literature~\citep[e.g.,][]{Frank2007,Anchukaitis2012,Mann2012,DArrigo2013,Stoffel2015,Anchukaitis2022,He2026}. However, models are also known to overestimate the cooling after large-magnitude eruptions, for example, due to the way volcanic aerosols are prescribed and volcanic forcing is calculated~\citep{Timmreck2009,Timmreck2012,Marotzke2015,LeGrande2016,Marshall2025}. The model--proxy discrepancy is further exacerbated by unforced variability and uncertainty in VSSI and timing~\citep{Zanchettin2019,Lucke2023}, although progress has been made in resolving large eruptions in ice cores~\citep{Burke2019} and reconciling simulated and reconstructed volcanism~\citep{Stoffel2015,Zhu2020}. Over 1960--2000, the timescale and magnitude of OHC loss in our reconstruction agree with instrumental datasets for three large eruptions (Fig.~\ref{fig:timeseries_historical_ohc}).

The loss of OHC is driven by a negative EEI at the TOA, peaking at $-0.2\,\mathrm{W\,m^{-2}}$ in the composite, and arising from increased RSR (Fig.~\ref{fig:composite_volcano}c,d). This is much weaker than expected for eruptions larger than Mt. Pinatubo in 1991, which had a peak EEI of around $-3\,\mathrm{W\,m^{-2}}$~\citep{Allan2014}. After 3--5 years, the EEI becomes marginally positive due to reduced RSR, which is also evident in coupled model simulations, although at much larger magnitudes (Fig.~S16c). By Eq.~(\ref{eq:eei-decomposition-ohc}), the difference between dOHC300/dt $ \approx C(dT/dt)$ and the EEI may be due to heat uptake $\gamma T$ by the deeper ocean. However, the difference here is likely due to the greatly underestimated volcanic forcing. The ocean cooling dOHC300/dt, with a peak of $-0.4\,\mathrm{W\,m^{-2}}$ in the composite, may be a better estimate of the volcanic forcing than the reconstructed EEI.

Previous modeling studies have found expanded sea ice persisting for more than a century following large eruptions due to sea-ice--ocean feedbacks in the Arctic Ocean and North Atlantic~\citep{Zhong2011,Miller2012,Slawinska2018}. In contrast, in our volcanic composite, the sea-ice area in both hemispheres returns to the pre-eruption value after 15--20 years (Fig.~S15c,d). However, \citet{Zhong2011} find that these sea ice changes following volcanic eruptions depend on the initial condition of the coupled atmosphere--ocean--sea-ice system. Centennial sea-ice expansion that sustains cold anomalies could thus be important for some eruptions and could have played a role in initiating the LIA~\citep{Miller2012}.

\subsection{Comparison with last-millennium simulations}

The majority of CMIP6 models do not show the millennial-scale cooling trend found in proxy reconstructions (Fig.~\ref{fig:trend_zm_panels} and Section~4\ref{sec:results-lmtrends}; \citealp{PAGESConsortium2019}), in contrast to CMIP5 models~\citep{Atwood2016}. The lack of a global-mean cooling trend over 850--1850 in models is the main reason for the difference in OHC and TOA radiation with our reconstruction. If we assume that trends of all fields scale linearly with the global-mean temperature trend and adjust the model trends to match the reconstructed temperature trend, agreement improves substantially, particularly for the global means (not shown). The benefit of DA is thus that variability on seasonal to millennial timescales is directly constrained to follow the climate trajectory as sampled by proxies. Model biases may be inherited by the LIMs and thus the reconstruction, which is partly mediated by our multi-model reconstruction.

Our reconstruction provides a new dataset to compare with energy budgets simulated in coupled models. First, since the annual-mean cooling trend is likely volcanically forced~\citep{Schurer2014,Buntgen2020,Wanner2022}, the lack of a simulated trend may imply that the coupled climate system response to volcanic eruptions, such as post-eruption OHC loss, is not well resolved in current models~\citep[e.g.,][]{Church2005}. Other model issues that can lead to discrepancies with proxy reconstructions include uncertainty in the forcing dataset and in the aerosol representation~\citep{Lucke2023,Marshall2022,Timmreck2012}. A false warming trend can also arise from impulsive positive radiative forcing when the prescribed background aerosols from the spin-up run are removed at the start of a transient last-millennium simulation~\citep{Gregory2013,Fyfe2021}. Finally, the initialization of coupled model simulations from equilibrium is likely a poor approximation of the coupled climate system. This critique applies to both historical simulations starting in 1850, which are preceded by strong volcanism and the last-millennium cooling trend~\citep{Ballinger2026,Gebbie2019}, and to last-millennium simulations starting in 850, when the deep ocean is potentially still adjusting to the last deglaciation~\citep{Zhu2019}. Proxy-constrained estimates of the pre-industrial deep ocean may help to provide initial conditions that better reflect earlier climate variability.

\section{Conclusion}
\label{sec:conclusion}

Earth's energy imbalance is a fundamental climate metric, but our understanding of its variability on long timescales is limited by a lack of long records and large uncertainty in coupled climate models. Our reconstruction over the last millennium extends the observational record by 1000 years, providing context for energy budget variability in the absence of anthropogenic forcing. By combining proxy data with a simplified climate model using data assimilation, we obtain a consistent dataset of global surface temperatures, sea ice, top-of-atmosphere radiation, and upper-ocean heat content. In contrast to climate models, the variability on all timescales is directly constrained by proxies to follow the observed climate trajectory. The reconstruction can be used to investigate seasonal to multidecadal variability, millennial trends, and episodic events like volcanic eruptions.

We demonstrate the feasibility of reconstructing top-of-atmosphere radiation from temperature-sensitive proxies, even though the quantification of the energy imbalance and ocean heat content remains challenging during the satellite period~\citep{Hakuba2024}. Through validation in pseudoproxy experiments and against satellite observations, we find that the global-mean energy imbalance, reflected shortwave radiation, and outgoing longwave radiation can be skillfully reconstructed; there is less skill in spatial patterns, especially outside of the tropics. Volcanic forcing events and a subsequent reduction in ocean heat content are captured, but with greatly underestimated magnitude. Since the simplified climate model we use is trained on last-millennium simulations, our reconstruction resolves natural forcing and the radiative response, but not anthropogenic forcing. Compared to the reconstruction of the last millennium by \citet{Gebbie2019}, we find agreement on the magnitude of sustained energy loss, but with large error bars around zero and uncorrelated multidecadal variability, likely due to different proxy networks and reconstruction methods. These uncertainties could potentially be be reduced by including constraints on the deep ocean from proxies, either in situ~\citep[e.g.,][]{Waelbroeck2002,Lear2000} or from globally integrating noble gases~\citep{Bereiter2018a}. These proxies have proven useful for reconstructions of the last deglaciation~\citep{Bereiter2018,Baggenstos2019,Shackleton2023} and the Pleistocene~\citep{Shackleton2026}. 

Our reconstruction provides context for the energy gain in recent decades, which we find to be unprecedented in the context of natural, pre-industrial variability. It also reaffirms the role of volcanism in the last-millennium cooling trend, which may have been continuously forced by volcanic clusters causing an accumulation of heat loss. The discrepancy of our reconstruction with coupled simulations may point to deficiencies in the energy budget representation in the same models used for future climate projections.

\paragraph*{Acknowledgements}
We thank Jess Tierney, Sylvia Dee, and an anonymous reviewer for their comments in review, which greatly improved this manuscript. We thank Kyle Armour, Zilu Meng, Eric Mei, Aaron Donohoe, and Vince Cooper for helpful conversations. Anna Black's master's thesis provided preliminary results that helped motivate this study. This material is based upon work supported by the National Science Foundation under Award No.~2202526. We would like to acknowledge computing support from the Casper system (https://ncar.pub/casper) provided by the NSF National Center for Atmospheric Research (NCAR), sponsored by the National Science Foundation. This work made use of pyleoclim~\citep{Khider2022} and climlab~\citep{Rose2018}.

\paragraph*{Code and data availability} Code is publicly available at \url{https://github.com/DominikStiller/energy-budget-from-proxies}. The reconstruction data will be published once this manuscript has been accepted.

\paragraph*{Copyright notice} This work has been submitted to Journal of Climate. Copyright in this work may be transferred without further notice.

\bibliographystyle{unsrtnat}
\bibliography{references}

@article{Allan2014,
  title = {Changes in Global Net Radiative Imbalance 1985--2012},
  author = {Allan, Richard P. and Liu, Chunlei and Loeb, Norman G. and Palmer, Matthew D. and Roberts, Malcolm and Smith, Doug and Vidale, Pier-Luigi},
  year = 2014,
  month = aug,
  journal = {Geophysical Research Letters},
  volume = {41},
  number = {15},
  pages = {5588--5597},
  issn = {0094-8276, 1944-8007},
  doi = {10.1002/2014GL060962},
  urldate = {2025-08-13},
  copyright = {http://creativecommons.org/licenses/by/3.0/},
  langid = {english}
}

@article{Allison2020,
  title = {Observations of Planetary Heating since the 1980s from Multiple Independent Datasets},
  author = {Allison, Lesley C and Palmer, Matthew D and Allan, Richard P and Hermanson, Leon and Liu, Chunlei and Smith, Doug M},
  year = 2020,
  month = oct,
  journal = {Environmental Research Communications},
  volume = {2},
  number = {10},
  pages = {101001},
  issn = {2515-7620},
  doi = {10.1088/2515-7620/abbb39},
  urldate = {2025-08-30},
  langid = {english}
}

@article{Anchukaitis2012,
  title = {Tree Rings and Volcanic Cooling},
  author = {Anchukaitis, Kevin J. and Breitenmoser, Petra and Briffa, Keith R. and Buchwal, Agata and B{\"u}ntgen, Ulf and Cook, Edward R. and D'Arrigo, Rosanne D. and Esper, Jan and Evans, Michael N. and Frank, David and Grudd, H{\aa}kan and Gunnarson, Bj{\"o}rn E. and Hughes, Malcolm K. and Kirdyanov, Alexander V. and K{\"o}rner, Christian and Krusic, Paul J. and Luckman, Brian and Melvin, Thomas M. and Salzer, Matthew W. and Shashkin, Alexander V. and Timmreck, Claudia and Vaganov, Eugene A. and Wilson, Rob J. S.},
  year = 2012,
  month = dec,
  journal = {Nature Geoscience},
  volume = {5},
  number = {12},
  pages = {836--837},
  issn = {1752-0894, 1752-0908},
  doi = {10.1038/ngeo1645},
  urldate = {2026-01-23},
  copyright = {http://www.springer.com/tdm},
  langid = {english}
}

@article{Anchukaitis2022,
  title = {Progress and Uncertainties in Global and Hemispheric Temperature Reconstructions of the {{Common Era}}},
  author = {Anchukaitis, Kevin J. and Smerdon, Jason E.},
  year = 2022,
  month = jun,
  journal = {Quaternary Science Reviews},
  volume = {286},
  pages = {107537},
  issn = {02773791},
  doi = {10.1016/j.quascirev.2022.107537},
  urldate = {2024-07-17},
  langid = {english}
}

@article{Anderson2012,
  title = {Localization and {{Sampling Error Correction}} in {{Ensemble Kalman Filter Data Assimilation}}},
  author = {Anderson, Jeffrey L.},
  year = 2012,
  month = jul,
  journal = {Monthly Weather Review},
  volume = {140},
  number = {7},
  pages = {2359--2371},
  issn = {0027-0644, 1520-0493},
  doi = {10.1175/MWR-D-11-00013.1},
  urldate = {2025-09-13},
  langid = {english}
}

@article{Atwood2016,
  title = {Quantifying {{Climate Forcings}} and {{Feedbacks}} over the {{Last Millennium}} in the {{CMIP5}}--{{PMIP3 Models}}*},
  author = {Atwood, A. R. and Wu, E. and Frierson, D. M. W. and Battisti, D. S. and Sachs, J. P.},
  year = 2016,
  month = feb,
  journal = {Journal of Climate},
  volume = {29},
  number = {3},
  pages = {1161--1178},
  issn = {0894-8755, 1520-0442},
  doi = {10.1175/JCLI-D-15-0063.1},
  urldate = {2024-06-20},
  langid = {english}
}

@article{Aubry2025a,
  title = {Stratospheric Aerosol Forcing for {{CMIP7}} (Part 1): {{Optical}} Properties for Pre-Industrial, Historical, and Scenario Simulations (Version 2.2.1)},
  author = {Aubry, T. J. and Toohey, M. and Khanal, S. and Chim, M. M. and Verkerk, M. and Johnson, B. and Schmidt, A. and Kovilakam, M. and Sigl, M. and Nicholls, Z. and Thomason, L. and Naik, V. and Rieger, L. and Stiller, D. and Ziegler, E. and Smith, I.},
  year = 2025,
  journal = {EGUsphere},
  volume = {2025},
  pages = {1--53},
  doi = {10.5194/egusphere-2025-4990},
  langid = {english}
}

@article{Baggenstos2019,
  title = {Earth's Radiative Imbalance from the {{Last Glacial Maximum}} to the Present},
  author = {Baggenstos, Daniel and H{\"a}berli, Marcel and Schmitt, Jochen and Shackleton, Sarah A. and Birner, Benjamin and Severinghaus, Jeffrey P. and Kellerhals, Thomas and Fischer, Hubertus},
  year = 2019,
  month = jul,
  journal = {Proceedings of the National Academy of Sciences},
  volume = {116},
  number = {30},
  pages = {14881--14886},
  issn = {0027-8424, 1091-6490},
  doi = {10.1073/pnas.1905447116},
  urldate = {2026-03-12},
  langid = {english}
}

@article{Ballinger2026,
  title = {Importance of Beginning Industrial-Era Climate Simulations in the Eighteenth Century},
  author = {Ballinger, Andrew P and Schurer, Andrew P and Hegerl, Gabriele C and Dittus, Andrea J and Hawkins, Ed and Cornes, Richard and Kent, Elizabeth C and Marshall, Lauren R and Morice, Colin P and Osborn, Timothy J and Rayner, Nick A and Rumbold, Steven T},
  year = 2026,
  month = jan,
  journal = {Environmental Research Letters},
  volume = {21},
  number = {1},
  pages = {14022},
  issn = {1748-9326},
  doi = {10.1088/1748-9326/ae1bbc},
  urldate = {2026-03-21},
  copyright = {https://creativecommons.org/licenses/by/4.0/},
  langid = {english}
}

@article{Bereiter2018,
  title = {Mean Global Ocean Temperatures during the Last Glacial Transition},
  author = {Bereiter, Bernhard and Shackleton, Sarah and Baggenstos, Daniel and Kawamura, Kenji and Severinghaus, Jeff},
  year = 2018,
  month = jan,
  journal = {Nature},
  volume = {553},
  number = {7686},
  pages = {39--44},
  issn = {0028-0836, 1476-4687},
  doi = {10.1038/nature25152},
  urldate = {2025-02-04},
  langid = {english}
}

@article{Bereiter2018a,
  title = {New Methods for Measuring Atmospheric Heavy Noble Gas Isotope and Elemental Ratios in Ice Core Samples},
  author = {Bereiter, Bernhard and Kawamura, Kenji and Severinghaus, Jeffrey P.},
  year = 2018,
  month = may,
  journal = {Rapid Communications in Mass Spectrometry},
  volume = {32},
  number = {10},
  pages = {801--814},
  issn = {0951-4198, 1097-0231},
  doi = {10.1002/rcm.8099},
  urldate = {2026-03-12},
  langid = {english}
}

@article{Berger1991,
  title = {Insolation Values for the Climate of the Last 10 Million Years},
  author = {Berger, A. and Loutre, M.F.},
  year = 1991,
  month = jan,
  journal = {Quaternary Science Reviews},
  volume = {10},
  number = {4},
  pages = {297--317},
  issn = {02773791},
  doi = {10.1016/0277-3791(91)90033-Q},
  urldate = {2025-08-26},
  copyright = {https://www.elsevier.com/tdm/userlicense/1.0/},
  langid = {english}
}

@article{Bhend2012,
  title = {An Ensemble-Based Approach to Climate Reconstructions},
  author = {Bhend, J. and Franke, J. and Folini, D. and Wild, M. and Br{\"o}nnimann, S.},
  year = 2012,
  month = may,
  journal = {Climate of the Past},
  volume = {8},
  number = {3},
  pages = {963--976},
  issn = {1814-9332},
  doi = {10.5194/cp-8-963-2012},
  urldate = {2025-07-20},
  copyright = {https://creativecommons.org/licenses/by/3.0/},
  langid = {english}
}

@article{Brennan2022,
  title = {Reconstructing {{Arctic Sea Ice}} over the {{Common Era Using Data Assimilation}}},
  author = {Brennan, M. Kathleen and Hakim, Gregory J.},
  year = 2022,
  month = feb,
  journal = {Journal of Climate},
  volume = {35},
  number = {4},
  pages = {1231--1247},
  publisher = {American Meteorological Society},
  doi = {10.1175/jcli-d-21-0099.1}
}

@article{Bretherton1999,
  title = {The Effective Number of Spatial Degrees of Freedom of a Time-Varying Field},
  author = {Bretherton, Christopher S. and Widmann, Martin and Dymnikov, Valentin P. and Wallace, John M. and Blad{\'e}, Ileana},
  year = 1999,
  month = jul,
  journal = {Journal of Climate},
  volume = {12},
  number = {7},
  pages = {1990--2009},
  publisher = {American Meteorological Society},
  issn = {1520-0442},
  doi = {10.1175/1520-0442(1999)012<1990:tenosd>2.0.co;2}
}

@article{Bronnimann2019,
  title = {Last Phase of the {{Little Ice Age}} Forced by Volcanic Eruptions},
  author = {Br{\"o}nnimann, Stefan and Franke, J{\"o}rg and Nussbaumer, Samuel U. and Zumb{\"u}hl, Heinz J. and Steiner, Daniel and Trachsel, Mathias and Hegerl, Gabriele C. and Schurer, Andrew and Worni, Matthias and Malik, Abdul and Fl{\"u}ckiger, Julian and Raible, Christoph C.},
  year = 2019,
  month = aug,
  journal = {Nature Geoscience},
  volume = {12},
  number = {8},
  pages = {650--656},
  issn = {1752-0894, 1752-0908},
  doi = {10.1038/s41561-019-0402-y},
  urldate = {2025-03-26},
  langid = {english}
}

@article{Brown2014,
  title = {Top-of-atmosphere Radiative Contribution to Unforced Decadal Global Temperature Variability in Climate Models},
  author = {Brown, Patrick T. and Li, Wenhong and Li, Laifang and Ming, Yi},
  year = 2014,
  month = jul,
  journal = {Geophysical Research Letters},
  volume = {41},
  number = {14},
  pages = {5175--5183},
  issn = {0094-8276, 1944-8007},
  doi = {10.1002/2014GL060625},
  urldate = {2024-07-16},
  copyright = {http://onlinelibrary.wiley.com/termsAndConditions\#vor},
  langid = {english}
}

@article{Buntgen2020,
  title = {Prominent Role of Volcanism in {{Common Era}} Climate Variability and Human History},
  author = {B{\"u}ntgen, Ulf and Arseneault, Dominique and Boucher, {\'E}tienne and Churakova (Sidorova), Olga V. and Gennaretti, Fabio and Crivellaro, Alan and Hughes, Malcolm K. and Kirdyanov, Alexander V. and Klippel, Lara and Krusic, Paul J. and Linderholm, Hans W. and Ljungqvist, Fredrik C. and Ludescher, Josef and McCormick, Michael and Myglan, Vladimir S. and Nicolussi, Kurt and Piermattei, Alma and Oppenheimer, Clive and Reinig, Frederick and Sigl, Michael and Vaganov, Eugene A. and Esper, Jan},
  year = 2020,
  month = dec,
  journal = {Dendrochronologia},
  volume = {64},
  pages = {125757},
  issn = {11257865},
  doi = {10.1016/j.dendro.2020.125757},
  urldate = {2025-08-23},
  langid = {english}
}

@article{Buntgen2021,
  title = {The Influence of Decision-Making in Tree Ring-Based Climate Reconstructions},
  author = {B{\"u}ntgen, Ulf and Allen, Kathy and Anchukaitis, Kevin J. and Arseneault, Dominique and Boucher, {\'E}tienne and Br{\"a}uning, Achim and Chatterjee, Snigdhansu and Cherubini, Paolo and Churakova, Olga V. and Corona, Christophe and Gennaretti, Fabio and Grie{\ss}inger, Jussi and Guillet, Sebastian and Guiot, Joel and Gunnarson, Bj{\"o}rn and Helama, Samuli and Hochreuther, Philipp and Hughes, Malcolm K. and Huybers, Peter and Kirdyanov, Alexander V. and Krusic, Paul J. and Ludescher, Josef and Meier, Wolfgang J.-H. and Myglan, Vladimir S. and Nicolussi, Kurt and Oppenheimer, Clive and Reinig, Frederick and Salzer, Matthew W. and Seftigen, Kristina and Stine, Alexander R. and Stoffel, Markus and St. George, Scott and Tejedor, Ernesto and Trevino, Aleyda and Trouet, Valerie and Wang, Jianglin and Wilson, Rob and Yang, Bao and Xu, Guobao and Esper, Jan},
  year = 2021,
  month = jun,
  journal = {Nature Communications},
  volume = {12},
  number = {1},
  pages = {3411},
  issn = {2041-1723},
  doi = {10.1038/s41467-021-23627-6},
  urldate = {2025-08-28},
  langid = {english}
}

@article{Burke2019,
  title = {Stratospheric Eruptions from Tropical and Extra-Tropical Volcanoes Constrained Using High-Resolution Sulfur Isotopes in Ice Cores},
  author = {Burke, Andrea and Moore, Kathryn A. and Sigl, Michael and Nita, Dan C. and McConnell, Joseph R. and Adkins, Jess F.},
  year = 2019,
  month = sep,
  journal = {Earth and Planetary Science Letters},
  volume = {521},
  pages = {113--119},
  issn = {0012821X},
  doi = {10.1016/j.epsl.2019.06.006},
  urldate = {2026-03-20},
  langid = {english}
}

@article{Ceppi2017,
  title = {Relationship of Tropospheric Stability to Climate Sensitivity and {{Earth}}'s Observed Radiation Budget},
  author = {Ceppi, Paulo and Gregory, Jonathan M.},
  year = 2017,
  month = dec,
  journal = {Proceedings of the National Academy of Sciences},
  volume = {114},
  number = {50},
  pages = {13126--13131},
  issn = {0027-8424, 1091-6490},
  doi = {10.1073/pnas.1714308114},
  urldate = {2025-12-04},
  langid = {english}
}

@article{Chan2024,
  title = {A {{Dynamically Consistent ENsemble}} of {{Temperature}} at the {{Earth}} Surface since 1850 from the {{DCENT}} Dataset},
  author = {Chan, Duo and Gebbie, Geoffrey and Huybers, Peter and Kent, Elizabeth C.},
  year = 2024,
  month = aug,
  journal = {Scientific Data},
  volume = {11},
  number = {1},
  pages = {953--969},
  issn = {2052-4463},
  doi = {10.1038/s41597-024-03742-x},
  urldate = {2026-01-25},
  langid = {english}
}

@article{Chan2025,
  title = {Re-{{Evaluating Historical Sea Surface Temperature Data Sets}}: {{Insights From}} the {{Diurnal Cycle}}, {{Coral Proxy Data}}, and {{Radiative Forcing}}},
  shorttitle = {Re-{{Evaluating Historical Sea Surface Temperature Data Sets}}},
  author = {Chan, Duo and Gebbie, Geoffrey and Huybers, Peter},
  year = 2025,
  month = jul,
  journal = {Geophysical Research Letters},
  volume = {52},
  number = {13},
  pages = {e2025GL116615},
  issn = {0094-8276, 1944-8007},
  doi = {10.1029/2025GL116615},
  urldate = {2025-10-02},
  langid = {english}
}

@misc{Chan2025a,
  title = {{{DCENT-I}}: {{A Globally Infilled Extension}} of the {{Dynamically Consistent ENsemble}} of {{Temperature Dataset}}},
  shorttitle = {Dcent-i},
  author = {Chan, Duo and Chan, Steven and Siddons, Joseph and Cable, Archie and Faulkner, Agnieszka and Cornes, Richard and Kent, Elizabeth C. and Gebbie, Geoffrey and Huybers, Peter},
  year = 2025,
  month = sep,
  publisher = {Atmospheric Sciences},
  doi = {10.31223/X5V16S},
  urldate = {2026-01-25},
  archiveprefix = {Atmospheric Sciences},
  copyright = {https://creativecommons.org/licenses/by/4.0/legalcode},
  langid = {english}
}

@article{Cheng2017,
  title = {Improved Estimates of Ocean Heat Content from 1960 to 2015},
  author = {Cheng, Lijing and Trenberth, Kevin E. and Fasullo, John and Boyer, Tim and Abraham, John and Zhu, Jiang},
  year = 2017,
  month = mar,
  journal = {Science Advances},
  volume = {3},
  number = {3},
  publisher = {American Association for the Advancement of Science (AAAS)},
  issn = {2375-2548},
  doi = {10.1126/sciadv.1601545}
}

@article{Church2005,
  title = {Significant Decadal-Scale Impact of Volcanic Eruptions on Sea Level and Ocean Heat Content},
  author = {Church, John A. and White, Neil J. and Arblaster, Julie M.},
  year = 2005,
  month = nov,
  journal = {Nature},
  volume = {438},
  number = {7064},
  pages = {74--77},
  issn = {0028-0836, 1476-4687},
  doi = {10.1038/nature04237},
  urldate = {2025-08-26},
  copyright = {http://www.springer.com/tdm},
  langid = {english}
}

@article{Cooper2025,
  title = {Monthly {{Sea-Surface Temperature}}, {{Sea Ice}}, and {{Sea-Level Pressure}} over 1850--2023 from {{Coupled Data Assimilation}}},
  author = {Cooper, Vincent T. and Hakim, Gregory J. and Armour, Kyle C.},
  year = 2025,
  month = jul,
  journal = {Journal of Climate},
  issn = {0894-8755, 1520-0442},
  doi = {10.1175/JCLI-D-25-0021.1},
  urldate = {2025-08-25},
  langid = {english}
}

@article{Dai2004,
  title = {A {{Global Dataset}} of {{Palmer Drought Severity Index}} for 1870--2002: {{Relationship}} with {{Soil Moisture}} and {{Effects}} of {{Surface Warming}}},
  shorttitle = {A Global Dataset of Palmer Drought Severity Index for 1870--2002},
  author = {Dai, Aiguo and Trenberth, Kevin E. and Qian, Taotao},
  year = 2004,
  month = dec,
  journal = {Journal of Hydrometeorology},
  volume = {5},
  number = {6},
  pages = {1117--1130},
  issn = {1525-7541, 1525-755X},
  doi = {10.1175/JHM-386.1},
  urldate = {2026-03-17},
  langid = {english}
}

@article{Dalaiden2023,
  title = {An {{Unprecedented Sea Ice Retreat}} in the {{Weddell Sea Driving}} an {{Overall Decrease}} of the {{Antarctic Sea}}-{{Ice Extent Over}} the 20th {{Century}}},
  author = {Dalaiden, Quentin and Rezs{\"o}hazy, Jeanne and Goosse, Hugues and Thomas, Elizabeth R. and Vladimirova, Diana O. and Tetzner, Dieter},
  year = 2023,
  month = nov,
  journal = {Geophysical Research Letters},
  volume = {50},
  number = {21},
  pages = {e2023GL104666},
  issn = {0094-8276, 1944-8007},
  doi = {10.1029/2023GL104666},
  urldate = {2025-10-10},
  langid = {english}
}

@article{Danabasoglu2020,
  title = {The {{Community Earth System Model Version}} 2 ({{CESM2}})},
  author = {Danabasoglu, G. and Lamarque, J.-F. and Bacmeister, J. and Bailey, D. A. and DuVivier, A. K. and Edwards, J. and Emmons, L. K. and Fasullo, J. and Garcia, R. and Gettelman, A. and Hannay, C. and Holland, M. M. and Large, W. G. and Lauritzen, P. H. and Lawrence, D. M. and Lenaerts, J. T. M. and Lindsay, K. and Lipscomb, W. H. and Mills, M. J. and Neale, R. and Oleson, K. W. and Otto-Bliesner, B. and Phillips, A. S. and Sacks, W. and Tilmes, S. and Van Kampenhout, L. and Vertenstein, M. and Bertini, A. and Dennis, J. and Deser, C. and Fischer, C. and Fox-Kemper, B. and Kay, J. E. and Kinnison, D. and Kushner, P. J. and Larson, V. E. and Long, M. C. and Mickelson, S. and Moore, J. K. and Nienhouse, E. and Polvani, L. and Rasch, P. J. and Strand, W. G.},
  year = 2020,
  month = feb,
  journal = {Journal of Advances in Modeling Earth Systems},
  volume = {12},
  number = {2},
  pages = {e2019MS001916},
  issn = {1942-2466, 1942-2466},
  doi = {10.1029/2019MS001916},
  urldate = {2024-08-17},
  langid = {english}
}

@article{DArrigo2008,
  title = {On the `{{Divergence Problem}}' in {{Northern Forests}}: {{A}} Review of the Tree-Ring Evidence and Possible Causes},
  shorttitle = {On the `Divergence Problem' in Northern Forests},
  author = {D'Arrigo, Rosanne and Wilson, Rob and Liepert, Beate and Cherubini, Paolo},
  year = 2008,
  month = feb,
  journal = {Global and Planetary Change},
  volume = {60},
  number = {3-4},
  pages = {289--305},
  issn = {09218181},
  doi = {10.1016/j.gloplacha.2007.03.004},
  urldate = {2026-03-10},
  copyright = {https://www.elsevier.com/tdm/userlicense/1.0/},
  langid = {english}
}

@article{DArrigo2013,
  title = {Volcanic Cooling Signal in Tree Ring Temperature Records for the Past Millennium},
  author = {D'Arrigo, Rosanne and Wilson, Rob and Anchukaitis, Kevin J.},
  year = 2013,
  month = aug,
  journal = {Journal of Geophysical Research: Atmospheres},
  volume = {118},
  number = {16},
  pages = {9001--9010},
  issn = {2169-897X, 2169-8996},
  doi = {10.1002/jgrd.50692},
  urldate = {2026-03-13},
  copyright = {http://onlinelibrary.wiley.com/termsAndConditions\#vor},
  langid = {english}
}

@article{Dee2016,
  title = {On the Utility of Proxy System Models for Estimating Climate States over the Common Era},
  author = {Dee, Sylvia G. and Steiger, Nathan J. and Emile-Geay, Julien and Hakim, Gregory J.},
  year = 2016,
  month = sep,
  journal = {Journal of Advances in Modeling Earth Systems},
  volume = {8},
  number = {3},
  pages = {1164--1179},
  issn = {1942-2466, 1942-2466},
  doi = {10.1002/2016MS000677},
  urldate = {2024-11-25},
  copyright = {http://creativecommons.org/licenses/by-nc-nd/4.0/},
  langid = {english}
}

@article{Dee2020,
  title = {No Consistent {{ENSO}} Response to Volcanic Forcing over the Last Millennium},
  author = {Dee, Sylvia G. and Cobb, Kim M. and {Emile-Geay}, Julien and Ault, Toby R. and Edwards, R. Lawrence and Cheng, Hai and Charles, Christopher D.},
  year = 2020,
  month = mar,
  journal = {Science},
  volume = {367},
  number = {6485},
  pages = {1477--1481},
  issn = {0036-8075, 1095-9203},
  doi = {10.1126/science.aax2000},
  urldate = {2024-11-14},
  langid = {english}
}

@article{Dee2022,
  title = {{{ENSO}}'s {{Response}} to {{Volcanism}} in a {{Data Assimilation-Based Paleoclimate Reconstruction Over}} the {{Common Era}}},
  author = {Dee, Sylvia G. and Steiger, Nathan J.},
  year = 2022,
  month = mar,
  journal = {Paleoceanography and Paleoclimatology},
  volume = {37},
  number = {3},
  pages = {e2021PA004290},
  issn = {2572-4517, 2572-4525},
  doi = {10.1029/2021PA004290},
  urldate = {2026-03-11},
  langid = {english}
}

@misc{Doelling2025,
  title = {{{CERES Energy Balanced}} and {{Filled}} ({{EBAF}}) {{TOA Monthly}} Means Data in {{netCDF Edition4}}.2.1},
  author = {Doelling, David},
  year = 2025,
  publisher = {NASA Langley Atmospheric Science Data Center Distributed Active Archive Center},
  doi = {10.5067/TERRA-AQUA-NOAA20/CERES/EBAF-TOA_L3B004.2.1},
  urldate = {2026-03-18},
  langid = {english}
}

@article{Dong2019,
  title = {Attributing {{Historical}} and {{Future Evolution}} of {{Radiative Feedbacks}} to {{Regional Warming Patterns}} Using a {{Green}}'s {{Function Approach}}: {{The Preeminence}} of the {{Western Pacific}}},
  shorttitle = {Attributing {{Historical}} and {{Future Evolution}} of {{Radiative Feedbacks}} to {{Regional Warming Patterns}} Using a {{Green}}'s {{Function Approach}}},
  author = {Dong, Yue and Proistosescu, Cristian and Armour, Kyle C. and Battisti, David S.},
  year = 2019,
  month = sep,
  journal = {Journal of Climate},
  volume = {32},
  number = {17},
  pages = {5471--5491},
  issn = {0894-8755, 1520-0442},
  doi = {10.1175/JCLI-D-18-0843.1},
  urldate = {2025-08-13},
  langid = {english}
}

@article{Donohoe2020,
  title = {Seasonal {{Asymmetries}} in the {{Lag}} between {{Insolation}} and {{Surface Temperature}}},
  author = {Donohoe, Aaron and Dawson, Eliza and McMurdie, Lynn and Battisti, David S. and Rhines, Andy},
  year = 2020,
  month = may,
  journal = {Journal of Climate},
  volume = {33},
  number = {10},
  pages = {3921--3945},
  issn = {0894-8755, 1520-0442},
  doi = {10.1175/JCLI-D-19-0329.1},
  urldate = {2024-11-12},
  langid = {english}
}

@article{Doscher2022,
  title = {The {{EC-Earth3 Earth}} System Model for the {{Coupled Model Intercomparison Project}} 6},
  author = {D{\"o}scher, Ralf and Acosta, Mario and Alessandri, Andrea and Anthoni, Peter and Arsouze, Thomas and Bergman, Tommi and Bernardello, Raffaele and Boussetta, Souhail and Caron, Louis-Philippe and Carver, Glenn and Castrillo, Miguel and Catalano, Franco and Cvijanovic, Ivana and Davini, Paolo and Dekker, Evelien and {Doblas-Reyes}, Francisco J. and Docquier, David and Echevarria, Pablo and Fladrich, Uwe and {Fuentes-Franco}, Ramon and Gr{\"o}ger, Matthias and V. Hardenberg, Jost and Hieronymus, Jenny and Karami, M. Pasha and Keskinen, Jukka-Pekka and Koenigk, Torben and Makkonen, Risto and Massonnet, Fran{\c c}ois and M{\'e}n{\'e}goz, Martin and Miller, Paul A. and {Moreno-Chamarro}, Eduardo and Nieradzik, Lars and Van Noije, Twan and Nolan, Paul and O'Donnell, Declan and Ollinaho, Pirkka and Van Den Oord, Gijs and Ortega, Pablo and Prims, Oriol Tint{\'o} and Ramos, Arthur and Reerink, Thomas and Rousset, Clement and {Ruprich-Robert}, Yohan and Le Sager, Philippe and Schmith, Torben and Schr{\"o}dner, Roland and Serva, Federico and Sicardi, Valentina and Sloth Madsen, Marianne and Smith, Benjamin and Tian, Tian and Tourigny, Etienne and Uotila, Petteri and Vancoppenolle, Martin and Wang, Shiyu and W{\aa}rlind, David and Will{\'e}n, Ulrika and Wyser, Klaus and Yang, Shuting and {Yepes-Arb{\'o}s}, Xavier and Zhang, Qiong},
  year = 2022,
  month = apr,
  journal = {Geoscientific Model Development},
  volume = {15},
  number = {7},
  pages = {2973--3020},
  issn = {1991-9603},
  doi = {10.5194/gmd-15-2973-2022},
  urldate = {2025-07-05},
  copyright = {https://creativecommons.org/licenses/by/4.0/},
  langid = {english}
}

@article{Ebisuzaki1997,
  title = {A {{Method}} to {{Estimate}} the {{Statistical Significance}} of a {{Correlation When}} the {{Data Are Serially Correlated}}},
  author = {Ebisuzaki, Wesley},
  year = 1997,
  month = sep,
  journal = {Journal of Climate},
  volume = {10},
  number = {9},
  pages = {2147--2153},
  issn = {0894-8755, 1520-0442},
  doi = {10.1175/1520-0442(1997)010<2147:AMTETS>2.0.CO;2},
  urldate = {2025-06-10},
  langid = {english}
}

@article{Erb2022,
  title = {Reconstructing {{Holocene}} Temperatures in Time and Space Using Paleoclimate Data Assimilation},
  author = {Erb, Michael P. and McKay, Nicholas P. and Steiger, Nathan and Dee, Sylvia and Hancock, Chris and Ivanovic, Ruza F. and Gregoire, Lauren J. and Valdes, Paul},
  year = 2022,
  month = dec,
  journal = {Climate of the Past},
  volume = {18},
  number = {12},
  pages = {2599--2629},
  issn = {1814-9332},
  doi = {10.5194/cp-18-2599-2022},
  urldate = {2024-12-13},
  copyright = {https://creativecommons.org/licenses/by/4.0/},
  langid = {english}
}

@misc{ERBEScienceTeam2020a,
  title = {Earth {{Radiation}} Area Average Time Series through {{Wide-field-of-view}} Nonscanner Abroad {{Earth Radiation Budget Satellite Edition}} 4.1},
  author = {{ERBE Science Team}},
  year = 2020,
  publisher = {NASA Langley Atmospheric Science Data Center Distributed Active Archive Center},
  doi = {10.5067/ERBE/S10N_WFOV_SF_ERBS_TS_EDITION4.1},
  urldate = {2025-07-03}
}

@article{Esper2002,
  title = {Low-{{Frequency Signals}} in {{Long Tree-Ring Chronologies}} for {{Reconstructing Past Temperature Variability}}},
  author = {Esper, Jan and Cook, Edward R. and Schweingruber, Fritz H.},
  year = 2002,
  month = mar,
  journal = {Science},
  volume = {295},
  number = {5563},
  pages = {2250--2253},
  issn = {0036-8075, 1095-9203},
  doi = {10.1126/science.1066208},
  urldate = {2025-09-06},
  langid = {english}
}

@article{Esper2012,
  title = {Orbital Forcing of Tree-Ring Data},
  author = {Esper, Jan and Frank, David C. and Timonen, Mauri and Zorita, Eduardo and Wilson, Rob J. S. and Luterbacher, J{\"u}rg and Holzk{\"a}mper, Steffen and Fischer, Nils and Wagner, Sebastian and Nievergelt, Daniel and Verstege, Anne and B{\"u}ntgen, Ulf},
  year = 2012,
  month = dec,
  journal = {Nature Climate Change},
  volume = {2},
  number = {12},
  pages = {862--866},
  issn = {1758-678X, 1758-6798},
  doi = {10.1038/nclimate1589},
  urldate = {2024-09-24},
  copyright = {http://www.springer.com/tdm},
  langid = {english}
}

@article{Esper2015,
  title = {Signals and Memory in Tree-Ring Width and Density Data},
  author = {Esper, Jan and Schneider, Lea and Smerdon, Jason E. and Sch{\"o}ne, Bernd R. and B{\"u}ntgen, Ulf},
  year = 2015,
  month = oct,
  journal = {Dendrochronologia},
  volume = {35},
  pages = {62--70},
  issn = {11257865},
  doi = {10.1016/j.dendro.2015.07.001},
  urldate = {2026-03-13},
  langid = {english}
}

@article{Evans2013,
  title = {Applications of Proxy System Modeling in High Resolution Paleoclimatology},
  author = {Evans, M.N. and {Tolwinski-Ward}, S.E. and Thompson, D.M. and Anchukaitis, K.J.},
  year = 2013,
  month = sep,
  journal = {Quaternary Science Reviews},
  volume = {76},
  pages = {16--28},
  issn = {02773791},
  doi = {10.1016/j.quascirev.2013.05.024},
  urldate = {2024-04-11},
  langid = {english}
}

@article{Evensen1994,
  title = {Sequential Data Assimilation with a Nonlinear Quasi-Geostrophic Model Using {{Monte Carlo}} Methods to Forecast Error Statistics},
  author = {Evensen, Geir},
  year = 1994,
  month = may,
  journal = {Journal of Geophysical Research: Oceans},
  volume = {99},
  number = {C5},
  pages = {10143--10162},
  issn = {0148-0227},
  doi = {10.1029/94JC00572},
  urldate = {2026-03-13},
  copyright = {http://onlinelibrary.wiley.com/termsAndConditions\#vor},
  langid = {english}
}

@article{Eyring2016,
  title = {Overview of the {{Coupled Model Intercomparison Project Phase}} 6 ({{CMIP6}}) Experimental Design and Organization},
  author = {Eyring, Veronika and Bony, Sandrine and Meehl, Gerald A. and Senior, Catherine A. and Stevens, Bjorn and Stouffer, Ronald J. and Taylor, Karl E.},
  year = 2016,
  month = may,
  journal = {Geoscientific Model Development},
  volume = {9},
  number = {5},
  pages = {1937--1958},
  issn = {1991-9603},
  doi = {10.5194/gmd-9-1937-2016},
  urldate = {2024-05-28},
  copyright = {https://creativecommons.org/licenses/by/3.0/},
  langid = {english}
}

@article{Fernandez-Donado2013,
  title = {Large-Scale Temperature Response to External Forcing in Simulations and Reconstructions of the Last Millennium},
  author = {{Fern{\'a}ndez-Donado}, L. and {Gonz{\'a}lez-Rouco}, J. F. and Raible, C. C. and Ammann, C. M. and Barriopedro, D. and {Garc{\'i}a-Bustamante}, E. and Jungclaus, J. H. and Lorenz, S. J. and Luterbacher, J. and Phipps, S. J. and Servonnat, J. and Swingedouw, D. and Tett, S. F. B. and Wagner, S. and Yiou, P. and Zorita, E.},
  year = 2013,
  month = feb,
  journal = {Climate of the Past},
  volume = {9},
  number = {1},
  pages = {393--421},
  issn = {1814-9332},
  doi = {10.5194/cp-9-393-2013},
  urldate = {2024-06-19},
  copyright = {https://creativecommons.org/licenses/by/3.0/},
  langid = {english}
}

@incollection{Forster2021,
  title = {The {{Earth}}'s Energy Budget, Climate Feedbacks, and Climate Sensitivity},
  booktitle = {Climate {{Change}} 2021: The {{Physical Science Basis}}. {{Contribution}} of {{Working Group I}} to the {{Sixth Assessment Report}} of the {{Intergovernmental Panel}} on {{Climate Change}}},
  author = {Forster, Piers and Storelvmo, Trude and Armour, Kyle and Collins, William and Dufresne, Jean-Luis and Frame, David and Lunt, Daniel J. and Mauritsen, Thorsten and Palmer, Matthew D. and Watanabe, Masahiro and Wild, Martin and Zhang, Xuebin},
  editor = {{Masson-Delmotte}, Val{\'e}rie and Zhai, Panmao and Pirani, Anna and Connors, Sarah L. and P{\'e}an, Clotilde and Berger, Sophie and Caud, Nada and Chen, Yang and Goldfarb, Leah and Gomis, Melissa I. and Huang, Mengtian and Leitzell, Katherine and Lonnoy, Elisabeth and Matthews, J. B. Robin and Maycock, Thomas K. and Waterfield, Tim and Yelek{\c c}i, {\"O}zge and Yu, Rong and Zhou, Botao},
  year = 2021,
  pages = {923--1054},
  publisher = {Cambridge University Press},
  address = {Cambridge, United Kingdom and New York, NY, USA},
  doi = {10.1017/9781009157896.001},
  langid = {english}
}

@article{Forster2025,
  title = {Indicators of {{Global Climate Change}} 2024: Annual Update of Key Indicators of the State of the Climate System and Human Influence},
  shorttitle = {Indicators of {{Global Climate Change}} 2024},
  author = {Forster, Piers M. and Smith, Chris and Walsh, Tristram and Lamb, William F. and Lamboll, Robin and Cassou, Christophe and Hauser, Mathias and Hausfather, Zeke and Lee, June-Yi and Palmer, Matthew D. and Von Schuckmann, Karina and Slangen, Aim{\'e}e B. A. and Szopa, Sophie and Trewin, Blair and Yun, Jeongeun and Gillett, Nathan P. and Jenkins, Stuart and Matthews, H. Damon and Raghavan, Krishnan and Ribes, Aur{\'e}lien and Rogelj, Joeri and Rosen, Debbie and Zhang, Xuebin and Allen, Myles and Aleluia Reis, Lara and Andrew, Robbie M. and Betts, Richard A. and Borger, Alex and Broersma, Jiddu A. and Burgess, Samantha N. and Cheng, Lijing and Friedlingstein, Pierre and Domingues, Catia M. and Gambarini, Marco and Gasser, Thomas and G{\"u}tschow, Johannes and Ishii, Masayoshi and Kadow, Christopher and Kennedy, John and Killick, Rachel E. and Krummel, Paul B. and Lin{\'e}, Aur{\'e}lien and Monselesan, Didier P. and Morice, Colin and M{\"u}hle, Jens and Naik, Vaishali and Peters, Glen P. and Pirani, Anna and Pongratz, Julia and Minx, Jan C. and Rigby, Matthew and Rohde, Robert and Savita, Abhishek and Seneviratne, Sonia I. and Thorne, Peter and Wells, Christopher and Western, Luke M. and Van Der Werf, Guido R. and Wijffels, Susan E. and {Masson-Delmotte}, Val{\'e}rie and Zhai, Panmao},
  year = 2025,
  month = jun,
  journal = {Earth System Science Data},
  volume = {17},
  number = {6},
  pages = {2641--2680},
  issn = {1866-3516},
  doi = {10.5194/essd-17-2641-2025},
  urldate = {2025-08-30},
  copyright = {https://creativecommons.org/licenses/by/4.0/},
  langid = {english}
}

@article{Frank2007,
  title = {Warmer Early Instrumental Measurements versus Colder Reconstructed Temperatures: Shooting at a Moving Target},
  shorttitle = {Warmer Early Instrumental Measurements versus Colder Reconstructed Temperatures},
  author = {Frank, David and B{\"u}ntgen, Ulf and B{\"o}hm, Reinhard and Maugeri, Maurizio and Esper, Jan},
  year = 2007,
  month = dec,
  journal = {Quaternary Science Reviews},
  volume = {26},
  number = {25-28},
  pages = {3298--3310},
  issn = {02773791},
  doi = {10.1016/j.quascirev.2007.08.002},
  urldate = {2026-03-13},
  copyright = {https://www.elsevier.com/tdm/userlicense/1.0/},
  langid = {english}
}

@article{Franke2017,
  title = {A Monthly Global Paleo-Reanalysis of the Atmosphere from 1600 to 2005 for Studying Past Climatic Variations},
  author = {Franke, J{\"o}rg and Br{\"o}nnimann, Stefan and Bhend, Jonas and Brugnara, Yuri},
  year = 2017,
  month = jun,
  journal = {Scientific Data},
  volume = {4},
  number = {1},
  pages = {170076},
  issn = {2052-4463},
  doi = {10.1038/sdata.2017.76},
  urldate = {2026-03-16},
  langid = {english}
}

@book{Fritts1976,
  title = {Tree Rings and Climate},
  author = {Fritts, Harold C.},
  year = 1976,
  publisher = {Academic Press},
  address = {London ; New York},
  isbn = {978-0-12-268450-0},
  langid = {english},
  lccn = {QC884.2.D4 F74}
}

@article{Fyfe2013,
  title = {Overestimated Global Warming over the Past 20 Years},
  author = {Fyfe, John C. and Gillett, Nathan P. and Zwiers, Francis W.},
  year = 2013,
  month = sep,
  journal = {Nature Climate Change},
  volume = {3},
  number = {9},
  pages = {767--769},
  issn = {1758-678X, 1758-6798},
  doi = {10.1038/nclimate1972},
  urldate = {2026-03-15},
  copyright = {http://www.springer.com/tdm},
  langid = {english}
}

@article{Fyfe2021,
  title = {Significant Impact of Forcing Uncertainty in a Large Ensemble of Climate Model Simulations},
  author = {Fyfe, John C. and Kharin, Viatcheslav V. and Santer, Benjamin D. and Cole, Jason N. S. and Gillett, Nathan P.},
  year = 2021,
  month = jun,
  journal = {Proceedings of the National Academy of Sciences},
  volume = {118},
  number = {23},
  pages = {e2016549118},
  issn = {0027-8424, 1091-6490},
  doi = {10.1073/pnas.2016549118},
  urldate = {2025-09-07},
  langid = {english}
}

@article{Gebbie2019,
  title = {The {{Little Ice Age}} and 20th-Century Deep {{Pacific}} Cooling},
  author = {Gebbie, G. and Huybers, P.},
  year = 2019,
  month = jan,
  journal = {Science (New York, N.Y.)},
  volume = {363},
  number = {6422},
  pages = {70--74},
  publisher = {American Association for the Advancement of Science (AAAS)},
  issn = {1095-9203},
  doi = {10.1126/science.aar8413}
}

@article{Gebbie2021,
  title = {Combining Modern and Paleoceanographic Perspectives on Ocean Heat Uptake},
  author = {Gebbie, Geoffrey},
  year = 2021,
  month = jan,
  journal = {Annual Review of Marine Science},
  volume = {13},
  number = {1},
  pages = {255--281},
  publisher = {Annual Reviews},
  doi = {10.1146/annurev-marine-010419-010844}
}

@article{Geoffroy2013,
  title = {Transient {{Climate Response}} in a {{Two-Layer Energy-Balance Model}}. {{Part I}}: {{Analytical Solution}} and {{Parameter Calibration Using CMIP5 AOGCM Experiments}}},
  shorttitle = {Transient {{Climate Response}} in a {{Two-Layer Energy-Balance Model}}. {{Part I}}},
  author = {Geoffroy, O. and {Saint-Martin}, D. and Olivi{\'e}, D. J. L. and Voldoire, A. and Bellon, G. and Tyt{\'e}ca, S.},
  year = 2013,
  month = mar,
  journal = {Journal of Climate},
  volume = {26},
  number = {6},
  pages = {1841--1857},
  issn = {0894-8755, 1520-0442},
  doi = {10.1175/JCLI-D-12-00195.1},
  urldate = {2025-07-05},
  langid = {english}
}

@article{Gillett2016,
  title = {The {{Detection}} and {{Attribution Model Intercomparison Project}} ({{DAMIP}} v1.0)Contribution to {{CMIP6}}},
  author = {Gillett, Nathan P. and Shiogama, Hideo and Funke, Bernd and Hegerl, Gabriele and Knutti, Reto and Matthes, Katja and Santer, Benjamin D. and Stone, Daithi and Tebaldi, Claudia},
  year = 2016,
  month = oct,
  journal = {Geoscientific Model Development},
  volume = {9},
  number = {10},
  pages = {3685--3697},
  issn = {1991-9603},
  doi = {10.5194/gmd-9-3685-2016},
  urldate = {2026-03-10},
  copyright = {https://creativecommons.org/licenses/by/3.0/},
  langid = {english}
}

@misc{GISTEMPTeam2025,
  title = {{{GISS Surface Temperature Analysis}} ({{GISTEMP}}), Version 4},
  author = {{GISTEMP Team}},
  year = 2025
}

@article{Goessling2025,
  title = {Recent Global Temperature Surge Intensified by Record-Low Planetary Albedo},
  author = {Goessling, Helge F. and Rackow, Thomas and Jung, Thomas},
  year = 2025,
  month = jan,
  journal = {Science},
  volume = {387},
  number = {6729},
  pages = {68--73},
  issn = {0036-8075, 1095-9203},
  doi = {10.1126/science.adq7280},
  urldate = {2025-01-25},
  langid = {english}
}

@article{Goosse2010,
  title = {Reconstructing Surface Temperature Changes over the Past 600 Years Using Climate Model Simulations with Data Assimilation},
  author = {Goosse, H. and Crespin, E. and {de Montety}, A. and Mann, M. E. and Renssen, H. and Timmermann, A.},
  year = 2010,
  month = may,
  journal = {Journal of Geophysical Research: Atmospheres},
  volume = {115},
  number = {D9},
  publisher = {American Geophysical Union (AGU)},
  issn = {0148-0227},
  doi = {10.1029/2009jd012737}
}

@article{Gregory2013,
  title = {Climate Models without Preindustrial Volcanic Forcing Underestimate Historical Ocean Thermal Expansion},
  author = {Gregory, J.M. and Bi, D. and Collier, M.A. and Dix, M.R. and Hirst, A.C. and Hu, A. and Huber, M. and Knutti, R. and Marsland, S.J. and Meinshausen, M. and Rashid, H.A. and Rotstayn, L.D. and Schurer, A. and Church, J.A.},
  year = 2013,
  month = apr,
  journal = {Geophysical Research Letters},
  volume = {40},
  number = {8},
  pages = {1600--1604},
  issn = {0094-8276, 1944-8007},
  doi = {10.1002/grl.50339},
  urldate = {2025-09-07},
  copyright = {http://onlinelibrary.wiley.com/termsAndConditions\#vor},
  langid = {english}
}

@article{Gregory2016,
  title = {Small Global-Mean Cooling Due to Volcanic Radiative Forcing},
  author = {Gregory, J. M. and Andrews, T. and Good, P. and Mauritsen, T. and Forster, P. M.},
  year = 2016,
  month = dec,
  journal = {Climate Dynamics},
  volume = {47},
  number = {12},
  pages = {3979--3991},
  publisher = {{Springer Science and Business Media LLC}},
  issn = {0930-7575, 1432-0894},
  doi = {10.1007/s00382-016-3055-1},
  urldate = {2025-07-11},
  copyright = {http://creativecommons.org/licenses/by/4.0},
  langid = {english}
}

@article{Guillet2017,
  title = {Climate Response to the {{Samalas}} Volcanic Eruption in 1257 Revealed by Proxy Records},
  author = {Guillet, S{\'e}bastien and Corona, Christophe and Stoffel, Markus and Khodri, Myriam and Lavigne, Franck and Ortega, Pablo and Eckert, Nicolas and Sielenou, Pascal Dkengne and Daux, Val{\'e}rie and Churakova~(Sidorova), Olga~V. and Davi, Nicole and Edouard, Jean-Louis and Zhang, Yong and Luckman, Brian~H. and Myglan, Vladimir S. and Guiot, Jo{\"e}l and Beniston, Martin and {Masson-Delmotte}, Val{\'e}rie and Oppenheimer, Clive},
  year = 2017,
  month = feb,
  journal = {Nature Geoscience},
  volume = {10},
  number = {2},
  pages = {123--128},
  issn = {1752-0894, 1752-0908},
  doi = {10.1038/ngeo2875},
  urldate = {2024-09-24},
  langid = {english}
}

@article{Gupta2018,
  title = {The {{Climate Response}} to {{Multiple Volcanic Eruptions Mediated}} by {{Ocean Heat Uptake}}: {{Damping Processes}} and {{Accumulation Potential}}},
  shorttitle = {The {{Climate Response}} to {{Multiple Volcanic Eruptions Mediated}} by {{Ocean Heat Uptake}}},
  author = {Gupta, Mukund and Marshall, John},
  year = 2018,
  month = nov,
  journal = {Journal of Climate},
  volume = {31},
  number = {21},
  pages = {8669--8687},
  issn = {0894-8755, 1520-0442},
  doi = {10.1175/JCLI-D-17-0703.1},
  urldate = {2024-07-12},
  copyright = {http://www.ametsoc.org/PUBSReuseLicenses},
  langid = {english}
}

@article{Hajima2020,
  title = {Development of the {{MIROC-ES2L Earth}} System Model and the Evaluation of Biogeochemical Processes and Feedbacks},
  author = {Hajima, Tomohiro and Watanabe, Michio and Yamamoto, Akitomo and Tatebe, Hiroaki and Noguchi, Maki A. and Abe, Manabu and Ohgaito, Rumi and Ito, Akinori and Yamazaki, Dai and Okajima, Hideki and Ito, Akihiko and Takata, Kumiko and Ogochi, Koji and Watanabe, Shingo and Kawamiya, Michio},
  year = 2020,
  month = may,
  journal = {Geoscientific Model Development},
  volume = {13},
  number = {5},
  pages = {2197--2244},
  issn = {1991-9603},
  doi = {10.5194/gmd-13-2197-2020},
  urldate = {2025-02-17},
  copyright = {https://creativecommons.org/licenses/by/4.0/},
  langid = {english}
}

@article{Hakim2016,
  title = {The Last Millennium Climate Reanalysis Project: {{Framework}} and First Results},
  author = {Hakim, Gregory J. and {Emile-Geay}, Julien and Steig, Eric J. and Noone, David and Anderson, David M. and Tardif, Robert and Steiger, Nathan and Perkins, Walter A.},
  year = 2016,
  month = jun,
  journal = {Journal of Geophysical Research: Atmospheres},
  volume = {121},
  number = {12},
  pages = {6745--6764},
  publisher = {American Geophysical Union (AGU)},
  doi = {10.1002/2016jd024751}
}

@article{Hakim2022,
  title = {Subseasonal Forecast Skill Improvement from Strongly Coupled Data Assimilation with a Linear Inverse Model},
  author = {Hakim, Gregory J. and Snyder, Chris and Penny, Stephen G. and Newman, Matthew},
  year = 2022,
  month = jun,
  journal = {Geophysical Research Letters},
  volume = {49},
  number = {11},
  publisher = {American Geophysical Union (AGU)},
  doi = {10.1029/2022gl097996}
}

@article{Hakuba2024,
  title = {Trends and {{Variability}} in {{Earth}}'s {{Energy Imbalance}} and {{Ocean Heat Uptake Since}} 2005},
  author = {Hakuba, Maria Z. and Fourest, S{\'e}bastien and Boyer, Tim and Meyssignac, Benoit and Carton, James A. and Forget, Ga{\"e}l and Cheng, Lijing and Giglio, Donata and Johnson, Gregory C. and Kato, Seiji and Killick, Rachel E. and Kolodziejczyk, Nicolas and Kuusela, Mikael and Landerer, Felix and Llovel, William and Locarnini, Ricardo and Loeb, Norman and Lyman, John M. and Mishonov, Alexey and Pilewskie, Peter and Reagan, James and Storto, Andrea and Sukianto, Thea and Von Schuckmann, Karina},
  year = 2024,
  month = dec,
  journal = {Surveys in Geophysics},
  volume = {45},
  number = {6},
  pages = {1721--1756},
  issn = {0169-3298, 1573-0956},
  doi = {10.1007/s10712-024-09849-5},
  urldate = {2025-12-19},
  langid = {english}
}

@book{Hartmann2016,
  title = {Global Physical Climatology},
  author = {Hartmann, Dennis L.},
  year = 2016,
  edition = {Second edition},
  publisher = {Elsevier},
  address = {Amsterdam ; Boston},
  isbn = {978-0-12-328531-7},
  lccn = {QC981 .H32 2016}
}

@article{He2026,
  title = {Comparison of {{Global Climatic Responses}} to {{Large Tropical Volcanic Eruptions}} over the {{Last Millennium}} in {{Paleoclimatic Reconstructions}} and {{Model Simulations}}},
  author = {He, Zhaoxiangrui and Tejedor, Ernesto and Smerdon, Jason E. and Vuille, Mathias and Polvani, Lorenzo M. and Seager, Richard and Sugiura, Ibuki},
  year = 2026,
  month = mar,
  journal = {Journal of Climate},
  volume = {39},
  number = {5},
  pages = {1295--1313},
  issn = {0894-8755, 1520-0442},
  doi = {10.1175/JCLI-D-25-0179.1},
  urldate = {2026-03-11},
  copyright = {http://www.ametsoc.org/PUBSReuseLicenses},
  langid = {english}
}

@article{Hodnebrog2024,
  title = {Recent Reductions in Aerosol Emissions Have Increased {{Earth}}'s Energy Imbalance},
  author = {Hodnebrog, {\O}ivind and Myhre, Gunnar and Jouan, Caroline and Andrews, Timothy and Forster, Piers M. and Jia, Hailing and Loeb, Norman G. and Olivi{\'e}, Dirk J. L. and Paynter, David and Quaas, Johannes and Raghuraman, Shiv Priyam and Schulz, Michael},
  year = 2024,
  month = apr,
  journal = {Communications Earth \& Environment},
  volume = {5},
  number = {1},
  pages = {166},
  issn = {2662-4435},
  doi = {10.1038/s43247-024-01324-8},
  urldate = {2025-09-03},
  langid = {english}
}

@article{Houtekamer2016,
  title = {Review of the Ensemble Kalman Filter for Atmospheric Data Assimilation},
  author = {Houtekamer, P. L. and Zhang, Fuqing},
  year = 2016,
  month = nov,
  journal = {Monthly Weather Review},
  volume = {144},
  number = {12},
  pages = {4489--4532},
  publisher = {American Meteorological Society},
  issn = {1520-0493},
  doi = {10.1175/mwr-d-15-0440.1}
}

@article{Huang2017,
  title = {Extended {{Reconstructed Sea Surface Temperature}}, {{Version}} 5 ({{ERSSTv5}}): {{Upgrades}}, {{Validations}}, and {{Intercomparisons}}},
  shorttitle = {Extended {{Reconstructed Sea Surface Temperature}}, {{Version}} 5 ({{ERSSTv5}})},
  author = {Huang, Boyin and Thorne, Peter W. and Banzon, Viva F. and Boyer, Tim and Chepurin, Gennady and Lawrimore, Jay H. and Menne, Matthew J. and Smith, Thomas M. and Vose, Russell S. and Zhang, Huai-Min},
  year = 2017,
  month = oct,
  journal = {Journal of Climate},
  volume = {30},
  number = {20},
  pages = {8179--8205},
  issn = {0894-8755, 1520-0442},
  doi = {10.1175/JCLI-D-16-0836.1},
  urldate = {2025-01-01},
  copyright = {http://www.ametsoc.org/PUBSReuseLicenses},
  langid = {english}
}

@article{Huang2025,
  title = {Extended {{Reconstructed Sea Surface Temperature}}, {{Version}} 6 ({{ERSSTv6}}). {{Part II}}: {{Upgrades}} on {{Quality Control}} and {{Large-Scale Filter}}},
  shorttitle = {Extended Reconstructed Sea Surface Temperature, Version 6 ({{ERSSTv6}}). {{Part II}}},
  author = {Huang, Boyin and Yin, Xungang and Boyer, Tim and Liu, Chunying and Menne, Matthew and Rao, Yuhan Douglas and Smith, Thomas and Vose, Russell and Zhang, Huai-Min},
  year = 2025,
  month = feb,
  journal = {Journal of Climate},
  volume = {38},
  number = {4},
  pages = {1123--1136},
  issn = {0894-8755, 1520-0442},
  doi = {10.1175/JCLI-D-24-0185.1},
  urldate = {2026-01-25},
  copyright = {http://www.ametsoc.org/PUBSReuseLicenses},
  langid = {english}
}

@article{Huang2025a,
  title = {Extended {{Reconstructed Sea Surface Temperature}}, {{Version}} 6 ({{ERSSTv6}}). {{Part I}}: {{An Artificial Neural Network Approach}}},
  shorttitle = {Extended {{Reconstructed Sea Surface Temperature}}, {{Version}} 6 ({{ERSSTv6}}). {{Part I}}},
  author = {Huang, Boyin and Yin, Xungang and Boyer, Tim and Liu, Chunying and Menne, Matthew and Rao, Yuhan Douglas and Smith, Thomas and Vose, Russell and Zhang, Huai-Min},
  year = 2025,
  month = feb,
  journal = {Journal of Climate},
  volume = {38},
  number = {4},
  pages = {1105--1121},
  issn = {0894-8755, 1520-0442},
  doi = {10.1175/JCLI-D-23-0707.1},
  urldate = {2026-01-25},
  copyright = {http://www.ametsoc.org/PUBSReuseLicenses},
  langid = {english}
}

@article{Huntley2010,
  title = {Assimilation of Time-Averaged Observations in a Quasi-Geostrophic Atmospheric Jet Model},
  author = {Huntley, Helga S. and Hakim, Gregory J.},
  year = 2010,
  month = nov,
  journal = {Climate Dynamics},
  volume = {35},
  number = {6},
  pages = {995--1009},
  issn = {0930-7575, 1432-0894},
  doi = {10.1007/s00382-009-0714-5},
  urldate = {2024-03-25},
  langid = {english}
}

@article{Jeevanjee2025,
  title = {A {{Holistic View}} of {{Climate Sensitivity}}},
  author = {Jeevanjee, Nadir and Paynter, David J. and Dunne, John P. and Sentman, Lori T. and Krasting, John P.},
  year = 2025,
  month = may,
  journal = {Annual Review of Earth and Planetary Sciences},
  volume = {53},
  number = {1},
  pages = {367--396},
  publisher = {Annual Reviews},
  issn = {0084-6597, 1545-4495},
  doi = {10.1146/annurev-earth-040523-014302},
  urldate = {2025-07-12},
  copyright = {http://creativecommons.org/licenses/by/4.0/},
  langid = {english}
}

@article{Jenkins2021,
  title = {The {{Impact}} of {{Sea}}-{{Ice Loss}} on {{Arctic Climate Feedbacks}} and {{Their Role}} for {{Arctic Amplification}}},
  author = {Jenkins, Matthew and Dai, Aiguo},
  year = 2021,
  month = aug,
  journal = {Geophysical Research Letters},
  volume = {48},
  number = {15},
  pages = {e2021GL094599},
  issn = {0094-8276, 1944-8007},
  doi = {10.1029/2021GL094599},
  urldate = {2025-06-10},
  langid = {english}
}

@article{Johnson2023,
  title = {Closure of {{Earth}}'s {{Global Seasonal Cycle}} of {{Energy Storage}}},
  author = {Johnson, Gregory C. and Landerer, Felix W. and Loeb, Norman G. and Lyman, John M. and Mayer, Michael and Swann, Abigail L. S. and Zhang, Jinlun},
  year = 2023,
  month = jul,
  journal = {Surveys in Geophysics},
  issn = {0169-3298, 1573-0956},
  doi = {10.1007/s10712-023-09797-6},
  urldate = {2024-06-03},
  langid = {english}
}

@article{Judd2024,
  title = {A 485-Million-Year History of {{Earth}}'s Surface Temperature},
  author = {Judd, Emily J. and Tierney, Jessica E. and Lunt, Daniel J. and Monta{\~n}ez, Isabel P. and Huber, Brian T. and Wing, Scott L. and Valdes, Paul J.},
  year = 2024,
  month = sep,
  journal = {Science},
  volume = {385},
  number = {6715},
  pages = {eadk3705},
  issn = {0036-8075, 1095-9203},
  doi = {10.1126/science.adk3705},
  urldate = {2024-09-20},
  langid = {english}
}

@article{Jungclaus2017,
  title = {The {{PMIP4}} Contribution to {{CMIP6}} -- {{Part}} 3: {{The}} Last Millennium, Scientific Objective, and Experimental Design for the {{PMIP4}} Past1000 Simulations},
  author = {Jungclaus, Johann H. and Bard, Edouard and Baroni, M{\'e}lanie and Braconnot, Pascale and Cao, Jian and Chini, Louise P. and Egorova, Tania and Evans, Michael and {Gonz{\'a}lez-Rouco}, J. Fidel and Goosse, Hugues and Hurtt, George C. and Joos, Fortunat and Kaplan, Jed O. and Khodri, Myriam and Klein Goldewijk, Kees and Krivova, Natalie and LeGrande, Allegra N. and Lorenz, Stephan J. and Luterbacher, J{\"u}rg and Man, Wenmin and Maycock, Amanda C. and Meinshausen, Malte and Moberg, Anders and Muscheler, Raimund and {Nehrbass-Ahles}, Christoph and {Otto-Bliesner}, Bette I. and Phipps, Steven J. and Pongratz, Julia and Rozanov, Eugene and Schmidt, Gavin A. and Schmidt, Hauke and Schmutz, Werner and Schurer, Andrew and Shapiro, Alexander I. and Sigl, Michael and Smerdon, Jason E. and Solanki, Sami K. and Timmreck, Claudia and Toohey, Matthew and Usoskin, Ilya G. and Wagner, Sebastian and Wu, Chi-Ju and Yeo, Kok Leng and Zanchettin, Davide and Zhang, Qiong and Zorita, Eduardo},
  year = 2017,
  month = nov,
  journal = {Geoscientific Model Development},
  volume = {10},
  number = {11},
  pages = {4005--4033},
  publisher = {Copernicus GmbH},
  issn = {1991-9603},
  doi = {10.5194/gmd-10-4005-2017}
}

@book{Kalnay2024,
  title = {Earth {{System Modeling}}, {{Data Assimilation}} and {{Predictability}}: {{Atmosphere}}, {{Oceans}}, {{Land}} and {{Human Systems}}},
  shorttitle = {Earth System Modeling, Data Assimilation and Predictability},
  author = {Kalnay, Eugenia and Mote, Safa and Da, Cheng},
  year = 2024,
  month = oct,
  edition = {2},
  publisher = {Cambridge University Press},
  doi = {10.1017/9780511920608},
  urldate = {2025-09-06},
  copyright = {https://www.cambridge.org/core/terms},
  isbn = {978-0-511-92060-8},
  langid = {english}
}

@article{Khider2022,
  title = {Pyleoclim: Paleoclimate Timeseries Analysis and Visualization with Python},
  shorttitle = {Pyleoclim},
  author = {Khider, Deborah and Emile-Geay, Julien and Zhu, Feng and James, Alexander and Landers, Jordan and Ratnakar, Varun and Gil, Yolanda},
  year = 2022,
  month = oct,
  journal = {Paleoceanography and Paleoclimatology},
  volume = {37},
  number = {10},
  publisher = {American Geophysical Union (AGU)},
  issn = {2572-4517, 2572-4525},
  doi = {10.1029/2022pa004509},
  urldate = {2025-07-21},
  copyright = {http://creativecommons.org/licenses/by-nc/4.0/},
  langid = {english}
}

@article{Lean2018,
  title = {Observation-based Detection and Attribution of 21st Century Climate Change},
  author = {Lean, Judith L.},
  year = 2018,
  month = mar,
  journal = {WIREs Climate Change},
  volume = {9},
  number = {2},
  pages = {e511},
  issn = {1757-7780, 1757-7799},
  doi = {10.1002/wcc.511},
  urldate = {2025-08-26},
  copyright = {http://onlinelibrary.wiley.com/termsAndConditions\#vor},
  langid = {english}
}

@article{Lear2000,
  title = {Cenozoic {{Deep-Sea Temperatures}} and {{Global Ice Volumes}} from {{Mg}}/{{Ca}} in {{Benthic Foraminiferal Calcite}}},
  author = {Lear, C. H. and Elderfield, H. and Wilson, P. A.},
  year = 2000,
  month = jan,
  journal = {Science},
  volume = {287},
  number = {5451},
  pages = {269--272},
  issn = {0036-8075, 1095-9203},
  doi = {10.1126/science.287.5451.269},
  urldate = {2026-03-12},
  langid = {english}
}

@article{LeGrande2016,
  title = {Role of Atmospheric Chemistry in the Climate Impacts of Stratospheric Volcanic Injections},
  author = {LeGrande, Allegra N. and Tsigaridis, Kostas and Bauer, Susanne E.},
  year = 2016,
  month = sep,
  journal = {Nature Geoscience},
  volume = {9},
  number = {9},
  pages = {652--655},
  issn = {1752-0894, 1752-0908},
  doi = {10.1038/ngeo2771},
  urldate = {2026-03-13},
  langid = {english}
}

@article{Lenssen2024,
  title = {A {{NASA GISTEMPv4 Observational Uncertainty Ensemble}}},
  author = {Lenssen, Nathan and Schmidt, Gavin A. and Hendrickson, Michael and Jacobs, Peter and Menne, Matthew J. and Ruedy, Reto},
  year = 2024,
  month = sep,
  journal = {Journal of Geophysical Research: Atmospheres},
  volume = {129},
  number = {17},
  pages = {e2023JD040179},
  issn = {2169-897X, 2169-8996},
  doi = {10.1029/2023JD040179},
  urldate = {2026-01-24},
  langid = {english}
}

@article{Liu2020,
  title = {Variability in the Global Energy Budget and Transports 1985--2017},
  author = {Liu, Chunlei and Allan, Richard P. and Mayer, Michael and Hyder, Patrick and Desbruy{\`e}res, Damien and Cheng, Lijing and Xu, Jianjun and Xu, Feng and Zhang, Yu},
  year = 2020,
  month = dec,
  journal = {Climate Dynamics},
  volume = {55},
  number = {11-12},
  pages = {3381--3396},
  publisher = {{Springer Science and Business Media LLC}},
  issn = {0930-7575, 1432-0894},
  doi = {10.1007/s00382-020-05451-8},
  urldate = {2025-07-15},
  copyright = {https://creativecommons.org/licenses/by/4.0},
  langid = {english}
}

@misc{Liu2022,
  title = {Reconstructions of the Radiation Fluxes at the Top of Atmosphere and Net Surface Energy Flux: {{DEEP-C Version}} 5.0},
  shorttitle = {Reconstructions of the Radiation Fluxes at the Top of Atmosphere and Net Surface Energy Flux},
  author = {Liu, Chunlei and Allan, Richard},
  year = 2022,
  publisher = {University of Reading},
  doi = {10.17864/1947.000347},
  urldate = {2025-07-15},
  copyright = {Creative Commons Attribution 4.0 International}
}

@article{Ljungqvist2019,
  title = {Centennial-{{Scale Temperature Change}} in {{Last Millennium Simulations}} and {{Proxy-Based Reconstructions}}},
  author = {Ljungqvist, Fredrik Charpentier and Zhang, Qiong and Brattstr{\"o}m, Gudrun and Krusic, Paul J. and Seim, Andrea and Li, Qiang and Zhang, Qiang and Moberg, Anders},
  year = 2019,
  month = may,
  journal = {Journal of Climate},
  volume = {32},
  number = {9},
  pages = {2441--2482},
  issn = {0894-8755, 1520-0442},
  doi = {10.1175/JCLI-D-18-0525.1},
  urldate = {2024-06-19},
  copyright = {http://creativecommons.org/licenses/by/4.0/},
  langid = {english}
}

@article{Loeb2018,
  title = {Clouds and the {{Earth}}'s {{Radiant Energy System}} ({{CERES}}) {{Energy Balanced}} and {{Filled}} ({{EBAF}}) {{Top-of-Atmosphere}} ({{TOA}}) {{Edition-4}}.0 {{Data Product}}},
  author = {Loeb, Norman G. and Doelling, David R. and Wang, Hailan and Su, Wenying and Nguyen, Cathy and Corbett, Joseph G. and Liang, Lusheng and Mitrescu, Cristian and Rose, Fred G. and Kato, Seiji},
  year = 2018,
  month = jan,
  journal = {Journal of Climate},
  volume = {31},
  number = {2},
  pages = {895--918},
  issn = {0894-8755, 1520-0442},
  doi = {10.1175/JCLI-D-17-0208.1},
  urldate = {2025-05-21},
  copyright = {http://www.ametsoc.org/PUBSReuseLicenses},
  langid = {english}
}

@article{Loeb2020,
  title = {New {{Generation}} of {{Climate Models Track Recent Unprecedented Changes}} in {{Earth}}'s {{Radiation Budget Observed}} by {{CERES}}},
  author = {Loeb, Norman G. and Wang, Hailan and Allan, Richard P. and Andrews, Timothy and Armour, Kyle and Cole, Jason N. S. and Dufresne, Jean-Louis and Forster, Piers and Gettelman, Andrew and Guo, Huan and Mauritsen, Thorsten and Ming, Yi and Paynter, David and Proistosescu, Cristian and Stuecker, Malte F. and Will{\'e}n, Ulrika and Wyser, Klaus},
  year = 2020,
  month = mar,
  journal = {Geophysical Research Letters},
  volume = {47},
  number = {5},
  pages = {e2019GL086705},
  issn = {0094-8276, 1944-8007},
  doi = {10.1029/2019GL086705},
  urldate = {2024-05-01},
  langid = {english}
}

@article{Loeb2024,
  title = {Continuity in {{Top-of-Atmosphere Earth Radiation Budget Observations}}},
  author = {Loeb, Norman G. and Doelling, David R. and Kato, Seiji and Su, Wenying and Mlynczak, Pamela E. and Wilkins, Joshua C.},
  year = 2024,
  month = dec,
  journal = {Journal of Climate},
  volume = {37},
  number = {23},
  pages = {6093--6108},
  publisher = {American Meteorological Society},
  issn = {0894-8755, 1520-0442},
  doi = {10.1175/jcli-d-24-0180.1},
  urldate = {2025-07-22},
  copyright = {http://www.ametsoc.org/PUBSReuseLicenses},
  langid = {english}
}

@article{Lucke2019,
  title = {Effects of {{Memory Biases}} on {{Variability}} of {{Temperature Reconstructions}}},
  author = {L{\"u}cke, Lucie J. and Hegerl, Gabriele C. and Schurer, Andrew P. and Wilson, Rob},
  year = 2019,
  month = dec,
  journal = {Journal of Climate},
  volume = {32},
  number = {24},
  pages = {8713--8731},
  issn = {0894-8755, 1520-0442},
  doi = {10.1175/JCLI-D-19-0184.1},
  urldate = {2025-08-28},
  copyright = {http://www.ametsoc.org/PUBSReuseLicenses},
  langid = {english}
}

@article{Lucke2021,
  title = {Orbital Forcing Strongly Influences Seasonal Temperature Trends during the Last Millennium},
  author = {L{\"u}cke, Lucie J. and Schurer, Andrew P. and Wilson, Rob and Hegerl, Gabriele C.},
  year = 2021,
  month = feb,
  journal = {Geophysical Research Letters},
  volume = {48},
  number = {4},
  publisher = {American Geophysical Union (AGU)},
  issn = {1944-8007},
  doi = {10.1029/2020gl088776}
}

@article{Lucke2023,
  title = {The Effect of Uncertainties in Natural Forcing Records on Simulated Temperature during the Last Millennium},
  author = {L{\"u}cke, Lucie J. and Schurer, Andrew P. and Toohey, Matthew and Marshall, Lauren R. and Hegerl, Gabriele C.},
  year = 2023,
  month = may,
  journal = {Climate of the Past},
  volume = {19},
  number = {5},
  pages = {959--978},
  issn = {1814-9332},
  doi = {10.5194/cp-19-959-2023},
  urldate = {2024-11-12},
  copyright = {https://creativecommons.org/licenses/by/4.0/},
  langid = {english}
}

@article{Mackie2025,
  title = {Circulation and {{Cloud Responses}} to {{Patterned SST Warming}}},
  author = {Mackie, Anna and Byrne, Michael P. and Van De Koot, Emily K. and Williams, Andrew I. L.},
  year = 2025,
  month = apr,
  journal = {Geophysical Research Letters},
  volume = {52},
  number = {8},
  pages = {e2024GL112543},
  issn = {0094-8276, 1944-8007},
  doi = {10.1029/2024GL112543},
  urldate = {2026-03-16},
  langid = {english}
}

@article{Mann1999,
  title = {Northern Hemisphere Temperatures during the Past Millennium: {{Inferences}}, Uncertainties, and Limitations},
  shorttitle = {Northern Hemisphere Temperatures during the Past Millennium},
  author = {Mann, Michael E. and Bradley, Raymond S. and Hughes, Malcolm K.},
  year = 1999,
  month = mar,
  journal = {Geophysical Research Letters},
  volume = {26},
  number = {6},
  pages = {759--762},
  issn = {0094-8276, 1944-8007},
  doi = {10.1029/1999GL900070},
  urldate = {2024-09-23},
  copyright = {http://onlinelibrary.wiley.com/termsAndConditions\#vor},
  langid = {english}
}

@article{Mann2009,
  title = {Global {{Signatures}} and {{Dynamical Origins}} of the {{Little Ice Age}} and {{Medieval Climate Anomaly}}},
  author = {Mann, Michael E. and Zhang, Zhihua and Rutherford, Scott and Bradley, Raymond S. and Hughes, Malcolm K. and Shindell, Drew and Ammann, Caspar and Faluvegi, Greg and Ni, Fenbiao},
  year = 2009,
  month = nov,
  journal = {Science},
  volume = {326},
  number = {5957},
  pages = {1256--1260},
  issn = {0036-8075, 1095-9203},
  doi = {10.1126/science.1177303},
  urldate = {2025-09-06},
  langid = {english}
}

@article{Mann2012,
  title = {Underestimation of Volcanic Cooling in Tree-Ring-Based Reconstructions of Hemispheric Temperatures},
  author = {Mann, Michael E. and Fuentes, Jose D. and Rutherford, Scott},
  year = 2012,
  month = mar,
  journal = {Nature Geoscience},
  volume = {5},
  number = {3},
  pages = {202--205},
  issn = {1752-0894, 1752-0908},
  doi = {10.1038/ngeo1394},
  urldate = {2026-03-13},
  copyright = {http://www.springer.com/tdm},
  langid = {english}
}

@article{Marotzke2015,
  title = {Forcing, Feedback and Internal Variability in Global Temperature Trends},
  author = {Marotzke, Jochem and Forster, Piers M.},
  year = 2015,
  month = jan,
  journal = {Nature},
  volume = {517},
  number = {7536},
  pages = {565--570},
  issn = {0028-0836, 1476-4687},
  doi = {10.1038/nature14117},
  urldate = {2026-03-13},
  langid = {english}
}

@article{Marshall2022,
  title = {Volcanic Effects on Climate: Recent Advances and Future Avenues},
  shorttitle = {Volcanic Effects on Climate},
  author = {Marshall, Lauren R. and Maters, Elena C. and Schmidt, Anja and Timmreck, Claudia and Robock, Alan and Toohey, Matthew},
  year = 2022,
  month = may,
  journal = {Bulletin of Volcanology},
  volume = {84},
  number = {5},
  pages = {54},
  issn = {1432-0819},
  doi = {10.1007/s00445-022-01559-3},
  urldate = {2024-06-11},
  langid = {english}
}

@article{Marshall2025,
  title = {Last-Millennium Volcanic Forcing and Climate Response Using {{SO}}{\textsubscript{2}} Emissions},
  author = {Marshall, Lauren R. and Schmidt, Anja and Schurer, Andrew P. and Abraham, Nathan Luke and L{\"u}cke, Lucie J. and Wilson, Rob and Anchukaitis, Kevin J. and Hegerl, Gabriele C. and Johnson, Ben and {Otto-Bliesner}, Bette L. and Brady, Esther C. and Khodri, Myriam and Yoshida, Kohei},
  year = 2025,
  month = jan,
  journal = {Climate of the Past},
  volume = {21},
  number = {1},
  pages = {161--184},
  issn = {1814-9332},
  doi = {10.5194/cp-21-161-2025},
  urldate = {2025-01-30},
  copyright = {https://creativecommons.org/licenses/by/4.0/},
  langid = {english}
}

@book{Masson-Delmotte2021,
  title = {Climate {{Change}} 2021: {{The Physical Science Basis}}. {{Contribution}} of {{Working Group I}} to the {{Sixth Assessment Report}} of the {{Intergovernmental Panel}} on {{Climate Change}}},
  shorttitle = {Climate {{Change}} 2021},
  editor = {{Masson-Delmotte}, Val{\'e}rie and Zhai, Panmao and Pirani, Anna and Connors, Sarah L. and P{\'e}an, Clotilde and Berger, Sophie and Caud, Nada and Chen, Yang and Goldfarb, Leah and Gomis, Melissa I. and Huang, Mengtian and Leitzell, Katherine and Lonnoy, Elisabeth and Matthews, J. B. Robin and Maycock, Thomas K. and Waterfield, Tim and Yelek{\c c}i, {\"O}zge and Yu, Rong and Zhou, Botao},
  year = 2021,
  publisher = {Cambridge University Press},
  address = {Cambridge, United Kingdom and New York, NY, USA},
  doi = {10.1017/9781009157896},
  langid = {english}
}

@article{Mauritsen2019,
  title = {Developments in the {{MPI}}-{{M Earth System Model}} Version 1.2 ({{MPI}}-{{ESM1}}.2) and {{Its Response}} to {{Increasing CO}} {\textsubscript{2}}},
  author = {Mauritsen, Thorsten and Bader, J{\"u}rgen and Becker, Tobias and Behrens, J{\"o}rg and Bittner, Matthias and Brokopf, Renate and Brovkin, Victor and Claussen, Martin and Crueger, Traute and Esch, Monika and Fast, Irina and Fiedler, Stephanie and Fl{\"a}schner, Dagmar and Gayler, Veronika and Giorgetta, Marco and Goll, Daniel S. and Haak, Helmuth and Hagemann, Stefan and Hedemann, Christopher and Hohenegger, Cathy and Ilyina, Tatiana and Jahns, Thomas and Jimen{\'e}z-de-la-Cuesta, Diego and Jungclaus, Johann and Kleinen, Thomas and Kloster, Silvia and Kracher, Daniela and Kinne, Stefan and Kleberg, Deike and Lasslop, Gitta and Kornblueh, Luis and Marotzke, Jochem and Matei, Daniela and Meraner, Katharina and Mikolajewicz, Uwe and Modali, Kameswarrao and M{\"o}bis, Benjamin and M{\"u}ller, Wolfgang A. and Nabel, Julia E. M. S. and Nam, Christine C. W. and Notz, Dirk and Nyawira, Sarah-Sylvia and Paulsen, Hanna and Peters, Karsten and Pincus, Robert and Pohlmann, Holger and Pongratz, Julia and Popp, Max and Raddatz, Thomas J{\"u}rgen and Rast, Sebastian and Redler, Rene and Reick, Christian H. and Rohrschneider, Tim and Schemann, Vera and Schmidt, Hauke and Schnur, Reiner and Schulzweida, Uwe and Six, Katharina D. and Stein, Lukas and Stemmler, Irene and Stevens, Bjorn and Von Storch, Jin-Song and Tian, Fangxing and Voigt, Aiko and Vrese, Philipp and Wieners, Karl-Hermann and Wilkenskjeld, Stiig and Winkler, Alexander and Roeckner, Erich},
  year = 2019,
  month = apr,
  journal = {Journal of Advances in Modeling Earth Systems},
  volume = {11},
  number = {4},
  pages = {998--1038},
  issn = {1942-2466, 1942-2466},
  doi = {10.1029/2018MS001400},
  urldate = {2024-06-14},
  langid = {english}
}

@article{Mauritsen2025,
  title = {Earth's Energy Imbalance More than Doubled in Recent Decades},
  author = {Mauritsen, Thorsten and Tsushima, Yoko and Meyssignac, Benoit and Loeb, Norman G. and Hakuba, Maria and Pilewskie, Peter and Cole, Jason and Suzuki, Kentaroh and Ackerman, Thomas P. and Allan, Richard P. and Andrews, Timothy and Bender, Frida A.-M. and Bloch-Johnson, Jonah and Bodas-Salcedo, Alejandro and Brookshaw, Anca and Ceppi, Paulo and Clerbaux, Nicolas and Dessler, Andrew E. and Donohoe, Aaron and Dufresne, Jean-Louis and Eyring, Veronika and Findell, Kirsten L. and Gettelman, Andrew and Gristey, Jake J. and Hawkins, Ed and Heimbach, Patrick and Hewitt, Helene T. and Jeevanjee, Nadir and Jones, Colin and Kang, Sarah M. and Kato, Seiji and Kay, Jennifer E. and Klein, Stephen A. and Knutti, Reto and Kramer, Ryan and Lee, June-Yi and McCoy, Daniel T. and Medeiros, Brian and Megner, Linda and Modak, Angshuman and Ogura, Tomoo and Palmer, Matthew D. and Paynter, David and Quaas, Johannes and Ramanathan, Veerabhadran and Ringer, Mark and Von Schuckmann, Karina and Sherwood, Steven and Stevens, Bjorn and Tan, Ivy and Tselioudis, George and Sutton, Rowan and Voigt, Aiko and Watanabe, Masahiro and Webb, Mark J. and Wild, Martin and Zelinka, Mark D.},
  year = 2025,
  month = jun,
  journal = {AGU Advances},
  volume = {6},
  number = {3},
  publisher = {American Geophysical Union (AGU)},
  issn = {2576-604X, 2576-604X},
  doi = {10.1029/2024av001636},
  urldate = {2025-07-22},
  copyright = {http://creativecommons.org/licenses/by/4.0/},
  langid = {english}
}

@article{Mayer2026,
  title = {The {{Atmosphere}}'s {{Substantial Role}} in {{Interannual Variability}} of {{Earth}}'s {{Energy Imbalance}}},
  author = {Mayer, Michael and Loeb, Norman G. and Lyman, John M. and Johnson, Gregory C. and Winkelbauer, Susanna},
  year = 2026,
  month = feb,
  journal = {Geophysical Research Letters},
  volume = {53},
  number = {3},
  pages = {e2025GL119833},
  issn = {0094-8276, 1944-8007},
  doi = {10.1029/2025GL119833},
  urldate = {2026-02-17},
  langid = {english}
}

@article{McGregor2015,
  title = {Robust Global Ocean Cooling Trend for the Pre-Industrial {{Common Era}}},
  author = {McGregor, Helen V. and Evans, Michael N. and Goosse, Hugues and Leduc, Guillaume and Martrat, Belen and Addison, Jason A. and Mortyn, P. Graham and Oppo, Delia W. and Seidenkrantz, Marit-Solveig and Sicre, Marie-Alexandrine and Phipps, Steven J. and Selvaraj, Kandasamy and Thirumalai, Kaustubh and Filipsson, Helena L. and Ersek, Vasile},
  year = 2015,
  month = aug,
  journal = {Nature Geoscience},
  volume = {8},
  number = {9},
  pages = {671--677},
  publisher = {{Springer Science and Business Media LLC}},
  issn = {1752-0908},
  doi = {10.1038/ngeo2510}
}

@article{Medhaug2017,
  title = {Reconciling Controversies about the `Global Warming Hiatus'},
  author = {Medhaug, Iselin and Stolpe, Martin B. and Fischer, Erich M. and Knutti, Reto},
  year = 2017,
  month = may,
  journal = {Nature},
  volume = {545},
  number = {7652},
  pages = {41--47},
  issn = {0028-0836, 1476-4687},
  doi = {10.1038/nature22315},
  urldate = {2024-08-13},
  langid = {english}
}

@misc{Meier2024,
  title = {{{NOAA}}/{{NSIDC}} Climate Data Record of Passive Microwave Sea Ice Concentration, Version 5},
  author = {Meier, Walt and Fetterer, Florence and Windnagel, Ann and Stewart, James Scott and Stafford, Trey},
  year = 2024,
  publisher = {{National Snow and Ice Data Center}},
  doi = {10.7265/RJZB-PF78},
  urldate = {2025-07-22},
  langid = {english}
}

@article{Meng2025,
  title = {Coupled {{Seasonal Data Assimilation}} of {{Sea Ice}}, {{Ocean}}, and {{Atmospheric Dynamics}} over the {{Last Millennium}}},
  author = {Meng, Zilu and Hakim, Gregory J. and Steig, Eric J.},
  year = 2025,
  month = sep,
  journal = {Journal of Climate},
  issn = {0894-8755, 1520-0442},
  doi = {10.1175/JCLI-D-25-0048.1},
  urldate = {2025-09-13},
  langid = {english}
}

@article{Miller2012,
  title = {Abrupt Onset of the {{Little Ice Age}} Triggered by Volcanism and Sustained by Sea-ice/Ocean Feedbacks},
  author = {Miller, Gifford H. and Geirsd{\'o}ttir, {\'A}slaug and Zhong, Yafang and Larsen, Darren J. and Otto-Bliesner, Bette L. and Holland, Marika M. and Bailey, David A. and Refsnider, Kurt A. and Lehman, Scott J. and Southon, John R. and Anderson, Chance and Bj{\"o}rnsson, Helgi and Thordarson, Thorvaldur},
  year = 2012,
  month = jan,
  journal = {Geophysical Research Letters},
  volume = {39},
  number = {2},
  pages = {2011GL050168},
  issn = {0094-8276, 1944-8007},
  doi = {10.1029/2011GL050168},
  urldate = {2024-06-14},
  copyright = {http://onlinelibrary.wiley.com/termsAndConditions\#vor},
  langid = {english}
}

@misc{Miniere2026,
  title = {Consolidating {{Global Estimates}} of {{Ocean Heat Content}}: {{Toward}} a {{Consistent Earth Heat Inventory}}},
  shorttitle = {Consolidating Global Estimates of Ocean Heat Content},
  author = {Mini{\`e}re, Audrey and Von Schuckmann, Karina and Speich, Sabrina},
  year = 2026,
  month = feb,
  publisher = {ESSD -- Ocean/Physical oceanography},
  doi = {10.5194/essd-2026-106},
  urldate = {2026-03-08},
  archiveprefix = {ESSD -- Ocean/Physical oceanography},
  copyright = {https://creativecommons.org/licenses/by/4.0/},
  langid = {english}
}

@article{Morice2021,
  title = {An Updated Assessment of Near-surface Temperature Change from 1850: The {{HadCRUT5}} Data Set},
  shorttitle = {An Updated Assessment of Near-surface Temperature Change from 1850},
  author = {Morice, C. P. and Kennedy, J. J. and Rayner, N. A. and Winn, J. P. and Hogan, E. and Killick, R. E. and Dunn, R. J. H. and Osborn, T. J. and Jones, P. D. and Simpson, I. R.},
  year = 2021,
  month = feb,
  journal = {Journal of Geophysical Research: Atmospheres},
  volume = {126},
  number = {3},
  publisher = {American Geophysical Union (AGU)},
  issn = {2169-897X, 2169-8996},
  doi = {10.1029/2019jd032361},
  urldate = {2025-07-23},
  copyright = {http://creativecommons.org/licenses/by/4.0/},
  langid = {english}
}

@article{Nash1970,
  title = {River Flow Forecasting through Conceptual Models Part {{I}} --- {{A}} Discussion of Principles},
  author = {Nash, J.E. and Sutcliffe, J.V.},
  year = 1970,
  month = apr,
  journal = {Journal of Hydrology},
  volume = {10},
  number = {3},
  pages = {282--290},
  issn = {00221694},
  doi = {10.1016/0022-1694(70)90255-6},
  urldate = {2026-03-10},
  copyright = {https://www.elsevier.com/tdm/userlicense/1.0/},
  langid = {english}
}

@article{Neukom2014,
  title = {Inter-Hemispheric Temperature Variability over the Past Millennium},
  author = {Neukom, Raphael and Gergis, Jo{\"e}lle and Karoly, David J. and Wanner, Heinz and Curran, Mark and Elbert, Julie and {Gonz{\'a}lez-Rouco}, Fidel and Linsley, Braddock K. and Moy, Andrew D. and Mundo, Ignacio and Raible, Christoph C. and Steig, Eric J. and Van Ommen, Tas and Vance, Tessa and Villalba, Ricardo and Zinke, Jens and Frank, David},
  year = 2014,
  month = may,
  journal = {Nature Climate Change},
  volume = {4},
  number = {5},
  pages = {362--367},
  issn = {1758-678X, 1758-6798},
  doi = {10.1038/nclimate2174},
  urldate = {2026-02-23},
  copyright = {http://www.springer.com/tdm},
  langid = {english}
}

@article{Neukom2019,
  title = {No Evidence for Globally Coherent Warm and Cold Periods over the Preindustrial {{Common Era}}},
  author = {Neukom, Raphael and Steiger, Nathan and {G{\'o}mez-Navarro}, Juan Jos{\'e} and Wang, Jianghao and Werner, Johannes P.},
  year = 2019,
  month = jul,
  journal = {Nature},
  volume = {571},
  number = {7766},
  pages = {550--554},
  publisher = {{Springer Science and Business Media LLC}},
  issn = {1476-4687},
  doi = {10.1038/s41586-019-1401-2}
}

@article{Notz2014,
  title = {Sea-Ice Extent and Its Trend Provide Limited Metrics of Model Performance},
  author = {Notz, D.},
  year = 2014,
  month = feb,
  journal = {The Cryosphere},
  volume = {8},
  number = {1},
  pages = {229--243},
  issn = {1994-0424},
  doi = {10.5194/tc-8-229-2014},
  urldate = {2025-04-08},
  copyright = {https://creativecommons.org/licenses/by/3.0/},
  langid = {english}
}

@article{Olonscheck2024,
  title = {Coupled {{Climate Models Systematically Underestimate Radiation Response}} to {{Surface Warming}}},
  author = {Olonscheck, Dirk and Rugenstein, Maria},
  year = 2024,
  month = mar,
  journal = {Geophysical Research Letters},
  volume = {51},
  number = {6},
  pages = {e2023GL106909},
  issn = {0094-8276, 1944-8007},
  doi = {10.1029/2023GL106909},
  urldate = {2025-09-03},
  langid = {english}
}

@article{Osman2021,
  title = {Globally Resolved Surface Temperatures since the {{Last Glacial Maximum}}},
  author = {Osman, Matthew B. and Tierney, Jessica E. and Zhu, Jiang and Tardif, Robert and Hakim, Gregory J. and King, Jonathan and Poulsen, Christopher J.},
  year = 2021,
  month = nov,
  journal = {Nature},
  volume = {599},
  number = {7884},
  pages = {239--244},
  publisher = {{Springer Science and Business Media LLC}},
  issn = {1476-4687},
  doi = {10.1038/s41586-021-03984-4}
}

@article{Otto-Bliesner2016,
  title = {Climate {{Variability}} and {{Change}} since 850 {{CE}}: {{An Ensemble Approach}} with the {{Community Earth System Model}}},
  shorttitle = {Climate {{Variability}} and {{Change}} since 850 {{CE}}},
  author = {{Otto-Bliesner}, Bette L. and Brady, Esther C. and Fasullo, John and Jahn, Alexandra and Landrum, Laura and Stevenson, Samantha and Rosenbloom, Nan and Mai, Andrew and Strand, Gary},
  year = 2016,
  month = may,
  journal = {Bulletin of the American Meteorological Society},
  volume = {97},
  number = {5},
  pages = {735--754},
  issn = {0003-0007, 1520-0477},
  doi = {10.1175/BAMS-D-14-00233.1},
  urldate = {2024-06-19},
  langid = {english}
}

@article{PAGESConsortium2017,
  title = {A Global Multiproxy Database for Temperature Reconstructions of the {{Common Era}}},
  author = {{PAGES 2k Consortium}},
  year = 2017,
  month = jul,
  journal = {Scientific Data},
  volume = {4},
  number = {1},
  publisher = {{Springer Science and Business Media LLC}},
  doi = {10.1038/sdata.2017.88}
}

@article{PAGESConsortium2019,
  title = {Consistent Multidecadal Variability in Global Temperature Reconstructions and Simulations over the {{Common Era}}},
  author = {{PAGES 2k Consortium}},
  year = 2019,
  month = jul,
  journal = {Nature Geoscience},
  volume = {12},
  number = {8},
  pages = {643--649},
  publisher = {{Springer Science and Business Media LLC}},
  issn = {1752-0908},
  doi = {10.1038/s41561-019-0400-0}
}

@article{Palmer2014,
  title = {Internal Variability of {{Earth}}'s Energy Budget Simulated by {{CMIP5}} Climate Models},
  author = {Palmer, M D and McNeall, D J},
  year = 2014,
  month = mar,
  journal = {Environmental Research Letters},
  volume = {9},
  number = {3},
  pages = {34016},
  publisher = {IOP Publishing},
  issn = {1748-9326},
  doi = {10.1088/1748-9326/9/3/034016},
  urldate = {2025-07-22},
  copyright = {http://iopscience.iop.org/info/page/text-and-data-mining},
  langid = {english}
}

@article{Pan2025,
  title = {Ocean Heat Content in 2024},
  author = {Pan, Yuying and Mini{\`e}re, Audrey and Von Schuckmann, Karina and Li, Zhi and Li, Yuanlong and Cheng, Lijing and Zhu, Jiang},
  year = 2025,
  month = apr,
  journal = {Nature Reviews Earth \& Environment},
  volume = {6},
  number = {4},
  pages = {249--251},
  issn = {2662-138X},
  doi = {10.1038/s43017-025-00655-0},
  urldate = {2025-06-20},
  langid = {english}
}

@article{Pauling2021,
  title = {Robust {{Inter}}-{{Hemispheric Asymmetry}} in the {{Response}} to {{Symmetric Volcanic Forcing}} in {{Model Large Ensembles}}},
  author = {Pauling, Andrew G. and Bushuk, Mitchell and Bitz, Cecilia M.},
  year = 2021,
  month = may,
  journal = {Geophysical Research Letters},
  volume = {48},
  number = {9},
  publisher = {American Geophysical Union (AGU)},
  issn = {0094-8276, 1944-8007},
  doi = {10.1029/2021gl092558},
  urldate = {2025-07-17},
  copyright = {http://onlinelibrary.wiley.com/termsAndConditions\#am},
  langid = {english}
}

@article{Penland1994,
  title = {A {{Balance Condition}} for {{Stochastic Numerical Models}} with {{Application}} to the {{El Ni\~no-Southern Oscillation}}},
  author = {Penland, C{\'e}cile and Matrosova, Ludmila},
  year = 1994,
  month = sep,
  journal = {Journal of Climate},
  volume = {7},
  number = {9},
  pages = {1352--1372},
  publisher = {American Meteorological Society},
  address = {Boston MA, USA},
  issn = {0894-8755, 1520-0442},
  doi = {10.1175/1520-0442(1994)007<1352:ABCFSN>2.0.CO;2},
  urldate = {2025-07-30},
  langid = {english}
}

@article{Penland1995,
  title = {The Optimal Growth of Tropical Sea Surface Temperature Anomalies},
  author = {Penland, C{\'e}cile and Sardeshmukh, Prashant D.},
  year = 1995,
  journal = {Journal of Climate},
  volume = {8},
  number = {8},
  pages = {1999--2024},
  publisher = {American Meteorological Society},
  address = {Boston MA, USA},
  doi = {10.1175/1520-0442(1995)008<1999:togots>2.0.co;2}
}

@article{Penland1996,
  title = {A Stochastic Model of {{IndoPacific}} Sea Surface Temperature Anomalies},
  author = {Penland, C{\'e}cile},
  year = 1996,
  month = nov,
  journal = {Physica D: Nonlinear Phenomena},
  volume = {98},
  number = {2-4},
  pages = {534--558},
  publisher = {Elsevier BV},
  issn = {0167-2789},
  doi = {10.1016/0167-2789(96)00124-8},
  urldate = {2025-07-23},
  copyright = {https://www.elsevier.com/tdm/userlicense/1.0/},
  langid = {english}
}

@article{Perkins2021,
  title = {Coupled {{Atmosphere}}--{{Ocean}} Reconstruction of the Last Millennium Using Online Data Assimilation},
  author = {Perkins, W. Andre and Hakim, Gregory J.},
  year = 2021,
  month = may,
  journal = {Paleoceanography and Paleoclimatology},
  volume = {36},
  number = {5},
  publisher = {American Geophysical Union (AGU)},
  doi = {10.1029/2020pa003959}
}

@article{Pincus2016,
  title = {The {{Radiative Forcing Model Intercomparison Project}} ({{RFMIP}}): Experimental Protocol for {{CMIP6}}},
  shorttitle = {The {{Radiative Forcing Model Intercomparison Project}} ({{RFMIP}})},
  author = {Pincus, Robert and Forster, Piers M. and Stevens, Bjorn},
  year = 2016,
  month = sep,
  journal = {Geoscientific Model Development},
  volume = {9},
  number = {9},
  pages = {3447--3460},
  issn = {1991-9603},
  doi = {10.5194/gmd-9-3447-2016},
  urldate = {2024-06-18},
  copyright = {https://creativecommons.org/licenses/by/3.0/},
  langid = {english}
}

@article{Proistosescu2018,
  title = {Radiative {{Feedbacks From Stochastic Variability}} in {{Surface Temperature}} and {{Radiative Imbalance}}},
  author = {Proistosescu, Cristian and Donohoe, Aaron and Armour, Kyle C. and Roe, Gerard H. and Stuecker, Malte F. and Bitz, Cecilia M.},
  year = 2018,
  month = may,
  journal = {Geophysical Research Letters},
  volume = {45},
  number = {10},
  pages = {5082--5094},
  issn = {0094-8276, 1944-8007},
  doi = {10.1029/2018GL077678},
  urldate = {2025-10-20},
  langid = {english}
}

@article{Raghuraman2021,
  title = {Anthropogenic Forcing and Response Yield Observed Positive Trend in {{Earth}}'s Energy Imbalance},
  author = {Raghuraman, Shiv Priyam and Paynter, David and Ramaswamy, V.},
  year = 2021,
  month = jul,
  journal = {Nature Communications},
  volume = {12},
  number = {1},
  pages = {4577},
  issn = {2041-1723},
  doi = {10.1038/s41467-021-24544-4},
  urldate = {2025-01-28},
  langid = {english}
}

@article{Rantanen2022,
  title = {The {{Arctic}} Has Warmed Nearly Four Times Faster than the Globe since 1979},
  author = {Rantanen, Mika and Karpechko, Alexey Yu. and Lipponen, Antti and Nordling, Kalle and Hyv{\"a}rinen, Otto and Ruosteenoja, Kimmo and Vihma, Timo and Laaksonen, Ari},
  year = 2022,
  month = aug,
  journal = {Communications Earth \& Environment},
  volume = {3},
  number = {1},
  pages = {168},
  issn = {2662-4435},
  doi = {10.1038/s43247-022-00498-3},
  urldate = {2025-06-10},
  langid = {english}
}

@article{Rayner2003,
  title = {Global Analyses of Sea Surface Temperature, Sea Ice, and Night Marine Air Temperature since the Late Nineteenth Century},
  author = {Rayner, N. A. and Parker, D. E. and Horton, E. B. and Folland, C. K. and Alexander, L. V. and Rowell, D. P. and Kent, E. C. and Kaplan, A.},
  year = 2003,
  month = jul,
  journal = {Journal of Geophysical Research: Atmospheres},
  volume = {108},
  number = {D14},
  publisher = {American Geophysical Union (AGU)},
  issn = {0148-0227},
  doi = {10.1029/2002jd002670},
  urldate = {2025-07-22},
  copyright = {http://onlinelibrary.wiley.com/termsAndConditions\#vor},
  langid = {english}
}

@article{Rose2018,
  title = {{{CLIMLAB}}: A {{Python}} Toolkit for Interactive, Process-Oriented Climate Modeling},
  shorttitle = {{{CLIMLAB}}},
  author = {E. J. Rose, Brian},
  year = 2018,
  month = apr,
  journal = {Journal of Open Source Software},
  volume = {3},
  number = {24},
  pages = {659},
  issn = {2475-9066},
  doi = {10.21105/joss.00659},
  urldate = {2025-08-26},
  copyright = {http://creativecommons.org/licenses/by/4.0/}
}

@misc{Sanchez2025,
  title = {Paleo Data Assimilation of Coral {{$\delta$18O Part}} 1: {{Best}} Practices and Uncertainties},
  shorttitle = {Paleo Data Assimilation of Coral {{$\delta$18O}} Part 1},
  author = {Sanchez, Sara C and Zhu, Feng and Saenger, Casey and Thompson, Diane M.},
  year = 2025,
  month = dec,
  publisher = {Preprints},
  doi = {10.22541/au.176607163.31022212/v1},
  urldate = {2026-01-23},
  archiveprefix = {Preprints},
  langid = {english}
}

@article{Schmidt2023,
  title = {{{CERESMIP}}: A Climate Modeling Protocol to Investigate Recent Trends in the {{Earth}}'s {{Energy Imbalance}}},
  shorttitle = {{{CERESMIP}}},
  author = {Schmidt, Gavin A. and Andrews, Timothy and Bauer, Susanne E. and Durack, Paul J. and Loeb, Norman G. and Ramaswamy, V. and Arnold, Nathan P. and Bosilovich, Michael G. and Cole, Jason and Horowitz, Larry W. and Johnson, Gregory C. and Lyman, John M. and Medeiros, Brian and Michibata, Takuro and Olonscheck, Dirk and Paynter, David and Raghuraman, Shiv Priyam and Schulz, Michael and Takasuka, Daisuke and Tallapragada, Vijay and Taylor, Patrick C. and Ziehn, Tilo},
  year = 2023,
  month = jul,
  journal = {Frontiers in Climate},
  volume = {5},
  pages = {1202161},
  issn = {2624-9553},
  doi = {10.3389/fclim.2023.1202161},
  urldate = {2024-11-19}
}

@article{Schneider2015,
  title = {Revising Midlatitude Summer Temperatures Back to {{A}}.{{D}}. 600 Based on a Wood Density Network},
  author = {Schneider, Lea and Smerdon, Jason E. and B{\"u}ntgen, Ulf and Wilson, Rob J. S. and Myglan, Vladimir S. and Kirdyanov, Alexander V. and Esper, Jan},
  year = 2015,
  month = jun,
  journal = {Geophysical Research Letters},
  volume = {42},
  number = {11},
  pages = {4556--4562},
  issn = {0094-8276, 1944-8007},
  doi = {10.1002/2015GL063956},
  urldate = {2026-02-23},
  copyright = {http://onlinelibrary.wiley.com/termsAndConditions\#vor},
  langid = {english}
}

@article{Schuckmann2016,
  title = {An Imperative to Monitor {{Earth}}'s Energy Imbalance},
  author = {{von Schuckmann}, K. and Palmer, M. D. and Trenberth, K. E. and Cazenave, A. and Chambers, D. and Champollion, N. and Hansen, J. and Josey, S. A. and Loeb, N. and Mathieu, P.-P. and Meyssignac, B. and Wild, M.},
  year = 2016,
  month = jan,
  journal = {Nature Climate Change},
  volume = {6},
  number = {2},
  pages = {138--144},
  publisher = {{Springer Science and Business Media LLC}},
  issn = {1758-6798},
  doi = {10.1038/nclimate2876}
}

@article{Schuckmann2023,
  title = {Heat Stored in the {{Earth}} System 1960--2020: Where Does the Energy Go?},
  shorttitle = {Heat Stored in the {{Earth}} System 1960--2020},
  author = {{von Schuckmann}, Karina and Mini{\`e}re, Audrey and Gues, Flora and {Cuesta-Valero}, Francisco Jos{\'e} and Kirchengast, Gottfried and Adusumilli, Susheel and Straneo, Fiammetta and Ablain, Micha{\"e}l and Allan, Richard P. and Barker, Paul M. and Beltrami, Hugo and Blazquez, Alejandro and Boyer, Tim and Cheng, Lijing and Church, John and Desbruyeres, Damien and Dolman, Han and Domingues, Catia M. and {Garc{\'i}a-Garc{\'i}a}, Almudena and Giglio, Donata and Gilson, John E. and Gorfer, Maximilian and Haimberger, Leopold and Hakuba, Maria Z. and Hendricks, Stefan and Hosoda, Shigeki and Johnson, Gregory C. and Killick, Rachel and King, Brian and Kolodziejczyk, Nicolas and Korosov, Anton and Krinner, Gerhard and Kuusela, Mikael and Landerer, Felix W. and Langer, Moritz and Lavergne, Thomas and Lawrence, Isobel and Li, Yuehua and Lyman, John and Marti, Florence and Marzeion, Ben and Mayer, Michael and MacDougall, Andrew H. and McDougall, Trevor and Monselesan, Didier Paolo and Nitzbon, Jan and Otosaka, In{\`e}s and Peng, Jian and Purkey, Sarah and Roemmich, Dean and Sato, Kanako and Sato, Katsunari and Savita, Abhishek and Schweiger, Axel and Shepherd, Andrew and Seneviratne, Sonia I. and Simons, Leon and Slater, Donald A. and Slater, Thomas and Steiner, Andrea K. and Suga, Toshio and Szekely, Tanguy and Thiery, Wim and Timmermans, Mary-Louise and Vanderkelen, Inne and Wjiffels, Susan E. and Wu, Tonghua and Zemp, Michael},
  year = 2023,
  month = apr,
  journal = {Earth System Science Data},
  volume = {15},
  number = {4},
  pages = {1675--1709},
  issn = {1866-3516},
  doi = {10.5194/essd-15-1675-2023},
  urldate = {2025-08-30},
  copyright = {https://creativecommons.org/licenses/by/4.0/},
  langid = {english}
}

@article{Schurer2014,
  title = {Small Influence of Solar Variability on Climate over the Past Millennium},
  author = {Schurer, Andrew P. and Tett, Simon F. B. and Hegerl, Gabriele C.},
  year = 2014,
  month = feb,
  journal = {Nature Geoscience},
  volume = {7},
  number = {2},
  pages = {104--108},
  issn = {1752-0894, 1752-0908},
  doi = {10.1038/ngeo2040},
  urldate = {2025-06-26},
  langid = {english}
}

@article{Seland2020,
  title = {Overview of the {{Norwegian Earth System Model}} ({{NorESM2}}) and Key Climate Response of {{CMIP6 DECK}}, Historical, and Scenario Simulations},
  author = {Seland, {\O}yvind and Bentsen, Mats and Olivi{\'e}, Dirk and Toniazzo, Thomas and Gjermundsen, Ada and Graff, Lise Seland and Debernard, Jens Boldingh and Gupta, Alok Kumar and He, Yan-Chun and Kirkev{\aa}g, Alf and Schwinger, J{\"o}rg and Tjiputra, Jerry and Aas, Kjetil Schanke and Bethke, Ingo and Fan, Yuanchao and Griesfeller, Jan and Grini, Alf and Guo, Chuncheng and Ilicak, Mehmet and Karset, Inger Helene Hafsahl and Landgren, Oskar and Liakka, Johan and Moseid, Kine Onsum and Nummelin, Aleksi and Spensberger, Clemens and Tang, Hui and Zhang, Zhongshi and Heinze, Christoph and Iversen, Trond and Schulz, Michael},
  year = 2020,
  month = dec,
  journal = {Geoscientific Model Development},
  volume = {13},
  number = {12},
  pages = {6165--6200},
  issn = {1991-9603},
  doi = {10.5194/gmd-13-6165-2020},
  urldate = {2026-01-29},
  copyright = {https://creativecommons.org/licenses/by/4.0/},
  langid = {english}
}

@article{Shackleton2023,
  title = {Benthic {{$\delta$18O}} Records {{Earth}}'s Energy Imbalance},
  author = {Shackleton, Sarah and Seltzer, Alan and Baggenstos, Daniel and Lisiecki, Lorraine E.},
  year = 2023,
  month = aug,
  journal = {Nature Geoscience},
  volume = {16},
  number = {9},
  pages = {797--802},
  publisher = {{Springer Science and Business Media LLC}},
  issn = {1752-0908},
  doi = {10.1038/s41561-023-01250-y}
}

@article{Shackleton2026,
  title = {Global Ocean Heat Content over the Past 3 Million Years},
  author = {Shackleton, Sarah and Hishamunda, Valens and Yan, Yuzhen and Carter, Austin and Morgan, Jacob and Severinghaus, Jeff and Aarons, Sarah and {Marks-Peterson}, Julia and Epifanio, Jenna and Buizert, Christo and Brook, Edward and Kurbatov, Andrei V. and Bender, Michael L. and Higgins, John},
  year = 2026,
  month = mar,
  journal = {Nature},
  volume = {651},
  number = {8106},
  pages = {653--657},
  publisher = {In Review},
  issn = {0028-0836, 1476-4687},
  doi = {10.1038/s41586-026-10116-3},
  urldate = {2026-03-20},
  copyright = {https://creativecommons.org/licenses/by/4.0/},
  langid = {english}
}

@article{Sherwood2015,
  title = {Adjustments in the {{Forcing-Feedback Framework}} for {{Understanding Climate Change}}},
  author = {Sherwood, Steven C. and Bony, Sandrine and Boucher, Olivier and Bretherton, Chris and Forster, Piers M. and Gregory, Jonathan M. and Stevens, Bjorn},
  year = 2015,
  month = feb,
  journal = {Bulletin of the American Meteorological Society},
  volume = {96},
  number = {2},
  pages = {217--228},
  issn = {0003-0007, 1520-0477},
  doi = {10.1175/BAMS-D-13-00167.1},
  urldate = {2024-06-18},
  langid = {english}
}

@article{Sherwood2020,
  title = {An Assessment of Earth's Climate Sensitivity Using Multiple Lines of Evidence},
  author = {Sherwood, S. C. and Webb, M. J. and Annan, J. D. and Armour, K. C. and Forster, P. M. and Hargreaves, J. C. and Hegerl, G. and Klein, S. A. and Marvel, K. D. and Rohling, E. J. and Watanabe, M. and Andrews, T. and Braconnot, P. and Bretherton, C. S. and Foster, G. L. and Hausfather, Z. and {von der Heydt}, A. S. and Knutti, R. and Mauritsen, T. and Norris, J. R. and Proistosescu, C. and Rugenstein, M. and Schmidt, G. A. and Tokarska, K. B. and Zelinka, M. D.},
  year = 2020,
  month = sep,
  journal = {Reviews of Geophysics},
  volume = {58},
  number = {4},
  publisher = {American Geophysical Union (AGU)},
  issn = {1944-9208},
  doi = {10.1029/2019rg000678}
}

@misc{Sigl2024,
  title = {Volcanic Stratospheric Sulfur Injections from 500 {{BCE}} to 1900 {{CE}}, {{eVolv2k}}\_version4},
  author = {Sigl, Michael and Toohey, Matthew},
  year = 2024,
  pages = {3072 data points},
  publisher = {PANGAEA},
  doi = {10.1594/PANGAEA.971968},
  urldate = {2026-03-12},
  copyright = {Creative Commons Attribution 4.0 International},
  langid = {english}
}

@article{Slawinska2018,
  title = {Impact of {{Volcanic Eruptions}} on {{Decadal}} to {{Centennial Fluctuations}} of {{Arctic Sea Ice Extent}} during the {{Last Millennium}} and on {{Initiation}} of the {{Little Ice Age}}},
  author = {Slawinska, Joanna and Robock, Alan},
  year = 2018,
  month = mar,
  journal = {Journal of Climate},
  volume = {31},
  number = {6},
  pages = {2145--2167},
  issn = {0894-8755, 1520-0442},
  doi = {10.1175/JCLI-D-16-0498.1},
  urldate = {2025-07-05},
  langid = {english}
}

@article{Smerdon2012,
  title = {Climate Models as a Test Bed for Climate Reconstruction Methods: Pseudoproxy Experiments},
  shorttitle = {Climate Models as a Test Bed for Climate Reconstruction Methods},
  author = {Smerdon, Jason E.},
  year = 2012,
  month = jan,
  journal = {WIREs Climate Change},
  volume = {3},
  number = {1},
  pages = {63--77},
  issn = {1757-7780, 1757-7799},
  doi = {10.1002/wcc.149},
  urldate = {2024-03-15},
  langid = {english}
}

@article{Smerdon2023,
  title = {The {{Historical Development}} of {{Large}}-{{Scale Paleoclimate Field Reconstructions Over}} the {{Common Era}}},
  author = {Smerdon, Jason E. and Cook, Edward R. and Steiger, Nathan J.},
  year = 2023,
  month = dec,
  journal = {Reviews of Geophysics},
  volume = {61},
  number = {4},
  pages = {e2022RG000782},
  publisher = {American Geophysical Union (AGU)},
  issn = {8755-1209, 1944-9208},
  doi = {10.1029/2022RG000782},
  urldate = {2025-07-30},
  langid = {english}
}

@misc{Smith2025,
  title = {Indicators of {{Global Climate Change}} 2024},
  author = {Smith, Chris and Walsh, Tristram and Gillett, Nathan and Hauser, Mattias and Krummel, Paul and Lamb, William and Lamboll, Robin and M{\"u}hle, Jens and Palmer, Matthew and Ribes, Aur{\'e}lien and Schumacher, Dominik and Seneviratne, Sonia and Slangen, Aim{\'e}e and Trewin, Blair and {von Schuckmann}, Karina and Forster, Piers},
  year = 2025,
  month = jun,
  doi = {10.5281/ZENODO.15639576},
  urldate = {2025-08-31},
  copyright = {Creative Commons Attribution 4.0 International},
  howpublished = {Zenodo}
}

@article{Steiger2014,
  title = {Assimilation of Time-Averaged Pseudoproxies for Climate Reconstruction},
  author = {Steiger, Nathan J. and Hakim, Gregory J. and Steig, Eric J. and Battisti, David S. and Roe, Gerard H.},
  year = 2014,
  month = jan,
  journal = {Journal of Climate},
  volume = {27},
  number = {1},
  pages = {426--441},
  publisher = {American Meteorological Society},
  issn = {1520-0442},
  doi = {10.1175/jcli-d-12-00693.1}
}

@article{Steiger2018,
  title = {A Reconstruction of Global Hydroclimate and Dynamical Variables over the {{Common Era}}},
  author = {Steiger, Nathan J. and Smerdon, Jason E. and Cook, Edward R. and Cook, Benjamin I.},
  year = 2018,
  month = may,
  journal = {Scientific Data},
  volume = {5},
  number = {1},
  publisher = {{Springer Science and Business Media LLC}},
  issn = {2052-4463},
  doi = {10.1038/sdata.2018.86},
  urldate = {2025-07-20},
  copyright = {https://creativecommons.org/licenses/by/4.0},
  langid = {english}
}

@article{Stoffel2015,
  title = {Estimates of Volcanic-Induced Cooling in the {{Northern Hemisphere}} over the Past 1,500 Years},
  author = {Stoffel, Markus and Khodri, Myriam and Corona, Christophe and Guillet, S{\'e}bastien and Poulain, Virginie and Bekki, Slimane and Guiot, Jo{\"e}l and Luckman, Brian H. and Oppenheimer, Clive and Lebas, Nicolas and Beniston, Martin and {Masson-Delmotte}, Val{\'e}rie},
  year = 2015,
  month = oct,
  journal = {Nature Geoscience},
  volume = {8},
  number = {10},
  pages = {784--788},
  issn = {1752-0894, 1752-0908},
  doi = {10.1038/ngeo2526},
  urldate = {2026-03-13},
  langid = {english}
}

@article{Swart2019,
  title = {The {{Canadian Earth System Model}} Version 5 ({{CanESM5}}.0.3)},
  author = {Swart, Neil C. and Cole, Jason N. S. and Kharin, Viatcheslav V. and Lazare, Mike and Scinocca, John F. and Gillett, Nathan P. and Anstey, James and Arora, Vivek and Christian, James R. and Hanna, Sarah and Jiao, Yanjun and Lee, Warren G. and Majaess, Fouad and Saenko, Oleg A. and Seiler, Christian and Seinen, Clint and Shao, Andrew and Sigmond, Michael and Solheim, Larry and Von Salzen, Knut and Yang, Duo and Winter, Barbara},
  year = 2019,
  month = nov,
  journal = {Geoscientific Model Development},
  volume = {12},
  number = {11},
  pages = {4823--4873},
  issn = {1991-9603},
  doi = {10.5194/gmd-12-4823-2019},
  urldate = {2026-03-10},
  copyright = {https://creativecommons.org/licenses/by/4.0/},
  langid = {english}
}

@article{Tardif2019,
  title = {Last {{Millennium Reanalysis}} with an Expanded Proxy Database and Seasonal Proxy Modeling},
  author = {Tardif, Robert and Hakim, Gregory J. and Perkins, Walter A. and Horlick, Kaleb A. and Erb, Michael P. and {Emile-Geay}, Julien and Anderson, David M. and Steig, Eric J. and Noone, David},
  year = 2019,
  month = jul,
  journal = {Climate of the Past},
  volume = {15},
  number = {4},
  pages = {1251--1273},
  publisher = {Copernicus GmbH},
  issn = {1814-9332},
  doi = {10.5194/cp-15-1251-2019}
}

@article{Thomas2019,
  title = {Antarctic {{Sea Ice Proxies}} from {{Marine}} and {{Ice Core Archives Suitable}} for {{Reconstructing Sea Ice}} over the {{Past}} 2000 {{Years}}},
  author = {Thomas, Elizabeth R. and Allen, Claire S. and Etourneau, Johan and King, Amy C. F. and Severi, Mirko and Winton, V. Holly L. and Mueller, Juliane and Crosta, Xavier and Peck, Victoria L.},
  year = 2019,
  month = dec,
  journal = {Geosciences},
  volume = {9},
  number = {12},
  pages = {506},
  issn = {2076-3263},
  doi = {10.3390/geosciences9120506},
  urldate = {2025-10-10},
  copyright = {https://creativecommons.org/licenses/by/4.0/},
  langid = {english}
}

@misc{Thorne2026,
  title = {How Well Can We Quantify When 1.5 {$^\circ$}{{C}} of Global Warming Has Been Exceeded?},
  author = {Thorne, Peter W. and Nicklas, John M. and Kennedy, John J. and Calvert, Bruce and {Fox-Kemper}, Baylor and Richardson, Mark T. and Simmons, Adrian and Hawkins, Ed and Rhode, Robert and Cowtan, Kathryn and Abram, Nerilie J. and Andersson, Axel and Noone, Simon and Marbaix, Phillipe and Lenssen, Nathan and Olonscheck, Dirk and Walsh, Tristram and Outten, Stephen and Bethke, Ingo and Samset, Bjorn H. and Smith, Chris and Pirani, Anna and Fuglestvedt, Jan and Rajamani, Lavanya and Betts, Richard A. and Kent, Elizabeth C. and Trewin, Blair and Morice, Colin and Osborn, Tim and Burgess, Samantha N. and Geden, Oliver and Parnell, Andrew and Forster, Piers M. and Hewitt, Chris and Hausfather, Zeke and {Masson-Delmotte}, Valerie and Marotzke, Jochem and Gillett, Nathan and Seneviratne, Sonia I. and Schmidt, Gavin A. and Chan, Duo and Br{\"o}nnimann, Stefan and Reisinger, Andy and Menne, Matthew and Rojas Corradi, Maisa and Kadow, Christopher and Huybers, Peter and Stephenson, David B. and Wallis, Emily and Rogelj, Joeri and Schurer, Andrew and McKinnon, Karen and Zhai, Panmao and Driouech, Fatima and Moufouma Okia, Wilfran and Vazifehkhah, Saeed and Szopa, Sophie and Merchant, Christopher J. and Hirahara, Shoji and Ishii, Masayoshi and Engelbrecht, Francois A. and Li, Qingxiang and Lee, June-Yi and Cannon, Alex J. and Cassou, Christophe and Von Schuckmann, Karina and Delju, Amir H. and Murtagh, Ellie},
  year = 2026,
  month = jan,
  publisher = {ESSD -- Atmosphere/Meteorology},
  doi = {10.5194/essd-2025-825},
  urldate = {2026-03-12},
  archiveprefix = {ESSD -- Atmosphere/Meteorology},
  copyright = {https://creativecommons.org/licenses/by/4.0/},
  langid = {english}
}

@article{Tierney2015,
  title = {Tropical Sea Surface Temperatures for the Past Four Centuries Reconstructed from Coral Archives},
  author = {Tierney, Jessica E. and Abram, Nerilie J. and Anchukaitis, Kevin J. and Evans, Michael N. and Giry, Cyril and Kilbourne, K. Halimeda and Saenger, Casey P. and Wu, Henry C. and Zinke, Jens},
  year = 2015,
  month = mar,
  journal = {Paleoceanography},
  volume = {30},
  number = {3},
  pages = {226--252},
  issn = {0883-8305, 1944-9186},
  doi = {10.1002/2014PA002717},
  urldate = {2026-01-25},
  copyright = {http://creativecommons.org/licenses/by-nc-nd/4.0/},
  langid = {english}
}

@article{Timmreck2009,
  title = {Limited Temperature Response to the Very Large {{AD}} 1258 Volcanic Eruption},
  author = {Timmreck, Claudia and Lorenz, Stephan J. and Crowley, Thomas J. and Kinne, Stefan and Raddatz, Thomas J. and Thomas, Manu A. and Jungclaus, Johann H.},
  year = 2009,
  month = nov,
  journal = {Geophysical Research Letters},
  volume = {36},
  number = {21},
  pages = {2009GL040083},
  issn = {0094-8276, 1944-8007},
  doi = {10.1029/2009GL040083},
  urldate = {2026-03-13},
  copyright = {http://onlinelibrary.wiley.com/termsAndConditions\#vor},
  langid = {english}
}

@article{Timmreck2012,
  title = {Modeling the Climatic Effects of Large Explosive Volcanic Eruptions},
  author = {Timmreck, Claudia},
  year = 2012,
  month = nov,
  journal = {WIREs Climate Change},
  volume = {3},
  number = {6},
  pages = {545--564},
  issn = {1757-7780, 1757-7799},
  doi = {10.1002/wcc.192},
  urldate = {2024-06-24},
  copyright = {http://onlinelibrary.wiley.com/termsAndConditions\#vor},
  langid = {english}
}

@article{Tippett2003,
  title = {Ensemble Square Root Filters},
  author = {Tippett, Michael K. and Anderson, Jeffrey L. and Bishop, Craig H. and Hamill, Thomas M. and Whitaker, Jeffrey S.},
  year = 2003,
  month = jul,
  journal = {Monthly Weather Review},
  volume = {131},
  number = {7},
  pages = {1485--1490},
  publisher = {American Meteorological Society},
  issn = {1520-0493},
  doi = {10.1175/1520-0493(2003)131<1485:esrf>2.0.co;2}
}

@article{Toohey2017,
  title = {Volcanic Stratospheric Sulfur Injections and Aerosol Optical Depth from 500 {{BCE}} to 1900 {{CE}}},
  author = {Toohey, Matthew and Sigl, Michael},
  year = 2017,
  month = nov,
  journal = {Earth System Science Data},
  volume = {9},
  number = {2},
  pages = {809--831},
  issn = {1866-3516},
  doi = {10.5194/essd-9-809-2017},
  urldate = {2024-05-20},
  copyright = {https://creativecommons.org/licenses/by/3.0/},
  langid = {english}
}

@article{Trenberth2013a,
  title = {An Apparent Hiatus in Global Warming?},
  author = {Trenberth, Kevin E. and Fasullo, John T.},
  year = 2013,
  month = dec,
  journal = {Earth's Future},
  volume = {1},
  number = {1},
  pages = {19--32},
  issn = {2328-4277, 2328-4277},
  doi = {10.1002/2013EF000165},
  urldate = {2024-12-01},
  copyright = {http://creativecommons.org/licenses/by-nc-nd/3.0/},
  langid = {english}
}

@article{Trenberth2014,
  title = {Earth's {{Energy Imbalance}}},
  author = {Trenberth, Kevin E. and Fasullo, John T. and Balmaseda, Magdalena A.},
  year = 2014,
  month = may,
  journal = {Journal of Climate},
  volume = {27},
  number = {9},
  pages = {3129--3144},
  issn = {0894-8755, 1520-0442},
  doi = {10.1175/JCLI-D-13-00294.1},
  urldate = {2024-07-12},
  langid = {english}
}

@article{Trenberth2015,
  title = {Climate Variability and Relationships between Top-of-atmosphere Radiation and Temperatures on {{Earth}}},
  author = {Trenberth, Kevin E. and Zhang, Yongxin and Fasullo, John T. and Taguchi, Shoichi},
  year = 2015,
  month = may,
  journal = {Journal of Geophysical Research: Atmospheres},
  volume = {120},
  number = {9},
  pages = {3642--3659},
  issn = {2169-897X, 2169-8996},
  doi = {10.1002/2014JD022887},
  urldate = {2024-07-15},
  copyright = {http://onlinelibrary.wiley.com/termsAndConditions\#am},
  langid = {english}
}

@article{Valler2022,
  title = {An Updated Global Atmospheric Paleo-Reanalysis Covering the Last 400 Years},
  author = {Valler, Veronika and Franke, J{\"o}rg and Brugnara, Yuri and Br{\"o}nnimann, Stefan},
  year = 2022,
  month = jun,
  journal = {Geoscience Data Journal},
  volume = {9},
  number = {1},
  pages = {89--107},
  issn = {2049-6060, 2049-6060},
  doi = {10.1002/gdj3.121},
  urldate = {2026-03-16},
  langid = {english}
}

@article{Valler2024,
  title = {{{ModE-RA}}: A Global Monthly Paleo-Reanalysis of the Modern Era 1421 to 2008},
  author = {Valler, Veronika and Franke, J{\"o}rg and Brugnara, Yuri and Samakinwa, Eric and Hand, Ralf and Lundstad, Elin and Burgdorf, Angela-Maria and Lipfert, Laura and Friedman, Andrew Ronald and Br{\"o}nnimann, Stefan},
  year = 2024,
  month = jan,
  journal = {Scientific Data},
  volume = {11},
  number = {1},
  publisher = {{Springer Science and Business Media LLC}},
  issn = {2052-4463},
  doi = {10.1038/s41597-023-02733-8}
}

@article{VanDijk2024,
  title = {High-Frequency Climate Forcing Causes Prolonged Cold Periods in the {{Holocene}}},
  author = {Van Dijk, Evelien J. C. and Jungclaus, Johann and Sigl, Michael and Timmreck, Claudia and Kr{\"u}ger, Kirstin},
  year = 2024,
  month = may,
  journal = {Communications Earth \& Environment},
  volume = {5},
  number = {1},
  publisher = {{Springer Science and Business Media LLC}},
  issn = {2662-4435},
  doi = {10.1038/s43247-024-01380-0},
  urldate = {2025-07-20},
  copyright = {https://creativecommons.org/licenses/by/4.0},
  langid = {english}
}

@misc{VanLoon2025b,
  title = {Spatial {{Controls}} of {{Lower Tropospheric Stability}}},
  author = {Van Loon, Senne and Rugenstein, Maria},
  year = 2025,
  month = oct,
  number = {arXiv:2510.27575},
  eprint = {2510.27575},
  primaryclass = {physics},
  publisher = {arXiv},
  doi = {10.48550/arXiv.2510.27575},
  urldate = {2025-12-19},
  archiveprefix = {arXiv},
  langid = {english}
}

@article{Vidal2016,
  title = {The 1257 {{Samalas}} Eruption ({{Lombok}}, {{Indonesia}}): The Single Greatest Stratospheric Gas Release of the {{Common Era}}},
  shorttitle = {The 1257 Samalas Eruption (Lombok, Indonesia)},
  author = {Vidal, C{\'e}line M. and M{\'e}trich, Nicole and Komorowski, Jean-Christophe and Pratomo, Indyo and Michel, Agn{\`e}s and Kartadinata, Nugraha and Robert, Vincent and Lavigne, Franck},
  year = 2016,
  month = oct,
  journal = {Scientific Reports},
  volume = {6},
  number = {1},
  pages = {34868},
  issn = {2045-2322},
  doi = {10.1038/srep34868},
  urldate = {2026-03-14},
  langid = {english}
}

@article{Waelbroeck2002,
  title = {Sea-Level and Deep Water Temperature Changes Derived from Benthic Foraminifera Isotopic Records},
  author = {Waelbroeck, C. and Labeyrie, L. and Michel, E. and Duplessy, J.C. and McManus, J.F. and Lambeck, K. and Balbon, E. and Labracherie, M.},
  year = 2002,
  month = jan,
  journal = {Quaternary Science Reviews},
  volume = {21},
  number = {1-3},
  pages = {295--305},
  issn = {02773791},
  doi = {10.1016/S0277-3791(01)00101-9},
  urldate = {2025-01-16},
  copyright = {https://www.elsevier.com/tdm/userlicense/1.0/},
  langid = {english}
}

@article{Walter2023,
  title = {The {{CoralHydro2k}} Database: A Global, Actively Curated Compilation of Coral {\emph{{$\delta$}}} {\textsuperscript{18}} {{O}} and {{Sr}} / {{Ca}} Proxy Records of Tropical Ocean Hydrology and Temperature for the {{Common Era}}},
  shorttitle = {The {{CoralHydro2k}} Database},
  author = {Walter, Rachel M. and Sayani, Hussein R. and Felis, Thomas and Cobb, Kim M. and Abram, Nerilie J. and Arzey, Ariella K. and Atwood, Alyssa R. and Brenner, Logan D. and Dassi{\'e}, {\'E}milie P. and DeLong, Kristine L. and Ellis, Bethany and {Emile-Geay}, Julien and Fischer, Matthew J. and Goodkin, Nathalie F. and Hargreaves, Jessica A. and Kilbourne, K. Halimeda and Krawczyk, Hedwig and McKay, Nicholas P. and Moore, Andrea L. and Murty, Sujata A. and Ong, Maria Rosabelle and Ramos, Riovie D. and Reed, Emma V. and Samanta, Dhrubajyoti and Sanchez, Sara C. and Zinke, Jens and {the PAGES CoralHydro2k Project Members}},
  year = 2023,
  month = may,
  journal = {Earth System Science Data},
  volume = {15},
  number = {5},
  pages = {2081--2116},
  issn = {1866-3516},
  doi = {10.5194/essd-15-2081-2023},
  urldate = {2024-10-21},
  copyright = {https://creativecommons.org/licenses/by/4.0/},
  langid = {english}
}

@article{Wanner2022,
  title = {The Variable {{European Little Ice Age}}},
  author = {Wanner, Heinz and Pfister, Christian and Neukom, Raphael},
  year = 2022,
  month = jul,
  journal = {Quaternary Science Reviews},
  volume = {287},
  pages = {107531},
  issn = {02773791},
  doi = {10.1016/j.quascirev.2022.107531},
  urldate = {2025-09-06},
  langid = {english}
}

@article{Webb2017,
  title = {The {{Cloud Feedback Model Intercomparison Project}} ({{CFMIP}}) Contribution to {{CMIP6}}},
  author = {Webb, Mark J. and Andrews, Timothy and {Bodas-Salcedo}, Alejandro and Bony, Sandrine and Bretherton, Christopher S. and Chadwick, Robin and Chepfer, H{\'e}l{\`e}ne and Douville, Herv{\'e} and Good, Peter and Kay, Jennifer E. and Klein, Stephen A. and Marchand, Roger and Medeiros, Brian and Siebesma, A. Pier and Skinner, Christopher B. and Stevens, Bjorn and Tselioudis, George and Tsushima, Yoko and Watanabe, Masahiro},
  year = 2017,
  month = jan,
  journal = {Geoscientific Model Development},
  volume = {10},
  number = {1},
  pages = {359--384},
  issn = {1991-9603},
  doi = {10.5194/gmd-10-359-2017},
  urldate = {2024-11-21},
  copyright = {https://creativecommons.org/licenses/by/3.0/},
  langid = {english}
}

@article{Whitaker2002,
  title = {Ensemble Data Assimilation without Perturbed Observations},
  author = {Whitaker, Jeffrey S. and Hamill, Thomas M.},
  year = 2002,
  month = jul,
  journal = {Monthly Weather Review},
  volume = {130},
  number = {7},
  pages = {1913--1924},
  publisher = {American Meteorological Society},
  address = {Boston MA, USA},
  issn = {1520-0493},
  doi = {10.1175/1520-0493(2002)130<1913:edawpo>2.0.co;2}
}

@article{Wild2020,
  title = {The Global Energy Balance as Represented in {{CMIP6}} Climate Models},
  author = {Wild, Martin},
  year = 2020,
  month = aug,
  journal = {Climate Dynamics},
  volume = {55},
  number = {3-4},
  pages = {553--577},
  publisher = {{Springer Science and Business Media LLC}},
  issn = {0930-7575, 1432-0894},
  doi = {10.1007/s00382-020-05282-7},
  urldate = {2025-07-20},
  copyright = {https://creativecommons.org/licenses/by/4.0},
  langid = {english}
}

@article{Wills2021,
  title = {Slow Modes of Global Temperature Variability and Their Impact on Climate Sensitivity Estimates},
  author = {Wills, Robert C. J. and Armour, Kyle C. and Battisti, David S. and Proistosescu, Cristian and Parsons, Luke A.},
  year = 2021,
  month = nov,
  journal = {Journal of Climate},
  volume = {34},
  number = {21},
  pages = {8717--8738},
  publisher = {American Meteorological Society},
  issn = {0894-8755, 1520-0442},
  doi = {10.1175/jcli-d-20-1013.1},
  urldate = {2025-07-22},
  copyright = {http://www.ametsoc.org/PUBSReuseLicenses},
  langid = {english}
}

@article{Wilson2016,
  title = {Last Millennium Northern Hemisphere Summer Temperatures from Tree Rings: {{Part I}}: {{The}} Long Term Context},
  shorttitle = {Last Millennium Northern Hemisphere Summer Temperatures from Tree Rings},
  author = {Wilson, Rob and Anchukaitis, Kevin and Briffa, Keith R. and B{\"u}ntgen, Ulf and Cook, Edward and D'Arrigo, Rosanne and Davi, Nicole and Esper, Jan and Frank, Dave and Gunnarson, Bj{\"o}rn and Hegerl, Gabi and Helama, Samuli and Klesse, Stefan and Krusic, Paul J. and Linderholm, Hans W. and Myglan, Vladimir and Osborn, Timothy J. and Rydval, Milo{\v s} and Schneider, Lea and Schurer, Andrew and Wiles, Greg and Zhang, Peng and Zorita, Eduardo},
  year = 2016,
  month = feb,
  journal = {Quaternary Science Reviews},
  volume = {134},
  pages = {1--18},
  issn = {02773791},
  doi = {10.1016/j.quascirev.2015.12.005},
  urldate = {2026-02-23},
  langid = {english}
}

@article{Wu2025a,
  title = {Time-Varying Global Energy Budget since 1880 from a Reconstruction of Ocean Warming},
  author = {Wu, Quran and Gregory, Jonathan M. and Zanna, Laure and Khatiwala, Samar},
  year = 2025,
  month = may,
  journal = {Proceedings of the National Academy of Sciences},
  volume = {122},
  number = {20},
  pages = {e2408839122},
  issn = {0027-8424, 1091-6490},
  doi = {10.1073/pnas.2408839122},
  urldate = {2025-05-17},
  langid = {english}
}

@article{Xie2016,
  title = {Distinct Energy Budgets for Anthropogenic and Natural Changes during Global Warming Hiatus},
  author = {Xie, Shang-Ping and Kosaka, Yu and Okumura, Yuko M.},
  year = 2016,
  month = jan,
  journal = {Nature Geoscience},
  volume = {9},
  number = {1},
  pages = {29--33},
  publisher = {{Springer Science and Business Media LLC}},
  issn = {1752-0894, 1752-0908},
  doi = {10.1038/ngeo2581},
  urldate = {2025-07-22},
  copyright = {http://www.springer.com/tdm},
  langid = {english}
}

@article{Yukimoto2019,
  title = {The {{Meteorological Research Institute Earth System Model Version}} 2.0, {{MRI-ESM2}}.0: {{Description}} and {{Basic Evaluation}} of the {{Physical Component}}},
  shorttitle = {The {{Meteorological Research Institute Earth System Model Version}} 2.0, {{MRI-ESM2}}.0},
  author = {Yukimoto, Seiji and Kawai, Hideaki and Koshiro, Tsuyoshi and Oshima, Naga and Yoshida, Kohei and Urakawa, Shogo and Tsujino, Hiroyuki and Deushi, Makoto and Tanaka, Taichu and Hosaka, Masahiro and Yabu, Shokichi and Yoshimura, Hiromasa and Shindo, Eiki and Mizuta, Ryo and Obata, Atsushi and Adachi, Yukimasa and Ishii, Masayoshi},
  year = 2019,
  journal = {Journal of the Meteorological Society of Japan. Ser. II},
  volume = {97},
  number = {5},
  pages = {931--965},
  issn = {0026-1165, 2186-9057},
  doi = {10.2151/jmsj.2019-051},
  urldate = {2024-06-10},
  langid = {english}
}

@article{Zanchettin2019,
  title = {Clarifying the {{Relative Role}} of {{Forcing Uncertainties}} and {{Initial-Condition Unknowns}} in {{Spreading}} the {{Climate Response}} to {{Volcanic Eruptions}}},
  author = {Zanchettin, Davide and Timmreck, Claudia and Toohey, Matthew and Jungclaus, Johann H. and Bittner, Matthias and Lorenz, Stephan J. and Rubino, Angelo},
  year = 2019,
  month = feb,
  journal = {Geophysical Research Letters},
  volume = {46},
  number = {3},
  pages = {1602--1611},
  issn = {0094-8276, 1944-8007},
  doi = {10.1029/2018GL081018},
  urldate = {2026-03-13},
  langid = {english}
}

@article{Zanna2019,
  title = {Global Reconstruction of Historical Ocean Heat Storage and Transport},
  author = {Zanna, Laure and Khatiwala, Samar and Gregory, Jonathan M. and Ison, Jonathan and Heimbach, Patrick},
  year = 2019,
  month = jan,
  journal = {Proceedings of the National Academy of Sciences},
  volume = {116},
  number = {4},
  pages = {1126--1131},
  publisher = {Proceedings of the National Academy of Sciences},
  doi = {10.1073/pnas.1808838115}
}

@article{Zhong2011,
  title = {Centennial-Scale Climate Change from Decadally-Paced Explosive Volcanism: A Coupled Sea Ice-Ocean Mechanism},
  shorttitle = {Centennial-Scale Climate Change from Decadally-Paced Explosive Volcanism},
  author = {Zhong, Y. and Miller, G. H. and {Otto-Bliesner}, B. L. and Holland, M. M. and Bailey, D. A. and Schneider, D. P. and Geirsdottir, A.},
  year = 2011,
  month = dec,
  journal = {Climate Dynamics},
  volume = {37},
  number = {11-12},
  pages = {2373--2387},
  issn = {0930-7575, 1432-0894},
  doi = {10.1007/s00382-010-0967-z},
  urldate = {2024-06-14},
  copyright = {http://www.springer.com/tdm},
  langid = {english}
}

@article{Zhou2016,
  title = {Impact of Decadal Cloud Variations on the {{Earth}}'s Energy Budget},
  author = {Zhou, Chen and Zelinka, Mark D. and Klein, Stephen A.},
  year = 2016,
  month = dec,
  journal = {Nature Geoscience},
  volume = {9},
  number = {12},
  pages = {871--874},
  issn = {1752-0894, 1752-0908},
  doi = {10.1038/ngeo2828},
  urldate = {2024-07-19},
  langid = {english}
}

@article{Zhu2019,
  title = {Climate Models Can Correctly Simulate the Continuum of Global-Average Temperature Variability},
  author = {Zhu, Feng and {Emile-Geay}, Julien and McKay, Nicholas P. and Hakim, Gregory J. and Khider, Deborah and Ault, Toby R. and Steig, Eric J. and Dee, Sylvia and Kirchner, James W.},
  year = 2019,
  month = apr,
  journal = {Proceedings of the National Academy of Sciences},
  volume = {116},
  number = {18},
  pages = {8728--8733},
  issn = {0027-8424, 1091-6490},
  doi = {10.1073/pnas.1809959116},
  urldate = {2026-01-14},
  langid = {english}
}

@article{Zhu2020,
  title = {Resolving the {{Differences}} in the {{Simulated}} and {{Reconstructed Temperature Response}} to {{Volcanism}}},
  author = {Zhu, Feng and {Emile-Geay}, Julien and Hakim, Gregory J. and King, Jonathan and Anchukaitis, Kevin J.},
  year = 2020,
  journal = {Geophysical Research Letters},
  volume = {47},
  number = {8},
  pages = {e2019GL086908},
  issn = {1944-8007},
  doi = {10.1029/2019GL086908},
  urldate = {2024-06-09},
  langid = {english}
}

@article{Zhu2022,
  title = {A Re-Appraisal of the {{ENSO}} Response to Volcanism with Paleoclimate Data Assimilation},
  author = {Zhu, Feng and {Emile-Geay}, Julien and Anchukaitis, Kevin J. and Hakim, Gregory J. and Wittenberg, Andrew T. and Morales, Mariano S. and Toohey, Matthew and King, Jonathan},
  year = 2022,
  month = feb,
  journal = {Nature Communications},
  volume = {13},
  number = {1},
  pages = {747},
  issn = {2041-1723},
  doi = {10.1038/s41467-022-28210-1},
  urldate = {2024-11-11},
  langid = {english}
}

\end{document}


\narrowlayout

\maketitle

\SItext

\subsection{Linear Inverse Model}

A linear inverse model (LIM) is a dynamical system emulator that explicitly models the linear, deterministic dynamics while parameterizing the non-linear residual as noise that is uncorrelated in time but may be correlated in the state variables. The linear dynamics are stable, so that anomalies decay toward the mean in the long-time limit. The noise forcing maintains the system in statistical equilibrium against the persistent decay from the linear dynamics. In the context of online data assimilation (DA), the linear part propagates the posterior mean forward in time while the noise part contributes to the covariance structure needed for the Kalman filter. The LIM dynamics have the form
\begin{align*}
    \dv{\vb{x}}{t} = \matr{L}\vb{x} + \matr{S}\vb*{\eta} = \matr{L}\vb{x} + \vb*{\xi},
\end{align*}
where $\vb{x} \in \mathbb{R}^{N_x}$ is the state vector, $\matr{L} \in \mathbb{R}^{N_x \times N_x}$ encodes the linear dynamics, $\matr{S} \in \mathbb{R}^{N_x \times N_x}$ is the noise amplitude matrix, and $\vb*{\eta} \sim \mathcal{N}(\vb{0}, \matrrm{I}) \in \mathbb{R}^{N_x}$ is additive Gaussian white noise with unit variance. Alternatively, the noise term can be written as $\vb*{\xi} \sim \mathcal{N}(\vb{0}, \matr{Q}/dt) \in \mathbb{R}^{N_x}$, where $\matr{Q} = \matr{S}\matr{S}^\text{T}\! dt$ is the noise covariance matrix. The state vector represents anomalies about a mean state. The state dimension here is $N_x = 130$, which is the sum of the number of principal components retained after EOF truncation.

Integrating the LIM dynamics in time, and taking the expectation over the noise forcing yields a deterministic forecast equation between two time steps~\citep{Penland1995}:
\begin{align}
    \label{eq:lim-forecast-det}
    \vu{x}(t + \tau) = \exp (\matr{L} \, \tau) \, \vb{x}(t)  = \matr{G} \vb{x}(t) ,
\end{align}
where $\matr{G} \in \mathbb{R}^{N_x \times N_x}$ is the linear forecast operator and $\tau = \text{1 season}$ is the time step. The forecast is stable (i.e., decays to zero for $\tau \to \infty$) if all eigenvalues of $\matr{L}$ have negative real parts. For probabilistic ensemble forecasts during data assimilation, we use the two-step stochastic integration scheme described by \citet{Penland1994}:
\begin{align*}
    \vb{\tilde x}(t+\delta t) &= \vb{\tilde x}(t) + \left[ \matr{L}\vb{\tilde x}(t) + \matr{S} \vb*{\eta}_0\right]\! \delta t,
    \\
    \vb{x}(t + \delta t / 2) &= \left[\vb{\tilde x}(t) + \vb{\tilde x}(t+\delta t)\right]\! / 2,
\end{align*}
where $\vb{\tilde x}$ is an intermediate variable and $\vb*{\eta}_0$ is a sample of the noise $\vb*{\eta}$. The integration time step $\delta t$ must be chosen much smaller than the corresponding deterministic time step $\tau$. We use 360 integration steps per $\tau$, or $\delta t \approx \text{6 hr}$~\citep{Penland1994,Perkins2021}. In practice, the whole ensemble can be propagated simultaneously by adding an ensemble dimension to $\vb{x}$ and by sampling an $N_x \times N_e$ noise matrix $\vb*{\eta}_0$. We use $N_e = 400$.

The system dynamics $\matr{L}$ and $\matr{G}$ as well as the noise covariance $\matr{Q}$ are determined from training data. The procedure is based on the zero-lag and $\tau$-lag covariance matrices:
\begin{align*}
    \matr{C}(0) = \langle \vb{x}(t) \vb{x}^\transpose\!(t) \rangle \qquad \mathrm{and} \qquad \matr{C}(\tau) = \langle \vb{x}(t+\tau) \vb{x}^\transpose\!(t) \rangle,
\end{align*}
where $\langle \cdot \rangle$ denotes the time average over all training data $\vb{x}(t)$. The deterministic forecast operator is
\begin{align*}
    \matr{G} = \matr{C}(\tau) \matr{C}(0)^{-1},
\end{align*}
i.e., the linear regression between the state over a $\tau$-lag interval. The linear dynamics $\matr{L}$ required for stochastic integration is then found by rearranging Eq. (\ref{eq:lim-forecast-det}) as
\begin{align*}
    \matr{L} = \frac{\ln\matr{G}}{\tau}.
\end{align*}
Practically, the logarithm is evaluated using the eigendecomposition of $\matr{G}$ (i.e., $\ln\matr{G} = \matrtil{G} \left(\ln \matr{\Lambda}_{\matr{G}}\right) \matrtil{G}$). Finally, we can find the noise covariance matrix from the fluctuation--dissipation relation as~\citep{Penland1995}
\begin{align*}
    \matr{Q} = - \left( \matr{L} \matr{C}(0) + \matr{C}(0) \matr{L} \right).
\end{align*}
The noise amplitude matrix is then~\citep{Penland1994}
\begin{align*}
      \matr{S} = \matrtil{Q} \left(\frac{\matr{\Lambda}}{\delta t}\right)^{1/2}
\end{align*}
where $\matrtil{Q}$ and $\matr{\Lambda}$ are the eigenvectors and eigenvalues of the noise covariance matrix $\matr{Q} = \matrtil{Q} \matr{\Lambda}\matrtil{Q}^{-1}$. While covariance matrices are always positive semi-definite, the $\matr{Q}$ derived from data may have spurious negative eigenvalues as a result of a small training dataset or significant non-linear dynamics~\citep{Penland1994}. We remove these negative eigenvalues and their respective eigenvectors from $\matr{\Lambda}$ and $\matrtil{Q}$, then rescale the remaining eigenvalues to retain the total variance~\citep[e.g.,][]{Penland1995}.

\subsection{Ensemble Kalman Filter}

Data assimilation applies Bayes' theorem to update a prior (e.g., a LIM forecast) with observations (e.g., proxies), taking into account their respective uncertainties. Since this becomes intractable for arbitrary distributions in high dimensions, Kalman filters assume normally distributed state variables, so that only the means and covariances need to be updated. In ensemble Kalman filters (EnKFs), the covariances are represented by an ensemble, which comprises equally likely samples. We use a particular flavor of EnKF, the serial ensemble square root filter (EnSRF;~\citealp{Whitaker2002}). A serial filter assimilates observations one-by-one by assuming independent observation error. Square root filters factorize the covariance matrix into its matrix square roots, which simplifies some calculations~\citep{Tippett2003}. In combination with serial observations, this becomes a regular scalar square root. See \citet{Houtekamer2016} for an extensive review of EnKFs.

The ensemble update given a single observation $y$, denoted by $\text{EnSRF}(y, \vb{x}_b)$, works as follows. First, the prior ensemble is converted into perturbations about the ensemble mean $\vb{\bar x}_b$:
\begin{align}
      \label{eq:kf-perturbs}
    \matr{X}_b = \mqty[ \vb{x}_b^{1} - \vb{\bar x}_b &  \vb{x}_b^{2} - \vb{\bar x}_b & \ldots &  \vb{x}_b^{N_e} - \vb{\bar x}_b ] \in \mathbb{R}^{N_x \times N_e},
\end{align}
where $\vb{x}_b^i \in \mathbb{R}^{N_x \times 1}$ are the prior ensemble members and $\vb{\bar x}_b$ is their mean. We then estimate the expected proxy observation, expressed as perturbation about the ensemble mean observation estimate, based on the prior:
\begin{align}
      \label{eq:kf-obs-estimate}
    \matr{\hat Y} = \matr{H} \matr{X}_b  \in \mathbb{R}^{1 \times N_e},
\end{align}
which has a single row because we use serial observations and thus the observation operator is $\matr{H} \in \mathbb{R}^{1 \times N_x}$. The scalar variance $\hat{\sigma}_b^2$ of the observation estimate is
\begin{align*}
    \hat{\sigma}_b^2 = \frac{\matrhat{Y} \matrhat{Y}^\transpose}{N_e-1}
\end{align*}
and the covariance vector $\vb*{\hat{\sigma}}_{\vb{x}y}$ between the prior state and the observation estimate is
\begin{align*}
    \vb*{\hat{\sigma}}_{\vb{x}y} = \frac{\matr{X}_b \matrhat{Y}^\transpose}{N_e-1} \in \mathbb{R}^{N_x \times 1}.
\end{align*}
The proxy error variance, determined during PSM calibration, is $\sigma_o^2$. By the Kalman filter update equation, the posterior ensemble mean is then
\begin{align}
      \label{eq:kf-update-mean}
    \vb{\bar x}_a = \vb{\bar x} + \matr{K} \left( y - \matr{H} \vb{\bar x} \right),
\end{align}
where $y$ is the actual proxy observation. The Kalman gain
\begin{align*}
    \matr{K} = \frac{\vb*{\hat{\sigma}}_{\vb{x}y}}{\sigma_o^2 + \hat{\sigma}_b^2} \in \mathbb{R}^{N_x \times 1}
\end{align*}
maps back from the observation space into the state space, updating any component of the state that covaries with the observation as determined from the ensemble. The Kalman gain takes the respective uncertainties of the prior and the observation into account: if the observation or its estimate are uncertain ($\sigma_o^2 + \hat{\sigma}_b^2$ large), then the posterior will remain closer to the prior. Notice that, if the actual and estimated proxy are the same, the innovation $y - \matr{H} \vb{\bar x}$ is zero and the prior remains unchanged.

The perturbations about the ensemble mean need to be updated such that the posterior ensemble covariance correctly reflects the prior and observation uncertainties. This requires that the Kalman gain is reduced by a factor $\alpha$, otherwise the posterior variance is underestimated~\citep{Whitaker2002}:
\begin{align*}
    \alpha = \left( 1 + \sqrt{\frac{\sigma_o^2}{\sigma_o^2 + \hat{\sigma}_b^2}} \right)^{-1} < 1.
\end{align*}
The posterior perturbations are then
\begin{align*}
    \matr{X}_a = \matr{X}_b - \alpha \matr{K} \matrhat{Y},
\end{align*}
which has the same form as Eq. (\ref{eq:kf-update-mean}), except that $y=0$ since all observation information is absorbed into the mean, and that the reduced Kalman gain $\alpha \matr{K}$ is used. The perturbation update can also be expressed as
\begin{align*}
      \matr{X}_a = \left( \matrrm{I} - \alpha \matr{K} \matr{H} \right) \matr{X}_b,
\end{align*}
which demonstrates that the posterior perturbations are just a linear combination of the prior perturbations. Finally, we combine the updated mean and perturbations to form the posterior ensemble
\begin{align*}
    \vb{x}_a^i = \vb{\bar x}_a + \matr{X}_a^i,
\end{align*}
essentially reversing Eq. (\ref{eq:kf-perturbs}). This concludes the assimilation of a single observation. The algorithm is repeated for all observations available at the current time step, with the posterior after each iteration serving as prior for assimilating the next observation. The final posterior is then propagated forward in time using the LIM to form the prior of the next time step.

As described in the main text, we update a window $\mathcal{T}$ of multiple seasons at once using the algorithm from \citet{Huntley2010}. This algorithm acknowledges the time-averaged nature of annually-resolved proxies. For example, if a proxy is sensitive to MAMJJA, the MAMJJA average is updated directly, rather than each season individually. The covariance with the annual average is expected to be less noisy, and fewer covariances have to be estimated. The time-mean state is denoted as $\langle \vb{x} \rangle$, seasonal perturbations around this mean as $\vb{x}'$, and the EnSRF algorithm for a single observation $y$ and the prior ensemble $\vb{x}_b^0, \ldots, \vb{x}_b^{N_e}$ described above as $\text{EnSRF}(y, \vb{x}_b)$.

We start by determining all seasons $t \in \mathcal{T}$ that the current observation $y$ is averaged over (i.e., the proxy seasonality). We find the perturbations for each season around the prior time mean $\langle \vb{x} \rangle_b$ over $\mathcal{T}$:
\begin{align*}
      \vb{x}_b'(t) = \vb{x}_b(t) - \langle \vb{x} \rangle_b, \quad t \in \mathcal{T}.
\end{align*}
We then apply the EnSRF to the time mean:
\begin{align*}
      \langle \vb{x} \rangle_a = \text{EnSRF}\left(y, \langle \vb{x} \rangle_b\right).
\end{align*}
Note that, for $\matrhat{Y}$ in Eq. (\ref{eq:kf-obs-estimate}), \citet{Huntley2010} use $\langle \matr{H} \vb{X}_b \rangle$ rather than $\matr{H} \langle \vb{X}_b \rangle$. However, this only matters for a nonlinear observation operator, while our linear PSM commutes with the time averaging. Finally, we obtain the posterior ensemble for each season by adding back the prior perturbations to the updated time mean:
\begin{align*}
      \vb{x}_a(t) = \langle \vb{x} \rangle_a + \vb{x}_b'(t), \quad t \in \mathcal{T}.
\end{align*}

An alternative to time averaging for windowed DA is time stacking as done by \citet{Meng2025}. They stack the states from all seasons to form $\overline{\mathbf{x}}$, then apply the EnSRF to the stacked state, which requires the estimation of a larger covariance matrix. However, tests showed that this does not lead to different results in practice (not shown).

\subsection{Proxy system models}

For all proxies, we use a linear proxy system model (PSM):
\begin{align*}
      y = \matr{H} \vb{x} + y_0 + \epsilon = h \matr{H}^* \vb{x} + y_0 + \epsilon,
\end{align*}
where $\matr{H}^* \in \mathbb{R}^{1 \times N_x}$ maps from the state space, which consists of the global means and the principal components associated with the EOF basis, to the temperature at the nearest grid point (surface air temperature for terrestrial proxies, sea surface temperature for marine proxies). The linear relationship between that temperature and the proxy value $y$ is characterized by the slope $h$ and the intercept $y_0$. Only the slope contains useful information, while the intercept converts from anomalies to absolute proxy values. The parameters are determined using ordinary least squares, and the residuals are used to estimate the observation error $\epsilon \sim \mathcal{N}(0, \sigma_o^2)$, which is then used in the EnSRF. Note that we assume that all observation errors are independent, which is a prerequisite for serial data assimilation.

The PSMs are calibrated on GISTEMP v4~\citep{Lenssen2024,GISTEMPTeam2025} and ERSSTv6~\citep{Huang2025a,Huang2025}, regridded to the same 2°$\times$ 2° grid as the reconstruction, then truncated to the EOF space. We use 1900--2000 data for GISTEMP and 1850--2000 data for ERSST. Note that the reconstruction is somewhat sensitive to the exact calibration dataset and period used.

During calibration, we objectively determine the seasonality of annual proxies. We test ten candidate seasonalities that are common in proxies, ranging from a single season to all four. The candidate PSM with the lowest Bayesian information criterion, or essentially the maximum likelihood, is then chosen~\citep{Perkins2021}. Proxies are excluded if they have less than 25 years of overlap with the calibration data, have correlations below $0.10$ with the calibration data, or have residual ($\epsilon$) annual autocorrelations exceeding $0.9$, as in~\citet{Meng2025}.

\widelayout


\begin{figure}
    \centering
      \includegraphics{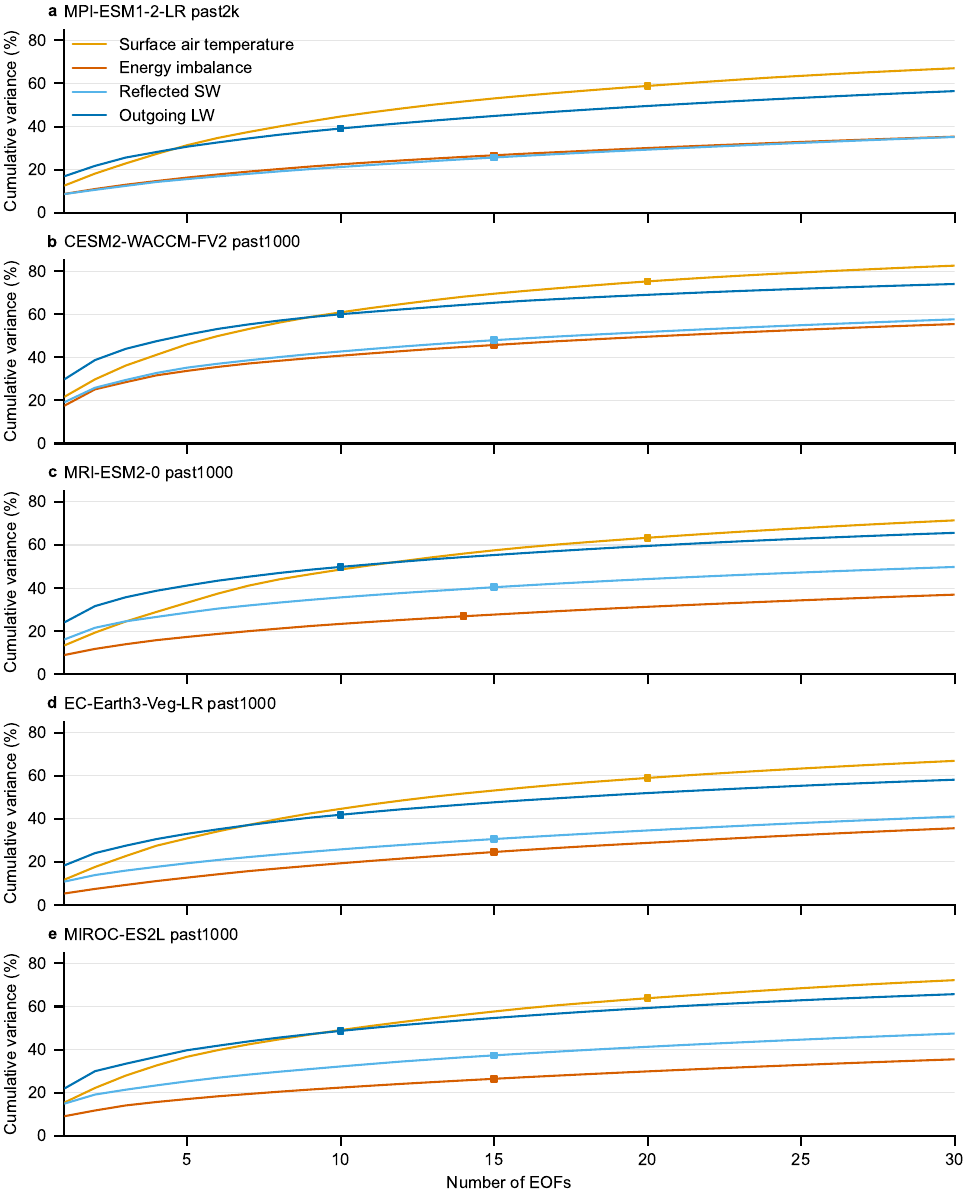}
      \caption{Fraction of explained variance as a function of the number of EOFs, for five last-millennium simulations (up to 1850) used to train LIMs. The variance due to the global mean is added as constant offset. We truncate the EOFs as follows, also indicated by the squares: 20 for SAT, 15 for EEI, 15 for RSR, 10 for OLR. For MRI-ESM2-0, we use 14 EEI PCs to minimize the LIM eigenvalue rescaling. SAT and OLR have a few dominant modes, particularly in CESM2. RSR and thus EEI are more noisy, and even with 20 EOFs for EEI, we explain only $30$--$50\%$ of the variance.}
      \label{fig:pc-spectra}
\end{figure}

\begin{figure}
    \centering
      \includegraphics{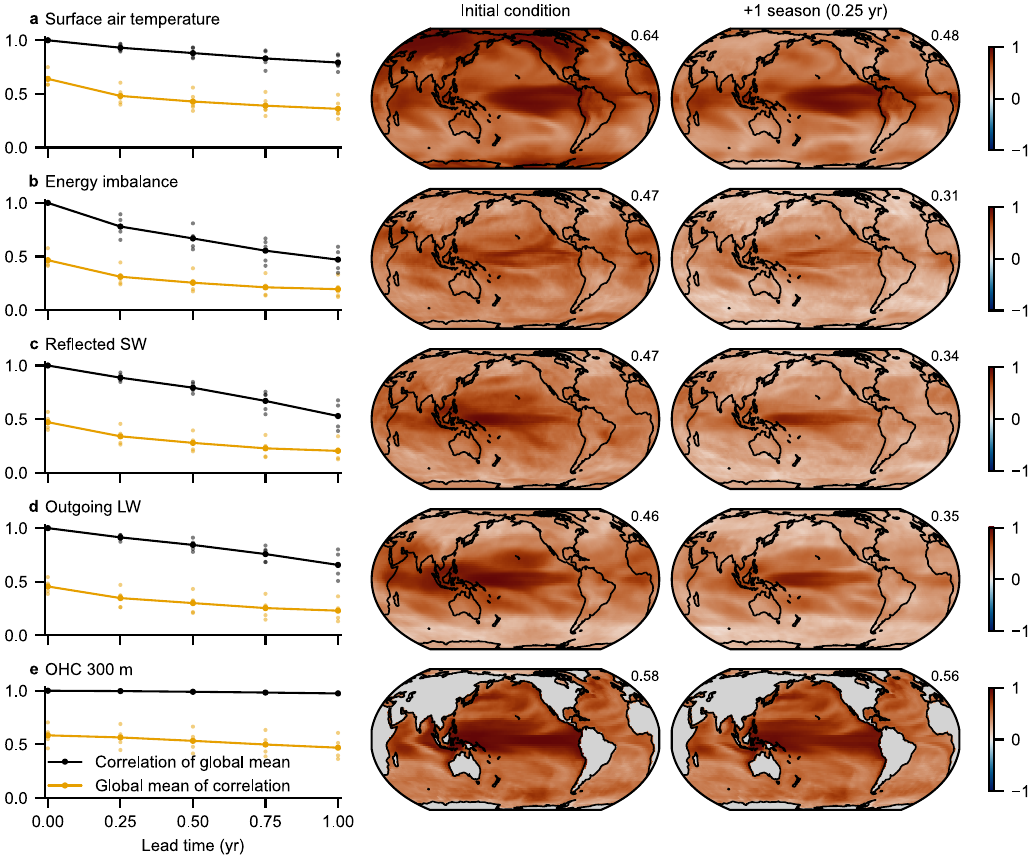}
      \caption{LIM forecast correlation at seasonal timescales. The correlation is based on deterministic forecasts for perfect-model 200 cases over 850--1850. (left) Correlation of the global mean (black) and global mean of the correlation (yellow; i.e., average over the maps on the right side). At each time step, the dots represent one of the five model priors. (right) Spatial correlations for the initial condition and at a one-season lead time (i.e., a single forecasting step), averaged over all model priors. Numbers in the top right corners represent global-mean values (i.e., the yellow data for lead times of 0.0 and 0.25 years on the left side). The initial condition correlation below one for spatial fields is due to the EOF truncation. The initial condition correlation is exactly one for the global means, demonstrating the benefit of separating the global mean in the state vector before EOF truncation to reconstruct the full global-mean variance.}
      \label{fig:lim-forecast}
\end{figure}

\begin{figure}
    \centering
      \includegraphics{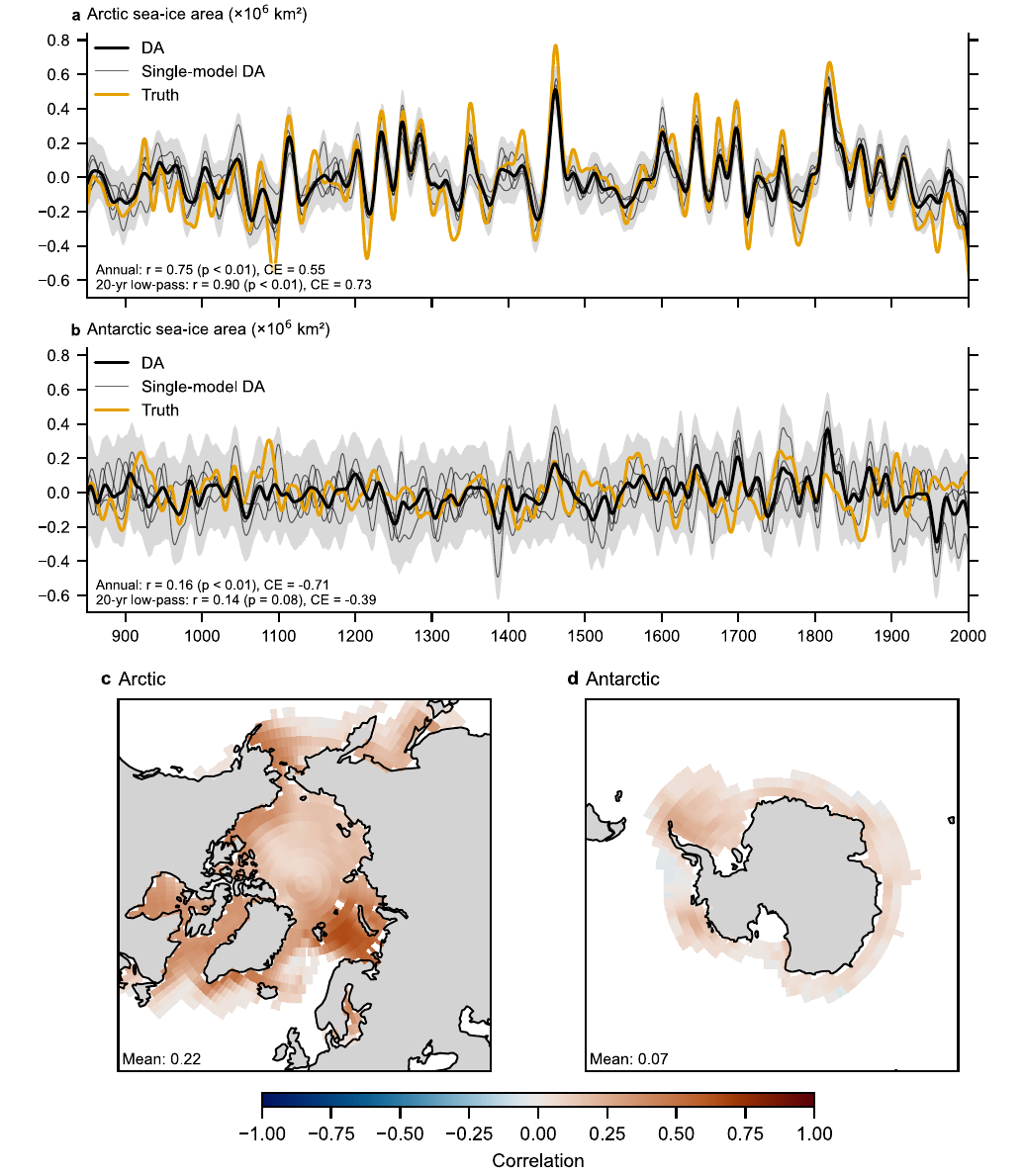}
      \caption{Comparison of the pseudoproxy imperfect-model reconstruction with the truth simulation, as in Fig.~2 but for sea-ice area and concentration. The truth model, MIROC6, has very little Antarctic sea-ice variability due to a Southern Ocean warm bias~\citep{Hajima2020}, leading to overestimated variability and thus a negative CE.}
      \label{fig:sic_ppe}
\end{figure}

\begin{figure}
    \centering
      \includegraphics{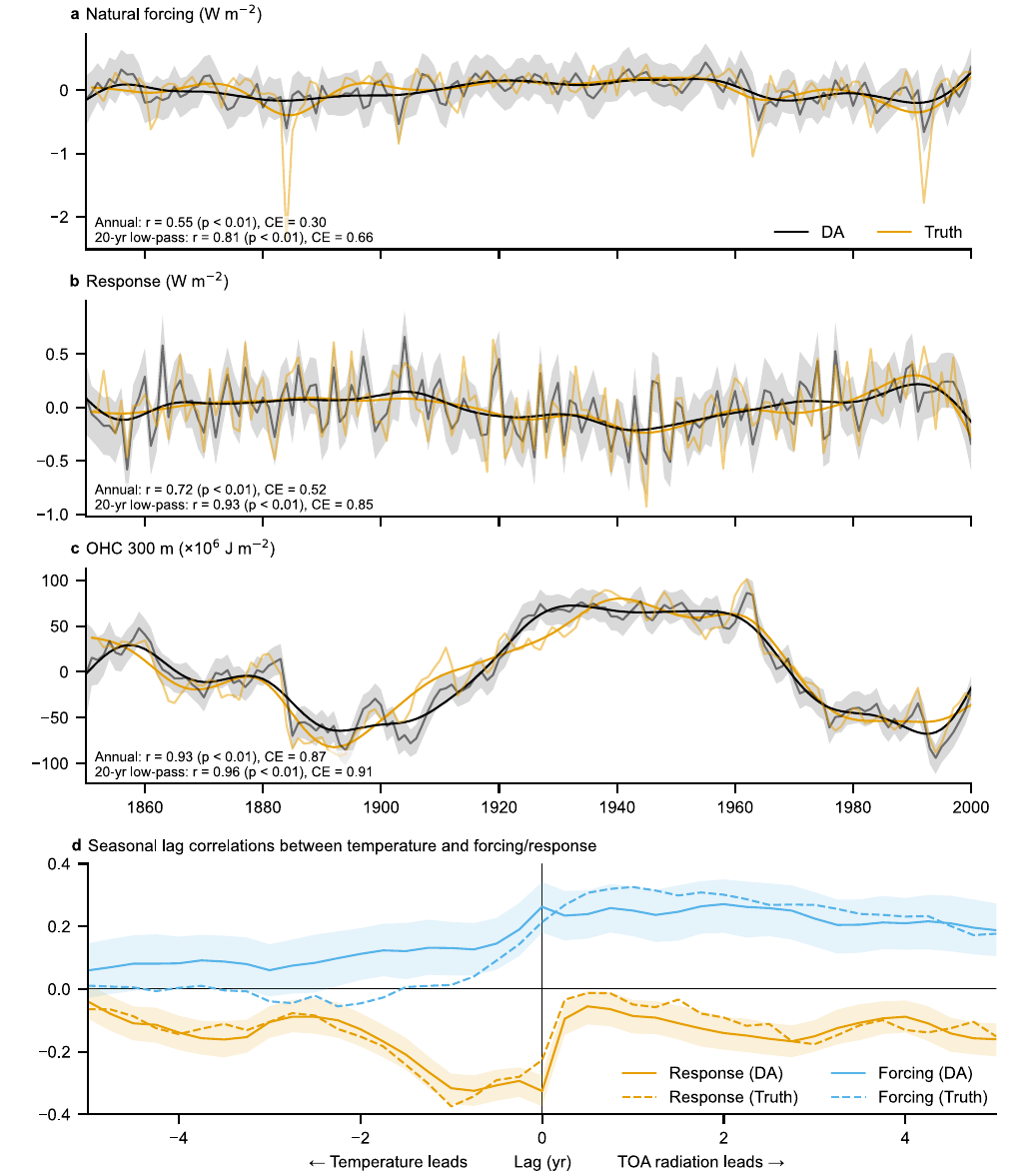}
      \caption{Example of pseudoproxy experiment with forcing (natural only) and response separated, to test if both are reconstructable. Both model prior and truth are from CanESM5. SAT, SST, and OHC300 are from the hist-nat simulation~\citep{Gillett2016}; the natural forcing is derived from the ensemble-mean difference on TOA radiation between the piClim-histnat and piClim-control simulation~\citep{Pincus2016}; the response is then the residual TOA radiation in the hist-nat simulation, making it consistent with the internal variability expressed in the SAT/SST. The pseudoproxy network emulates the real-proxy network used in the reconstruction but with constant time availability. (a--c) Timeseries of true and reconstructed forcing, radiative response, and OHC300. Correlations are significant at annual and 20-yr timescales. The magnitude of annual variability on TOA radiation is underestimated, particularly in the forcing and for volcanic eruptions (e.g., 1883, 1963, 1991). (d) Lag correlations between temperature and TOA radiation. Temperature leads the response, but forcing generally leads temperature, with good agreement between the reconstruction and the truth (i.e., the CanESM5 hist-nat simulation). For the reconstruction, however, there is a small but significant correlation when temperature leads the forcing, which is likely the reason that the forcing is reconstructable from temperature-sensitive proxies in the first place.}
      \label{fig:forcing_response_ppe}
\end{figure}

\begin{table}\centering
\caption{Statistics of pseudoproxy experiments with forcing (natural only) and response separated, as in Fig.~S4 but for additional prior/truth model combinations. The first value in each cell refers to the annual correlation or CE, the second value (after the /) refers to the correlation or CE of the 20-yr low-pass-filtered data. Asterisks denote the statistical significance (** if $p\leq 0.01$, * if $0.01 < p \geq 0.05$, and n.s. if $p > 0.05$) of correlations based on the random-phase test~\citep{Ebisuzaki1997}.}

\begin{tabular}{@{}lllllllll@{}}
\toprule
\multicolumn{1}{c}{\multirow{2}{*}{\textbf{Training}}} & \multicolumn{1}{c}{\multirow{2}{*}{\textbf{Truth}}} & \multicolumn{1}{c}{\multirow{2}{*}{\textbf{Type}}} & \multicolumn{2}{c}{\textbf{Forcing}} & \multicolumn{2}{c}{\textbf{Response}} & \multicolumn{2}{c}{\textbf{OHC 300 m}} \\ \cmidrule(l){4-9} 
\multicolumn{1}{c}{}                                   & \multicolumn{1}{c}{}                                & \multicolumn{1}{c}{}                               & r                  & CE              & r                  & CE               & r                    & CE              \\ \midrule

CanESM5                                                & CanESM5                                             & Perfect model                     & 0.55 (**) / 0.81 (**) & 0.30 (0.66)  & 0.72 (**) / 0.93 (**) & 0.52 (0.85)   & 0.93 (**) / 0.96 (**)   & 0.87 (0.91)  \\
NorESM2-LM                                             & NorESM2-LM                                          & Perfect model                     & 0.54 (**) / 0.74 (**) & 0.28 (0.55)  & 0.79 (**) / 0.72 (**) & 0.62 (0.51)   & 0.86 (**) / 0.92 (**)   & 0.72 (0.83)  \\
CanESM5                                                & NorESM2-LM                                          & Imperfect model                   & 0.22 (*) / 0.33 (n.s) & -0.01 (0.06) & 0.69 (**) / 0.56 (**) & 0.36 (0.27)   & 0.72 (**) / 0.75 (**)   & 0.47 (0.48)  \\
NorESM2-LM                                             & CanESM5                                             & Imperfect model                   & 0.16 (*) / 0.52 (**)  & -0.18 (0.27) & 0.46 (**) / 0.53 (*)  & -0.94 (-0.02) & 0.77 (**) / 0.83 (**)   & 0.59 (0.69)  \\ \bottomrule

\end{tabular}
\end{table}

\begin{figure}
    \centering
      \includegraphics{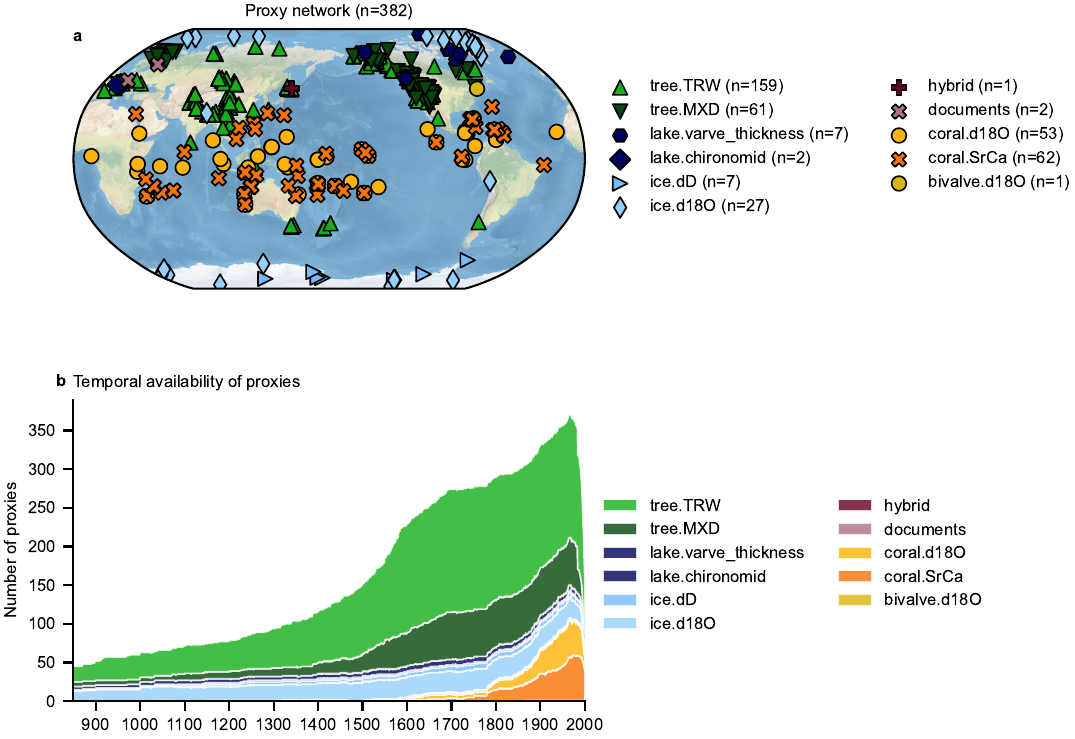}
      \caption{Proxy network after calibration comprising the PAGES2k~\citep{PAGESConsortium2017} and CoralHydro2k~\citep{Walter2023} databases, and the \cite{Dee2020} coral proxy. The proxy availability increases drastically after 1500, and most corals only start after 1700.}
      \label{fig:proxy_network}
\end{figure}


\begin{figure}
    \centering
      \includegraphics{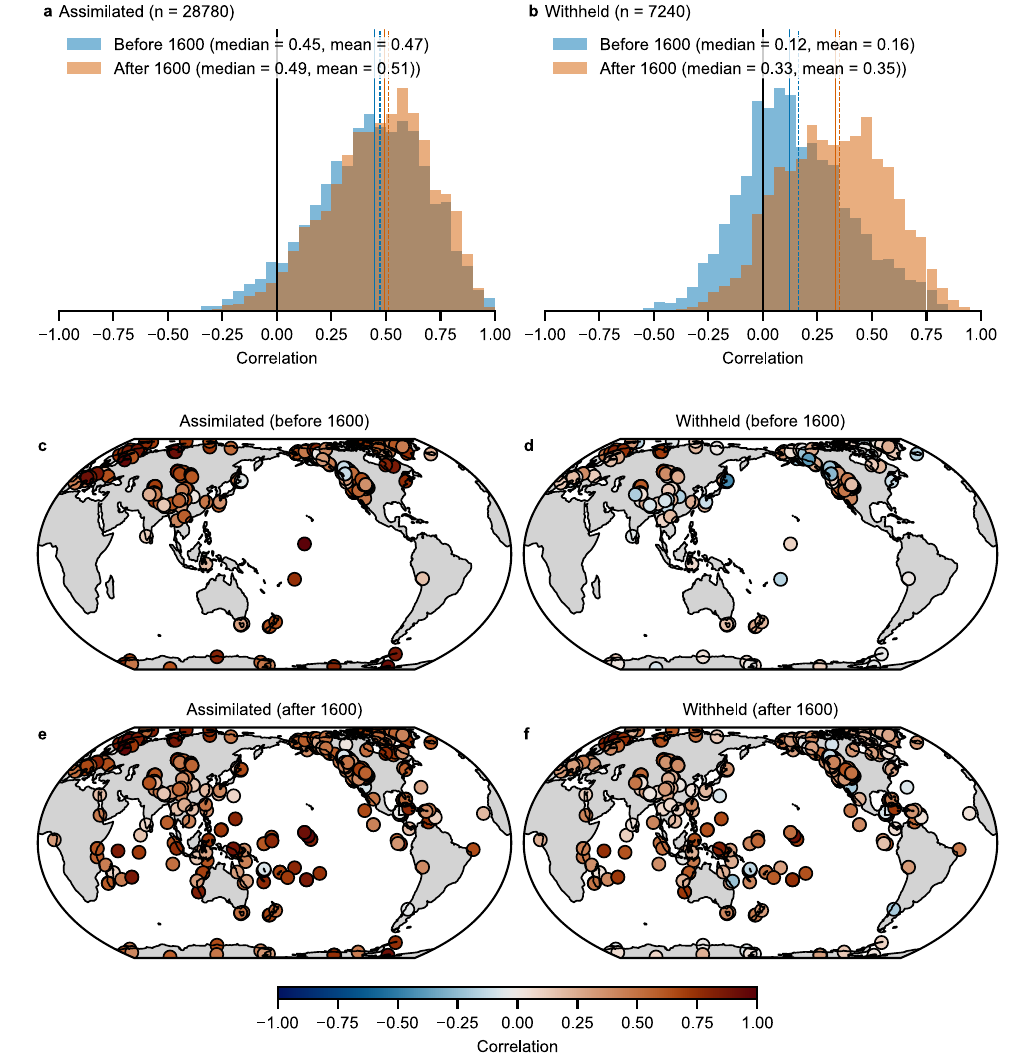}
      \caption{Correlation of assimilated and withheld proxies with their PSM estimate from the reconstruction. Colored vertical lines indicate the distribution median (solid) and mean (dashed, using Fisher z-transformation). In each of the 20 Monte Carlo iterations for each of the three model priors, we withhold 20\% of proxies. For seasonal proxies, we use annual averages since the seasonal cycle would otherwise artificially inflate the correlations. The correlations in (c--f) are calculated from the proxy estimate averaged over all iterations and model priors. The network-mean correlations in (a,b; i.e., the vertical lines) are statistically significant ($p \ll 0.01$) as determined using the random-phase test~\citep{Ebisuzaki1997}. Correlations are lower before 1600 since fewer proxies are available, particularly corals (see Fig.~S5).}
      \label{fig:independent-proxies}
\end{figure}

\begin{figure}
    \centering
      \includegraphics{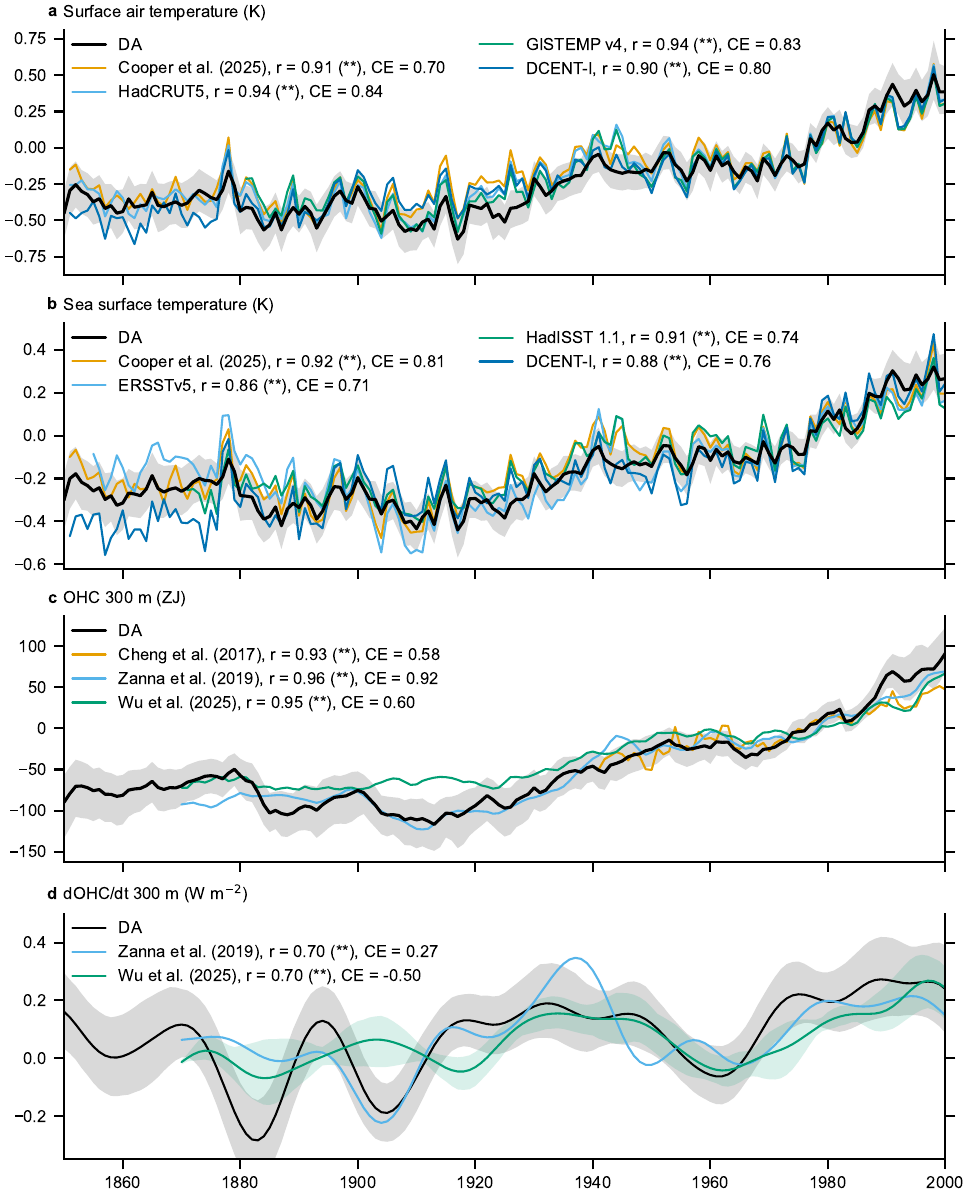}
      \caption{Annual-mean global-means over 1850--2000. All correlations are significant as denoted by (**). Panels (a--c) show anomalies relative to 1961--1990, (d) shows absolute values. Instrumental/observation-based products and their annual correlation with our reconstruction are shown in color. Shading denotes the very likely range. Compared to the instrumental time series, our proxy-based reconstruction generally has less interannual variability due to the smoothing effect of averaging over the ensemble and the low signal-to-noise ratio in proxies. However, the decadal variability is very similar. The instrumental datasets are \citet{Cooper2025}, HadCRUT5~\citep{Morice2021}, GISTEMP v4~\citep{GISTEMPTeam2025}, ERSST v5~\citep{Huang2017}, HadISST 1.1~\citep{Rayner2003}, \citet{Wu2025a}, \citet{Zanna2019}, and \citet{Cheng2017}.}
      \label{fig:timeseries_gm_instrumental}
\end{figure}

\begin{figure}
    \centering
      \includegraphics{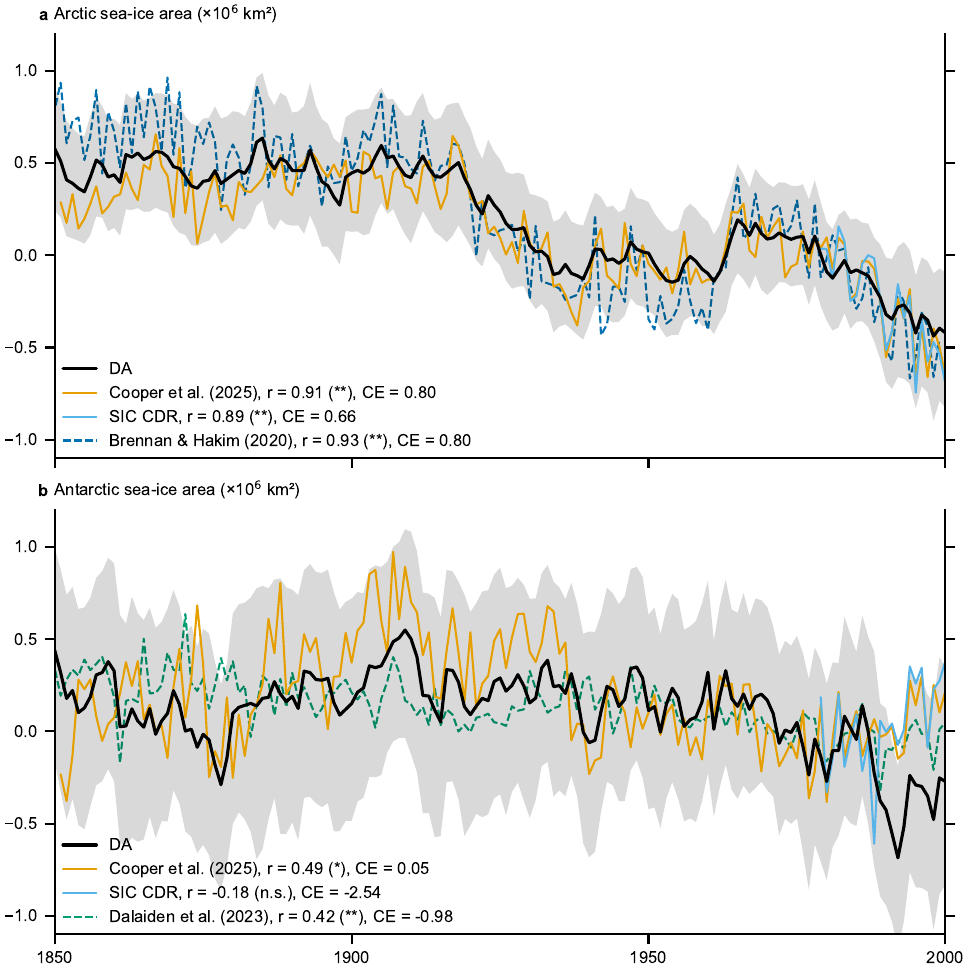}
      \caption{Annual-mean sea-ice area over 1850--2000, expressed as anomalies relative to 1961--1990. Reconstructed Arctic sea ice tracks closely the \citet{Cooper2025} reconstruction, based on DA of instrumental observations, and the NSIDC SIC CDR~\citep[assimilated by \citealp{Cooper2025}]{Meier2024}. We anchor SIC CDR anomalies to \citet{Cooper2025} since it starts in 1979. Antarctic sea ice does not have much interannual skill but has similar decadal variability to \citet{Cooper2025}, although their reconstruction relaxes to zero anomaly in the 1800s due to observation sparsity. We also show two other proxy-based reconstructions. The Arctic sea-ice area from \citet{Brennan2022} agrees well with the other reconstructions, while the Antarctic sea-ice area from \citet{Dalaiden2023} appears to be low-biased. Shading denotes the 5th--95th percentile range.}
      \label{fig:timeseries_siarea_annual}
\end{figure}

\begin{figure}
    \centering
      \includegraphics{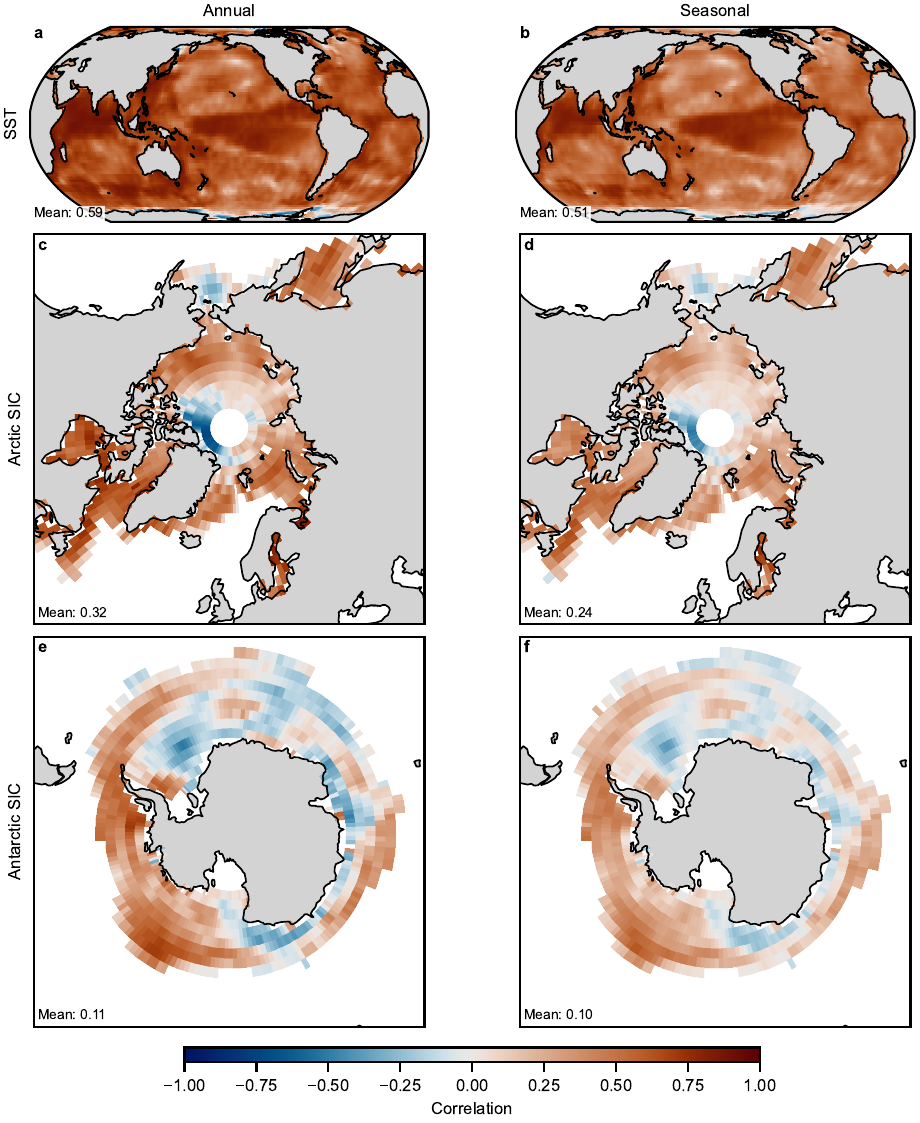}
      \caption{Correlations of seasonal and annual-mean SST and SIC reconstruction with instrumental products (HadISST 1.1 over 1870--2000, \citealp{Rayner2003}; NSIDC SIC CDR V5 over 1979--2000, \citealp{Meier2024}). No satellite SIC data is available for latitudes above 85° due to the pole hole. Lower SIC skill is found in Arctic regions of perennial sea ice and off the coast of East Antarctica and the Weddell Sea. Note that much of the negative correlations for Arctic SIC in (c,d) are in regions of perennial ice cover, for which anomalies are very small.}
      \label{fig:sst_sic_corr_map}
\end{figure}

\begin{figure}
    \centering
      \includegraphics{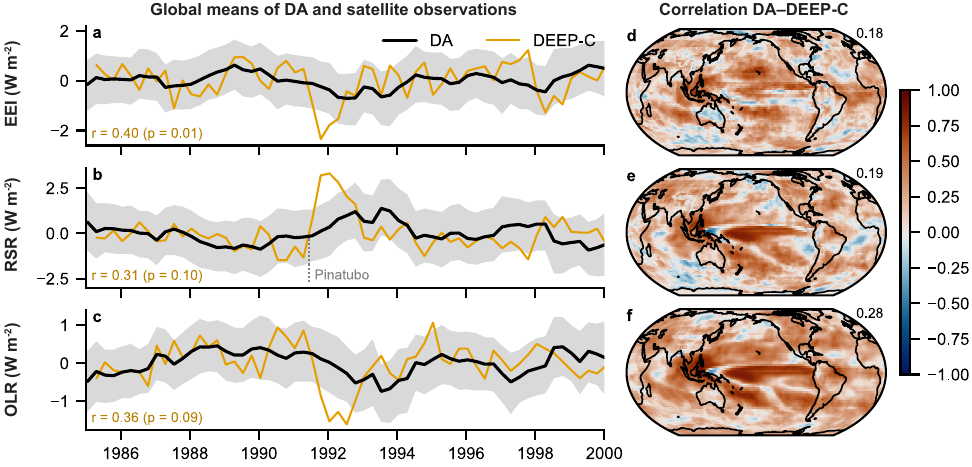}
      \caption{As in Fig.~3 but at seasonal resolution. The reconstruction agrees with DEEP-C at multiannual timescales but does not capture the 1991 Pinatubo eruption well. We note that proxy availability falls off quickly toward 2000 (see Fig.~S5) and that the seasonal correlations are relatively low and not statistically significant due to the short record and high temporal autocorrelation.}
      \label{fig:deepc-panels-seasonal}
\end{figure}


\begin{figure}
    \centering
      \includegraphics{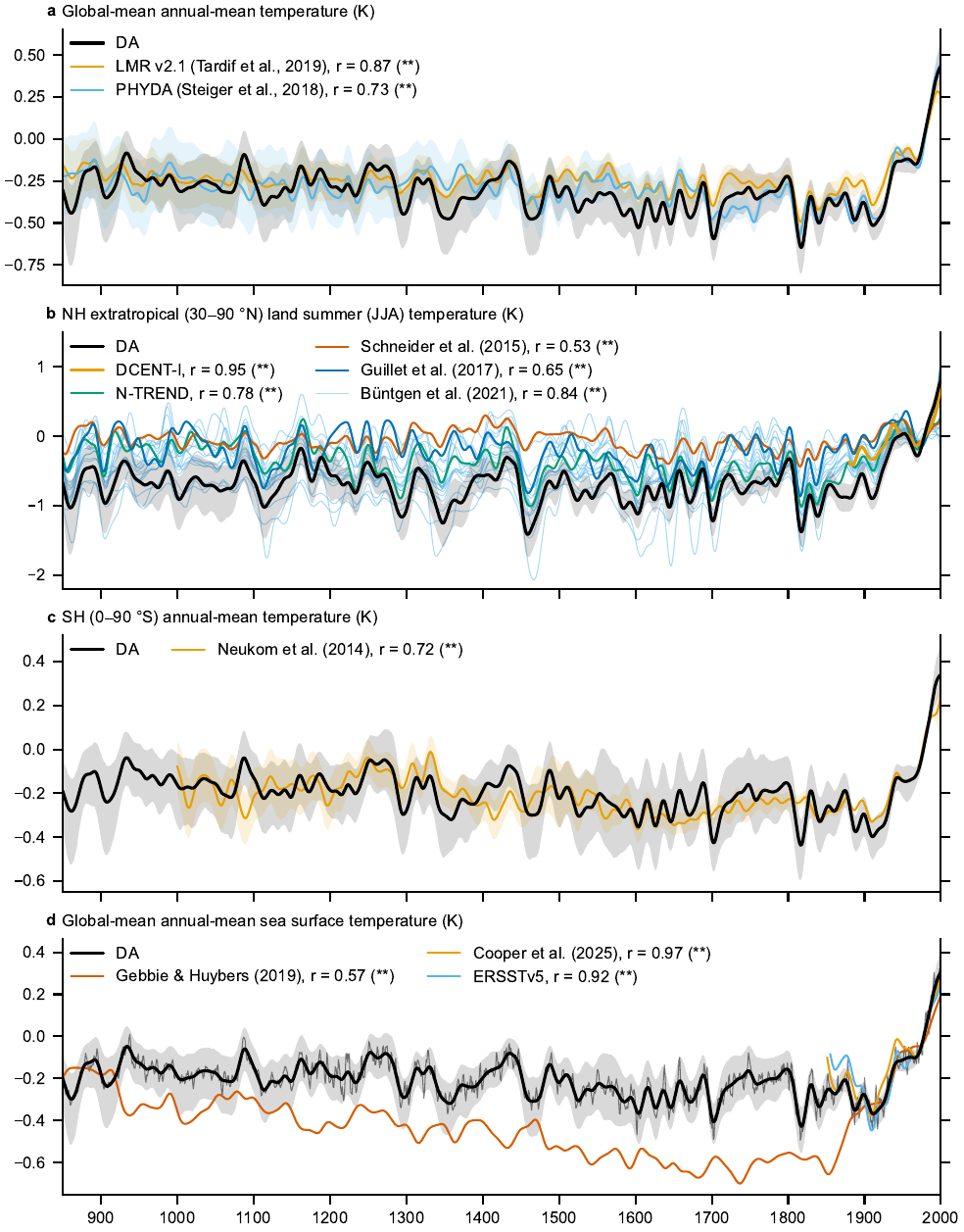}
      \caption{As in Fig.~5 but for large-scale temperatures means, shown as anomalies relative to 1961--1990. All correlations are for 20-yr low-pass-filtered values and are significant as denoted by (**). All comparison datasets except for DCENT-I and Cooper et al. (2025) are proxy reconstructions. (a) For annual-mean, global-mean temperatures, which are also shown in Fig.~5a, we additionally compare to LMR v2.1~\citep{Tardif2019} and PHYDA~\citep{Steiger2018}. (b) For NH extratropical land summer temperature, we compare to DCENT-I~\citep{Chan2025a}, N-TREND~\citep{Wilson2016}, \citet{Schneider2015}, \citet{Guillet2017}, and the multi-method ensemble from \citet{Buntgen2021}. (c) For SH annual-mean temperature, we compare to \citet{Neukom2014}. (d) The SST from \citet{Gebbie2019}, based on SSTs from Ocean2k~\citep{McGregor2015}, are systematically colder over the last millennium and start to warm about 150 years before our reconstructed SSTs (in 1750 rather than 1900). Over 1850--1900, \citet{Gebbie2019}'s warming disagrees with the instrumental datasets from \citet{Cooper2025} and ERSSTv5~\citep{Huang2017}. Possibly this is related to way they prescribe SSTs: they use instrumental winter SSTs when available, but annual-mean SSTs before that, with linear blending over 1870--1950.}
      \label{fig:timeseries_large_scale_temperature_supplement}
\end{figure}


\begin{figure}
    \centering
      \includegraphics{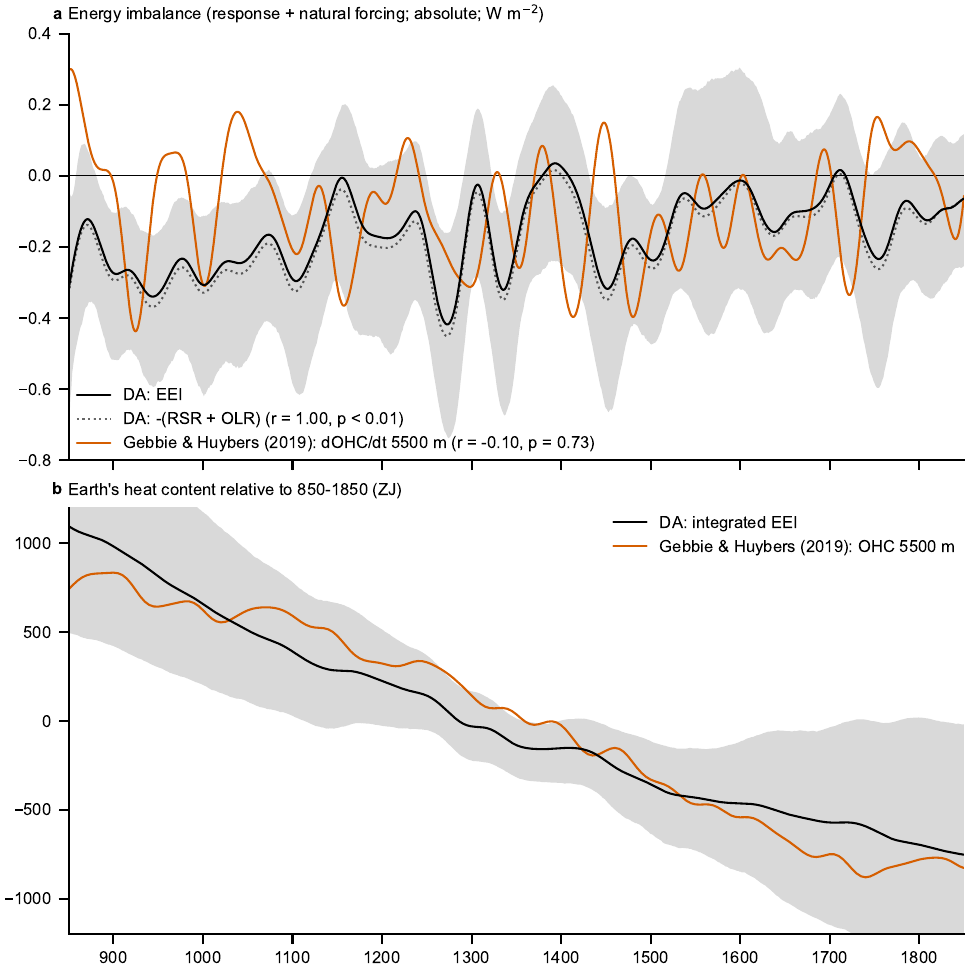}
      \caption{Estimates of Earth's energy budget over 850--1850. We compare to the OHC reconstruction by \citet{Gebbie2019}, who use a constant circulation model to propagate SSTs into the ocean interior. We divide the ocean heat content estimate from \citet{Gebbie2019} by 90\%, which is the fraction of energy imbalance absorbed by the ocean~\citep{Schuckmann2023}. We also rescale dOHC/dt values by Earth's ocean fraction (71\%) to obtain the equivalent EEI. (a) EEI, or rate of change of total energy content. For our reconstruction, we show the directly reconstructed EEI and $-$(RSR $+$ OLR). We compare them to the rate of OHC change from the \citet{Gebbie2019} reconstruction. Correlations are for 50-yr low-pass-filtered values, with p-values are based on the alternative hypothesis that $r > 0$. (b) Earth's heat content, which primarily comprises the full-depth ocean heat content, relative to 850--1850. We convert our reconstructed EEI anomalies to absolute values as described in the main text, then integrate numerically. An offset in the EEI amounts to adding a linear function to the cumulative energy content; small changes in the mean EEI across ensemble members thus result in a large ensemble spread upon integration.}
      \label{fig:timeseries_global_energy_budget}
\end{figure}

\begin{figure}
    \centering
      \includegraphics{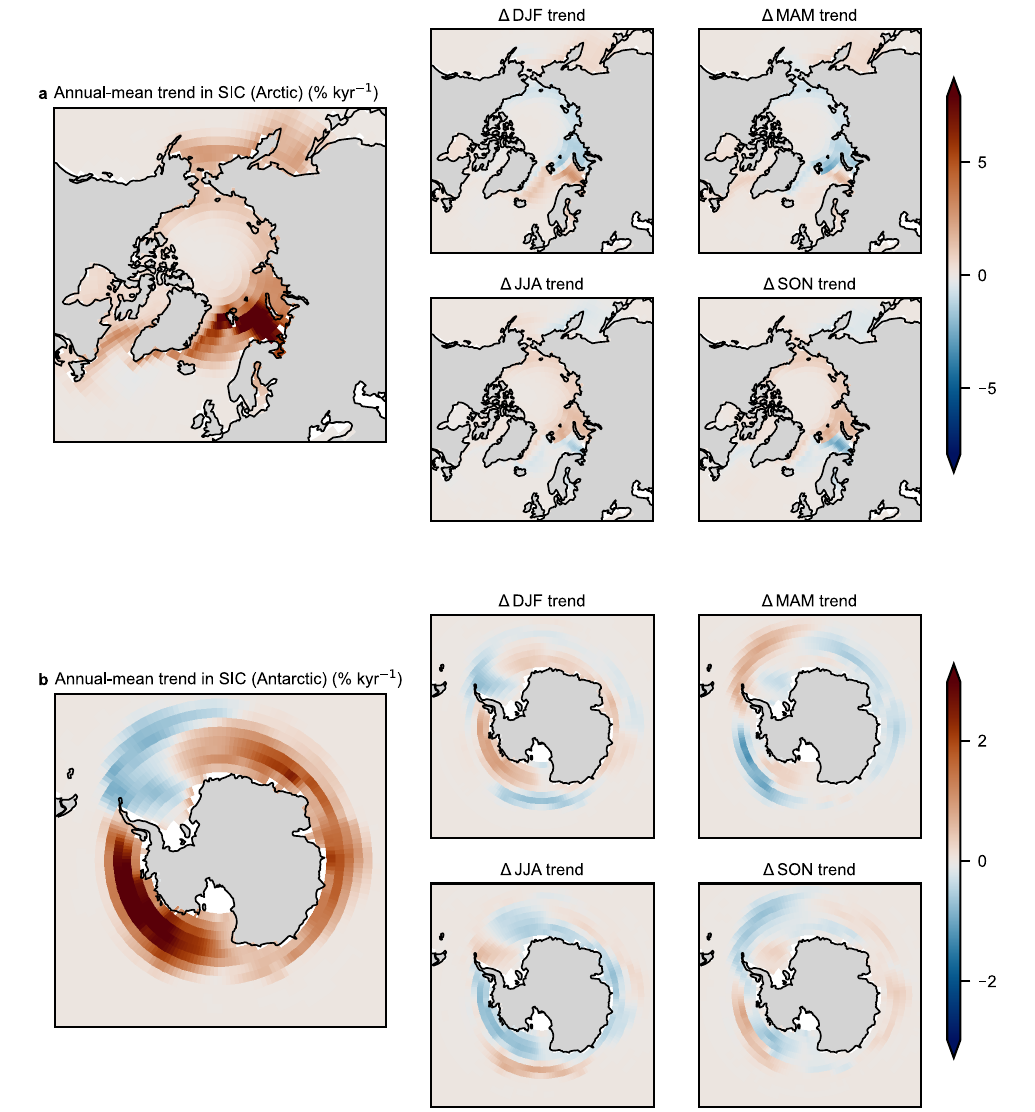}
      \caption{Linear sea-ice trend over 850--1850 in (a) Arctic and (b) Antarctic. (left) Annual-mean trend and (right) departures in seasonal trends from the annual-mean trend. Sea ice generally expanded in all seasons in both hemispheres. The reduction in the Weddell Sea may not be real, considering the low reconstruction skill there (cf. Fig.~S9), but the ensemble spread is also large in this region.}
      \label{fig:lm_trend_map_siconc}
\end{figure}

\begin{figure}
    \centering
      \includegraphics{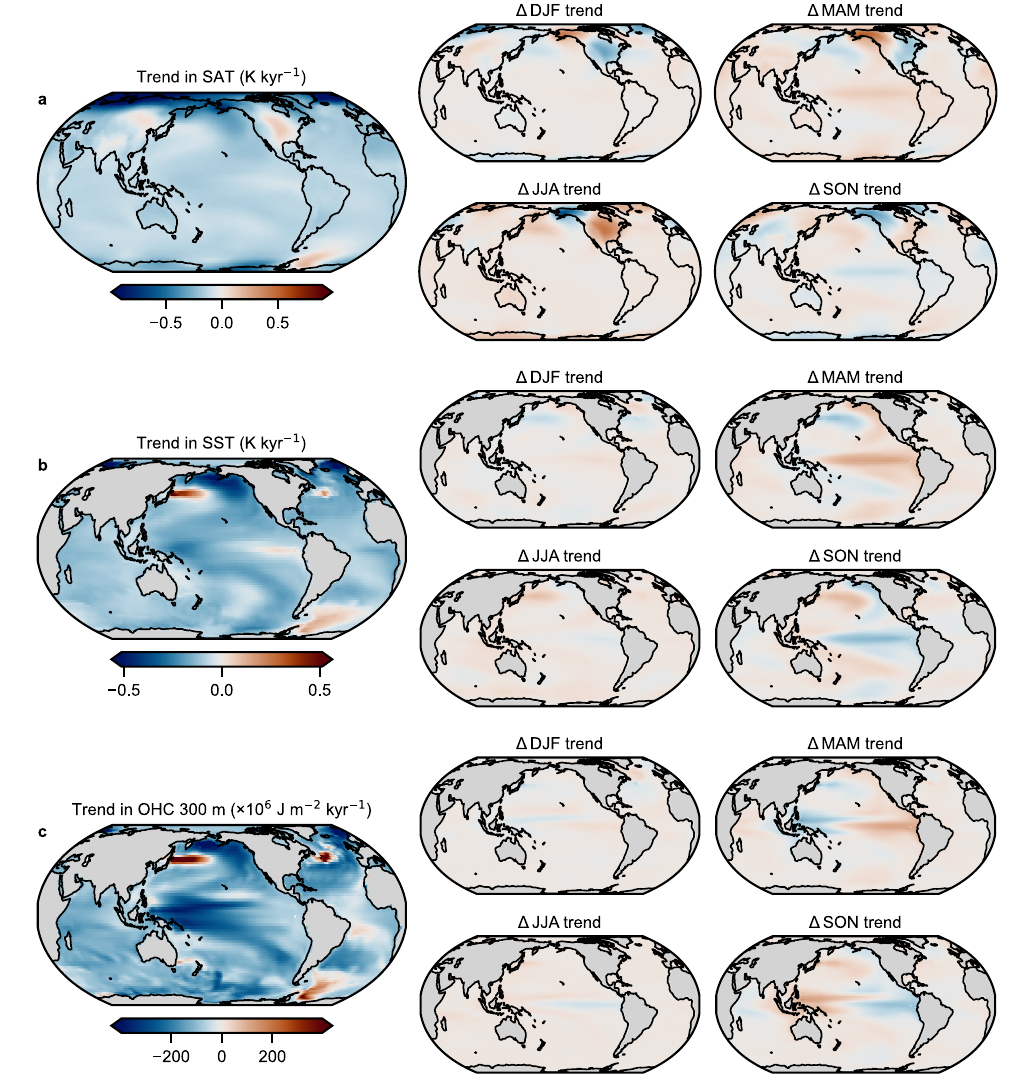}
      \caption{Linear trends in temperature and radiation fields over 850--1850. (left) Annual-mean trend and (right) departures in seasonal trends from the annual-mean trend. Much of the land and oceans cooled, but local warming occurs in North America, parts of Asia, and around the Kuroshio--Oyashio Extension and the Gulf Stream.}
      \label{fig:lm_trend_map}
\end{figure}

\renewcommand{\thefigure}{S\arabic{figure} (cont.)}
\addtocounter{figure}{-1}

\begin{figure}
    \centering
    \includegraphics{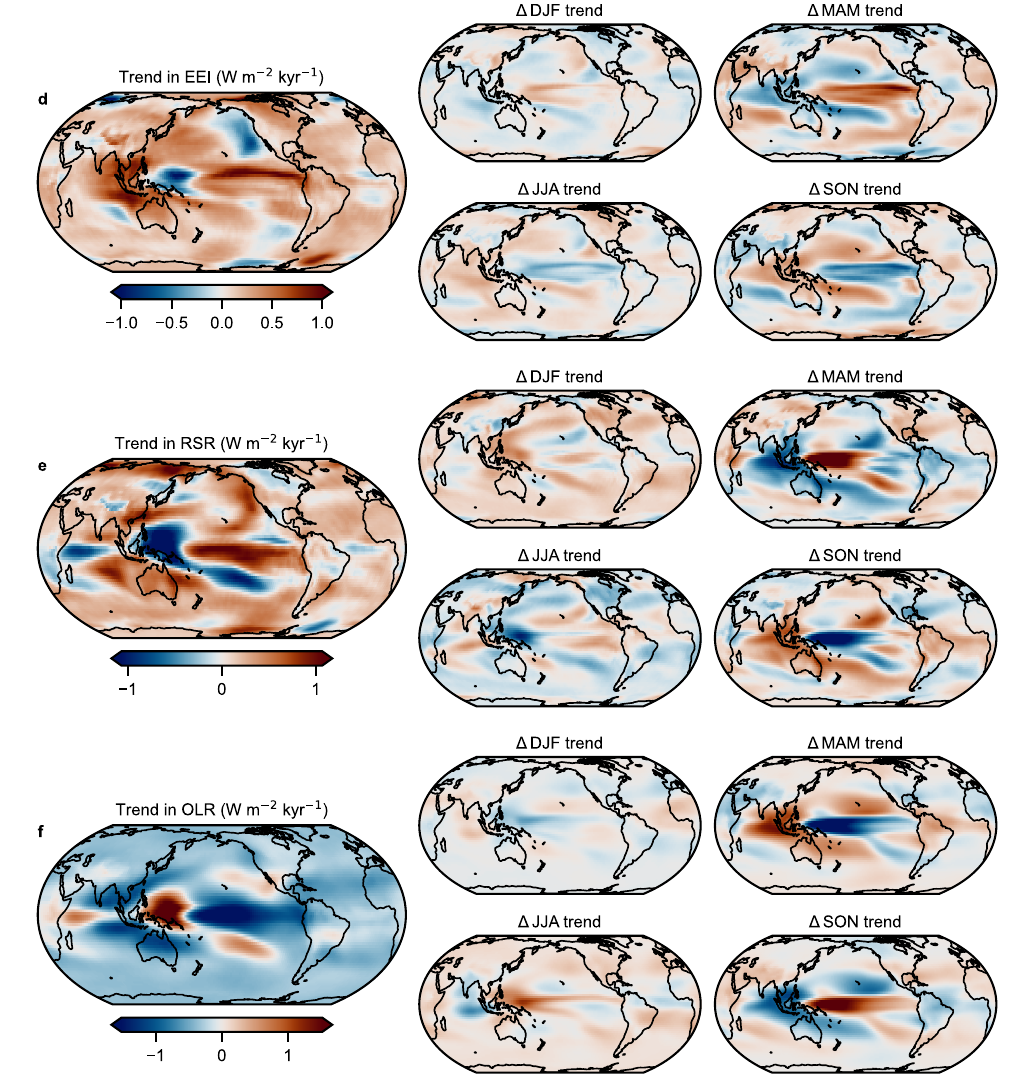}
    \caption{The radiation field trends give insight into the physical processes governing the feedbacks and forcings. The OLR trend suggests a eastward shift of the Indo--Pacific convective region, similar to a positive ENSO phase.}
    \label{fig:lm_trend_map2}
\end{figure}

\renewcommand{\thefigure}{S\arabic{figure}}

\begin{figure}[ht]
      \centering
      \includegraphics{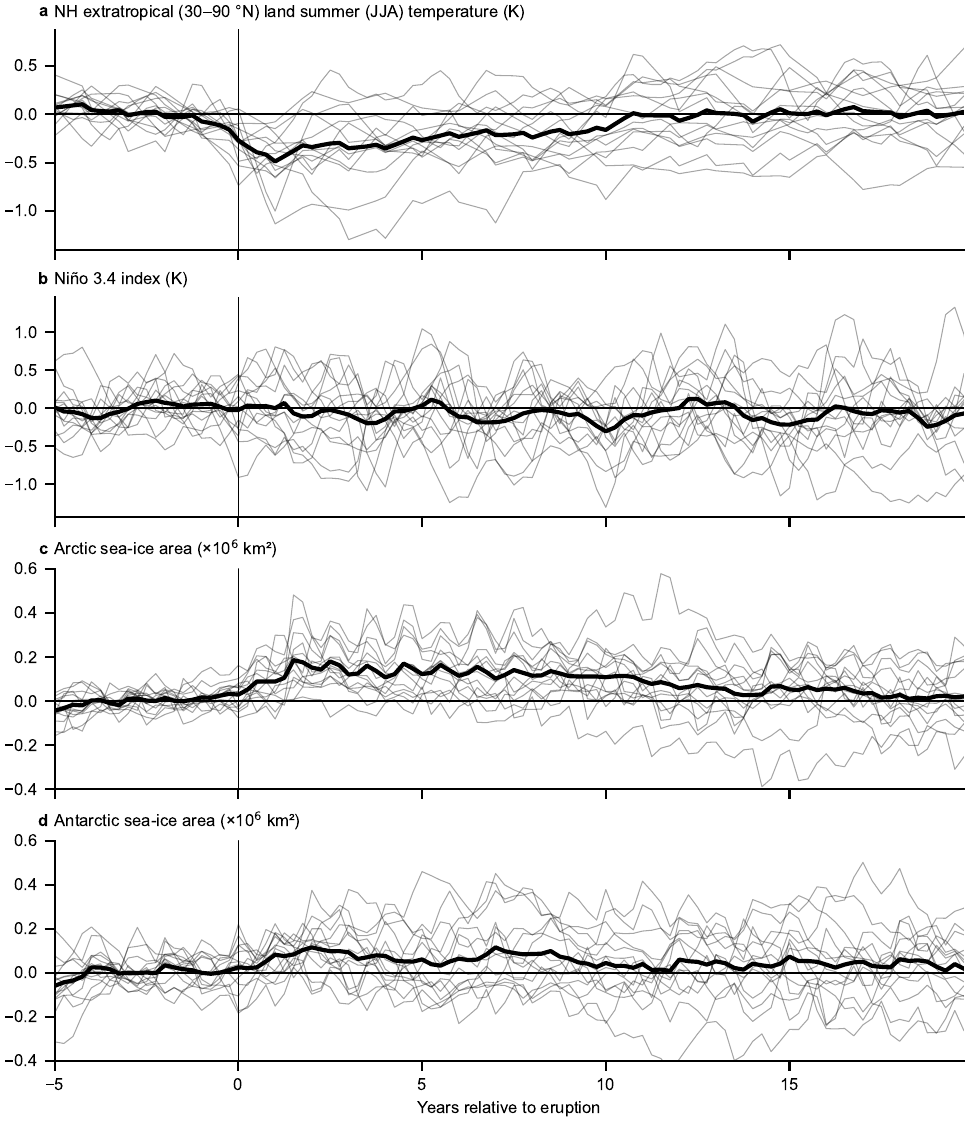}
      \caption{As in Fig.~10 but for miscellaneous fields.}
      \label{fig:composite-volcano-supplement}
\end{figure}


\begin{figure}[ht]
      \centering
      \includegraphics{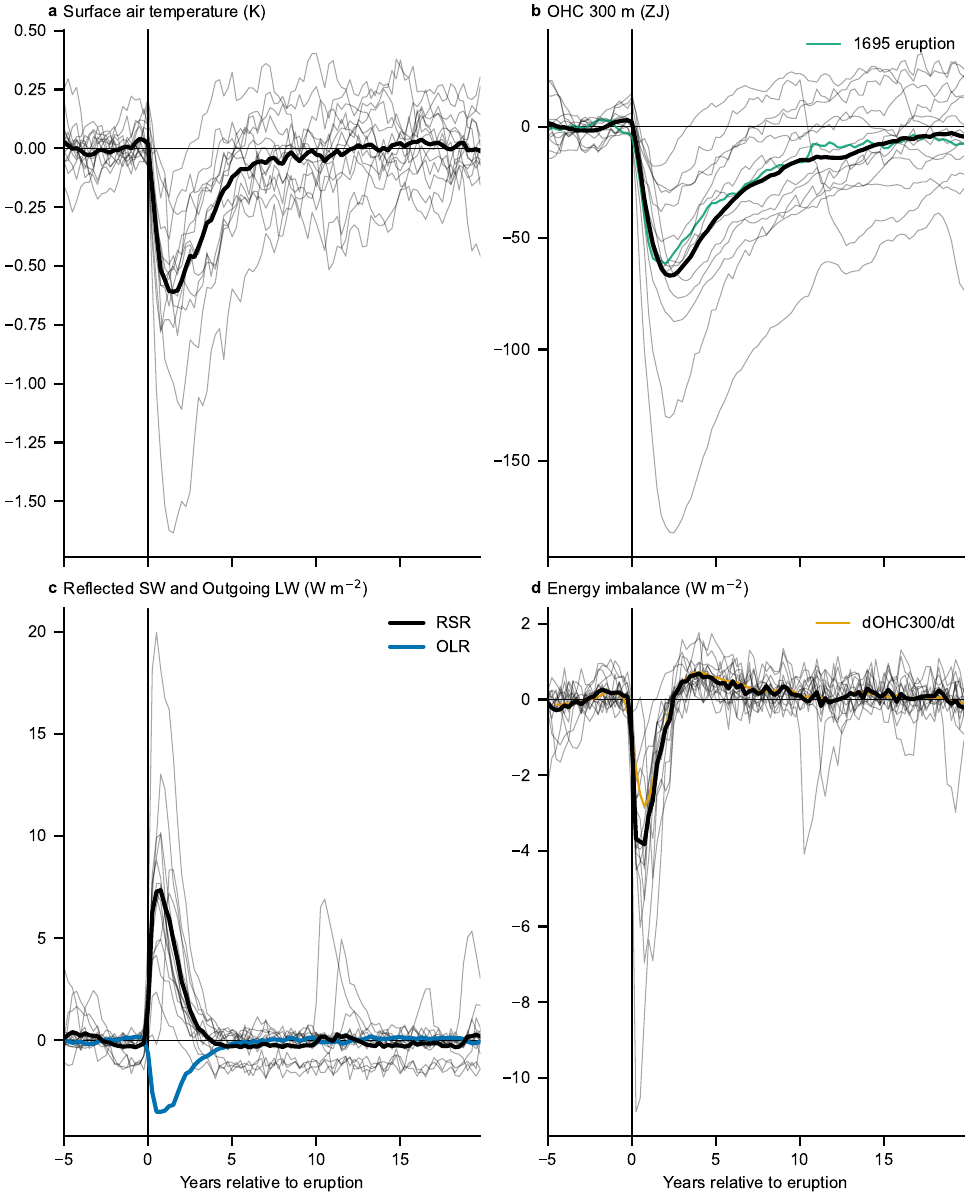}
      \caption{As in Fig.~10 but for CMIP6 last-millennium simulations, averaged over four models (MPI-ESM1-2-LR, CESM2-WACCM-FV2, and MRI-ESM2-0, MIROC-ES2L).}
      \label{fig:composite-volcano-models}
\end{figure}

\begin{figure}[ht]
      \centering
      \includegraphics{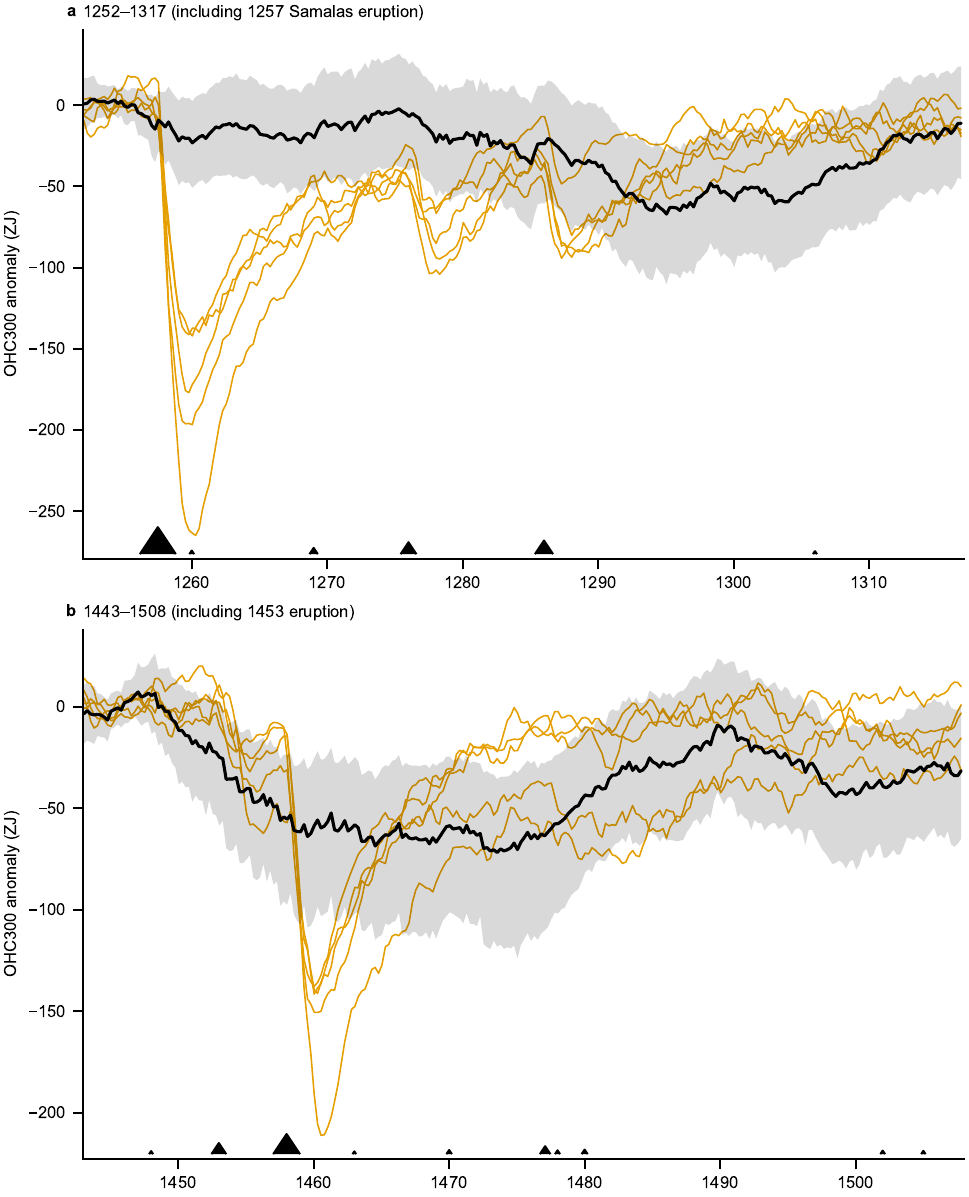}
      \caption{Upper-ocean OHC anomalies over periods with clusters of volcanic eruptions, relative to the first five years of the periods. Shown are the reconstruction (black) with the very likely range and the CMIP6 past1000 simulations from the MPI, CESM, MRI, and MIROC models (yellow). The carets indicate volcanic eruptions, scaled by their volcanic stratospheric sulfur injection. (a) During 1103-1168, four eruptions occur, leading to an OHC loss of $35\,\mathrm{ZJ}$. (b) During 1252-1317, six eruptions occur. The 1257 Samalas eruption is barely evident in our reconstruction (cf. \citealp{Zhu2020}), but the compound effect of the 1276 and 1286 eruptions, thought to have initiated the Little Ice Age, leads to large OHC loss. Figure is continued on the next page.}
\end{figure}

\renewcommand{\thefigure}{S\arabic{figure} (cont.)}
\addtocounter{figure}{-1}

\begin{figure}[ht]
      \centering
      \includegraphics{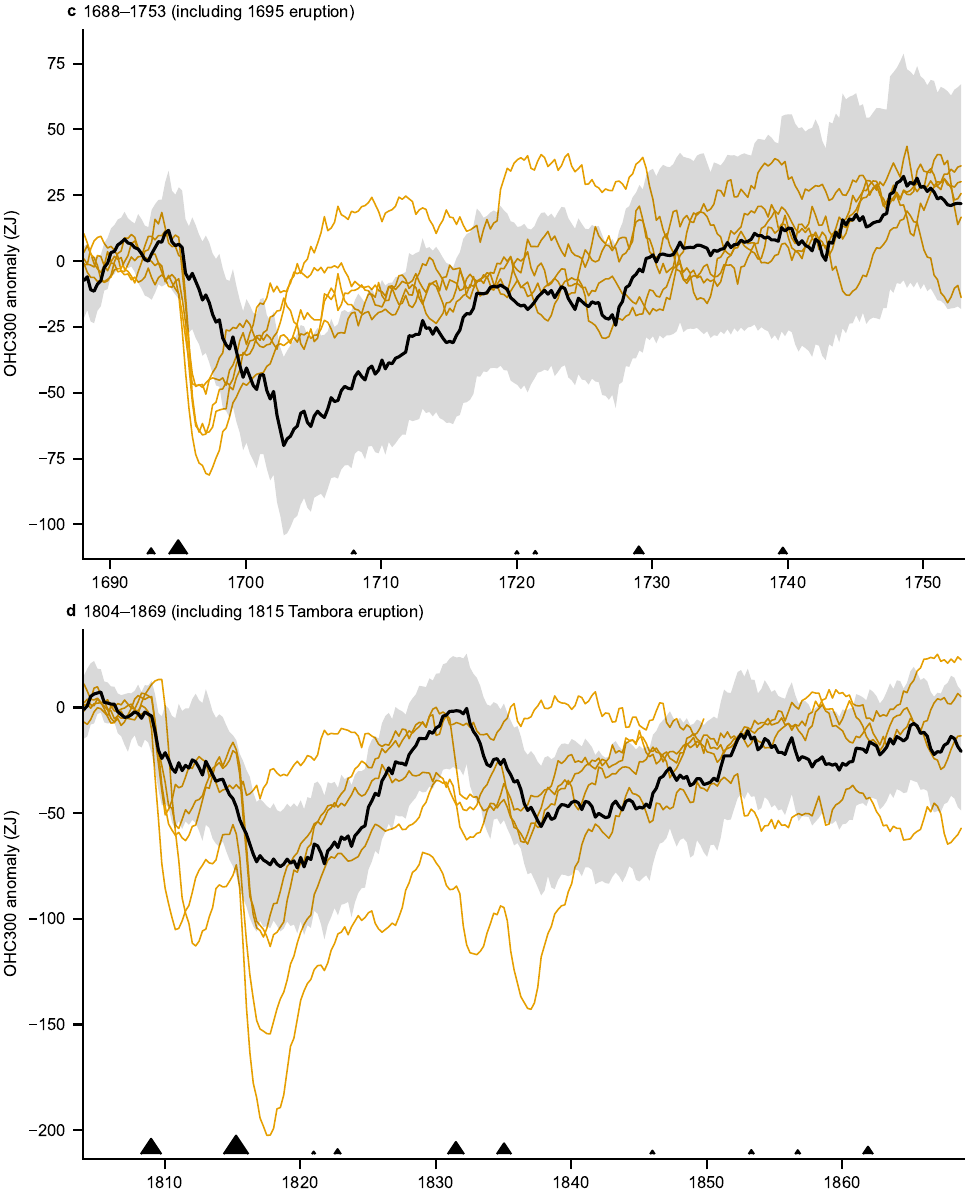}
      \caption{(c) During 1688--1753, seven volcanic eruptions occur, leading to an OHC loss of $55\,\mathrm{ZJ}$ over 10 years and a prolonged recovery period of 40 years. (d) During 1804--1869, ten eruptions occur, including an unidentified eruption in 1809 and the 1815 Tambora eruption. This leads to an OHC loss of $58\,\mathrm{ZJ}$ that recovers over 10 years, followed by an OHC loss of similar magnitude that recovers again over 40 years. Also see \citet{Bronnimann2019} for this period.}
\end{figure}

\renewcommand{\thefigure}{S\arabic{figure}}

\begin{figure}
  \centering
      \includegraphics{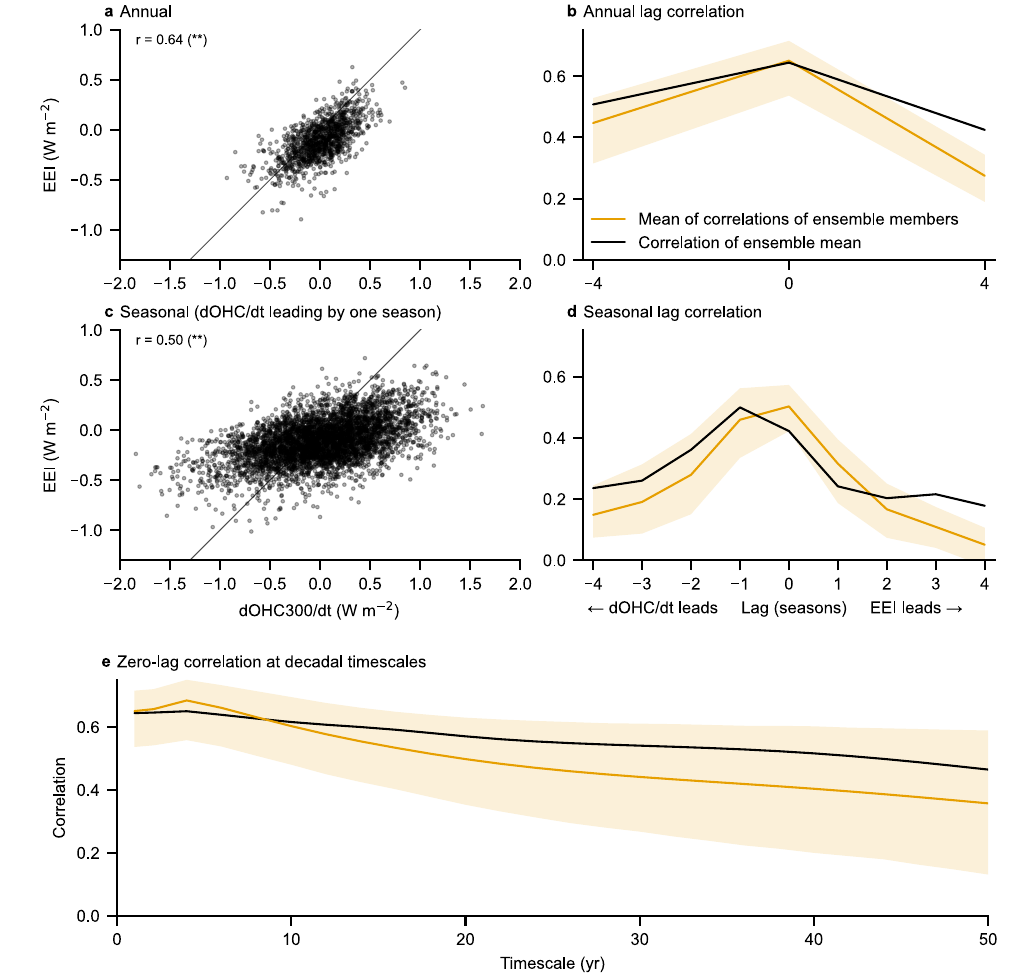}
      \caption{Comparison of global-mean EEI and dOHC300/dt. We rescale dOHC300/dt by Earth's ocean fraction (71\%) to obtain the equivalent EEI. Shading denotes the 5th--95th percentile range. (a,b) At annual timescales, they have a correlation of 0.60 and a similar standard deviation of around 0.25 W m$^{-2}$. (c,d) At seasonal timescales, dOHC300/dt leads by one season as shown by the maximum lag correlation of 0.50. The scatter plot (c) considers this lag. However, the standard deviations differ: 0.42 W m$^{-2}$ for dOHC300/dt, and 0.27 W m$^{-2}$ for EEI. (e) Zero-lag correlation at decadal timescales, determined by low-pass-filtering both time series. The correlation peaks at 4 years, then slowly decays for longer timescales.}
      \label{fig:ohc_eei}
\end{figure}




\FloatBarrier






\narrowlayout

\bibliographystyle{ametsocV6}
\bibliography{references}